\documentclass[prb,amsmath,amssymb,twocolumn,lengthcheck,showpacs,floatfix,groupedaddress,superscriptaddress,longbibliography]{revtex4-1}
\usepackage{graphicx}
\usepackage[english]{babel}
\usepackage{times}
\usepackage{dcolumn}
\usepackage[usenames,dvipsnames]{color}
\usepackage{units}
\usepackage{bm}

\begin{document}

\title{Numerical sub-gap spectroscopy of double quantum dots coupled to
superconductors}

\author{Rok \v{Z}itko}

\affiliation{Jo\v{z}ef Stefan Institute, Jamova 39, SI-1000 Ljubljana, Slovenia}
\affiliation{Faculty  of Mathematics and Physics, University of Ljubljana, 
Jadranska 19, SI-1000 Ljubljana, Slovenia}

\date{\today}

\begin{abstract}
Double quantum dot nanostructures embedded between two superconducting
leads or in a superconducting ring have complex excitation spectra
inside the gap which reveal the competition between different
many-body phenomena. We study the corresponding two-impurity Anderson
model using the non-perturbative numerical renormalization group (NRG)
technique and identify the characteristic features in the spectral
function in various parameter regimes. At half-filling, the system
always has a singlet ground state. For large hybridization, we observe
an inversion of excited inter-dot triplet and singlet states due to
the level-repulsion between two sub-gap singlet states. The Shiba
doublet states split in two cases: a) at non-zero superconducting
phase difference and b) away from half-filling. The most complex
structure of sub-gap states is found when one or both dots are in the
valence fluctuation regime. Doublet splitting can lead to a
parity-changing quantum phase transition to a doublet ground state in
some circumstances. In such cases, we observe very different spectral
weights for the transitions to singlet or triplet excited Shiba
states: the triplet state is best visible on the valence-fluctuating
dot, while the singlets are more pronounced on the half-filled dot.
\end{abstract}


\maketitle

\newcommand{\vc}[1]{{\mathbf{#1}}}
\renewcommand{\Im}{\mathrm{Im}}
\renewcommand{\Re}{\mathrm{Re}}
\newcommand{\expv}[1]{\langle #1 \rangle}
\newcommand{\corr}[1]{\langle\langle #1 \rangle\rangle}
\newcommand{\beq}[1]{\begin{equation} #1 \end{equation}}
\newcommand{\ket}[1]{|#1\rangle}

\section{Introduction}

The advances in fabrication and characterization of small electronic
devices have enabled new ways to perform spectroscopy of strongly
correlated electron systems. A prominent example is the tunneling
spectroscopy \cite{giaever1960,giaever1974,pan1999zn,balatsky2006} of
interacting quantum dots coupled to superconducting contacts as well
as magnetic adatoms on superconducting surfaces
\cite{ji2008,iavarone2010,ji2010,franke2011,franke2015}, where the
competition between the Kondo screening \cite{hewson} and
superconducting correlations can be studied in exquisite detail
because it engenders spectroscopically sharp resonances known as the
Andreev bound states (or Yu-Shiba-Rusinov states or simply Shiba
states)
\cite{shiba1968,sakurai1970,zmha,choi2004josephson,bauer2007,karrasch2008,
moca2008,meng2009,hybrid2010,maurand2012}. At low enough temperatures
even fine details can be resolved
\cite{pillet2010,franke2011,lee2014,franke2015} and in some cases
quantitative agreement is found between the experiment and accurate
theoretical modelling using non-perturbative numerical techniques
\cite{pillet2013}.

The most thoroughly studied problem is that of a single quantum dot in
the deep Kondo limit where charge fluctuations are frozen out and the
device behaves essentially as a local moment characterized by a single
bare parameter, the Kondo exchange coupling $J_K$, and at low
temperatures by a single scaling parameter, the Kondo temperature
$T_K$. When such a quantum dot (QD) is coupled to a superconducting
host with energy gap $\Delta$, the ground state of the system depends
on the ratio $T_K/\Delta$. For $T_K \ll \Delta$, the Kondo screening
is incomplete due to a lack of quasiparticle states at low energy
scales, thus the system behaves as a free local moment decoupled from
the BCS bath, which is an overall spin doublet state. For $T_K \gg
\Delta$, the Kondo screening is fully completed on energy scales much
above the onset of pairing; the superconducting state is thus formed
out of the local Fermi liquid state resulting from the Kondo effect,
and is an overall spin singlet. For $T_K \sim \Delta$ there is a
quantum phase transition between these two different ground states. In
this parameter range, the many-particle spectrum usually includes at
least three states (one singlet and one doublet) below the onset of
the continuum of quasiparticle states at $\Delta$. The singlet-doublet
excitations are spectroscopically visible as resonance pairs at
$\omega = \pm (E_S-E_D)$. Additional complexity in the problem is
brought about by non-zero difference of the superconducting phases
that leads to Josephson current. The direction of the current depends
on the sub-gap states and the singlet-doublet transitions can be
directly related to the physics of $0$ or $\pi$ junction behavior
\cite{hybrid2010,rodero2011,maurand2012}.

Recently, this research direction has intensified with a focus on more
complex systems. Double quantum dots (DQD) are the minimal non-trivial
systems that capture the essence of extended strongly-correlated
materials described by lattice models
\cite{georges1999,aguado2000,aono2001,lopez2002,simon2005,revisited}.
Coupled to superconducting leads, DQDs can serve to explore the
competing effects of exchange coupling, charge fluctuations, Kondo
screening, and superconductivity
\cite{bergeret2006,zitko2010,dqdscaniso,yao2014}. In this work, DQDs
are studied using a reliable non-perturbative numerical
renormalization group technique in a wide parameter range with the
goal of identifying the characteristic behavior of the sub-gap Shiba
states in various regimes, both their positions and spectral weights.
The states are analyzed in terms of the eigenstates of the
superconducting atomic (wide-gap) limit; the deviations between this
simple theory and the full numerical calculations are pointed out. In
the context of superconducting rings, an important question is the
variation of the sub-gap state energies as a function of the flux.
Strong flux dependence implies sizeable particle exchange between the
superconducting contacts on either side of the DQD structure, thus the
regimes of enhanced valence fluctuations away from the integer filling
limit are particularly interesting.

This work is structured as follows. In Sec.~\ref{sec2} we introduce
the model, the numerical method, and the wide-gap approximation. The
presentation of the results in Sec.~\ref{sec3} is divided in three
subsections: the left-right symmetric case at A) half-filling and B)
away from half-filling, and C) the fully generic case with unequal
quantum dots. Sec.~\ref{sec4} contains a short discussion of the
two-impurity Kondo quantum quantum phase transition in the presence of
superconductivity.

\section{Model and method}
\label{sec2}

\begin{figure}
\centering 
\includegraphics[width=0.3\textwidth]{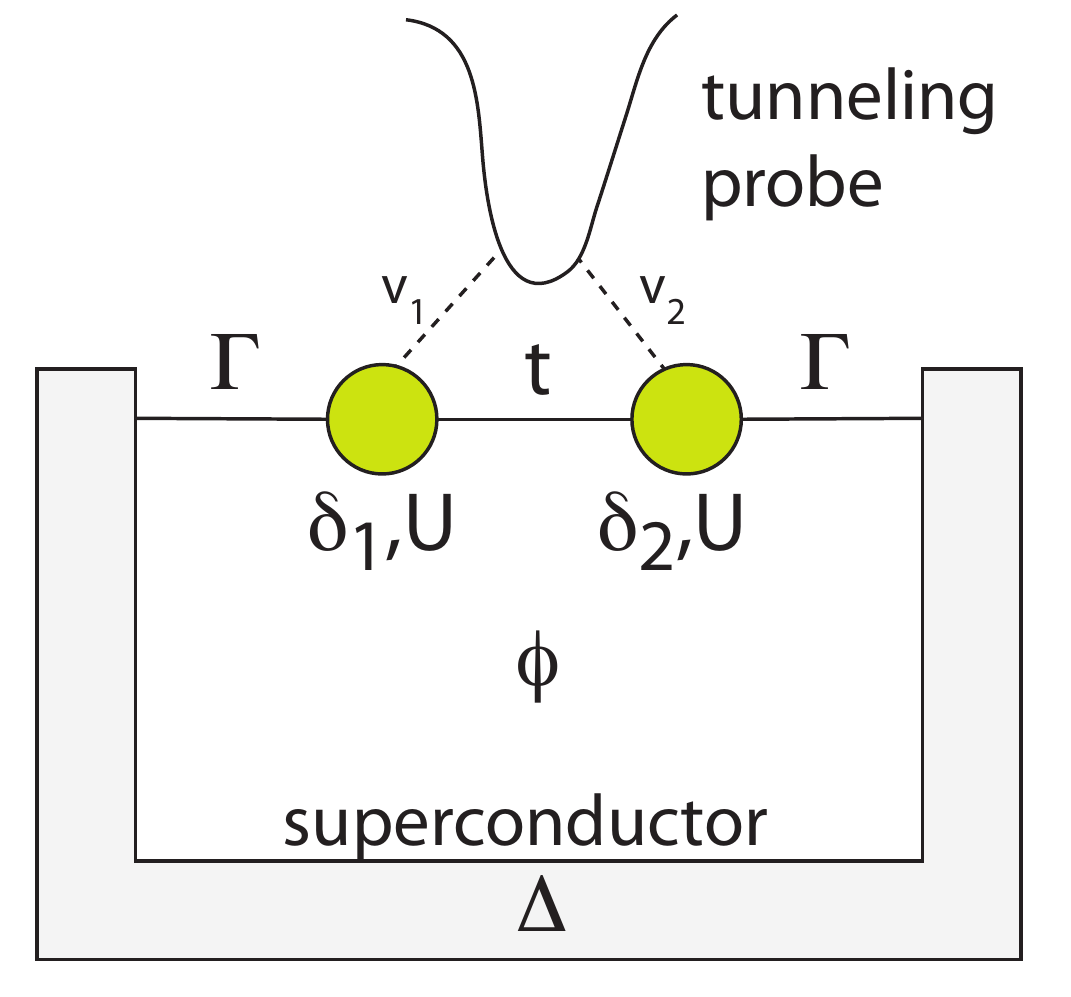}
\caption{(Color online) Schematic representation of the system: double
quantum dot coupled to superconducting leads forming a ring pierced by
the magnetic flux. The tunneling probe is very weakly coupled to
either or both quantum dots.} \label{fig0}
\end{figure}

The DQD system is schematically represented in
Fig.~\ref{fig0}. We consider two quantum dots ($i=1,2$) described by
the Anderson impurity model:
\begin{equation}
\begin{split}
H_i &= \epsilon_i n_i + U n_{i\uparrow} n_{i\downarrow} \\
&= \delta_i (n_i-1) + \frac{U}{2} (n_i-1)^2 + \text{const.}
\end{split}
\end{equation}
Here $n_i=n_{i\uparrow} + n_{i\downarrow}$ with $n_{i\sigma} =
d^\dag_{i\sigma} d_{i\sigma}$ the QD occupancy, $U$ is the Hubbard
charge repulsion, and $\delta_i$ is the impurity energy level measured
with respect to the particle-hole symmetric point (half filling),
i.e., $\delta_i=\epsilon_i+U/2$.
Each dot is coupled to a separate superconducting bath
\begin{equation}
H_{\mathrm{SC},i} = \sum_{k\sigma} \epsilon_{k,i} c^\dag_{k\sigma,i}
c_{k\sigma,i} - \sum_{k\sigma} \Delta_i e^{i \phi_i}
c^\dag_{k,\uparrow,i} c_{-k,\downarrow,i} + \text{H.c.},
\end{equation}
where $\epsilon_{k,i}$ is the band dispersion, $\Delta_i$ the gap
parameter and $\phi_i$ the superconducting phase. The difference of
phases will be denoted as $\phi=\phi_1-\phi_2$. The coupling terms are
\begin{equation}
H_{\mathrm{c},i} = \sum_{k\sigma} V_{k,i} d^\dag_{i\sigma} c_{k\sigma,i} 
+ \text{H.c.},
\end{equation}
where $V_{k,i}$ is the hopping parameter. We assume that the two bands
are flat: in the absence of superconductivity they have constant
density of states $\rho=1/2D$ on the interval $[-D:D]$, thus $D$ is
one half of the bandwidth ($D$ will henceforth be used as the energy
unit, $D=1$). The hybridization strengths are characterized by
functions 
\begin{equation}
\Gamma_i(\omega)=\sum_k |V_{k,i}|^2 \delta(\omega-\epsilon_{k,i})
=\pi \rho V_i^2,
\end{equation}
which will be taken as constants.  The two dots are interconnected by
a tunneling term
\begin{equation}
H_{12} = -t \sum_\sigma d^\dag_{1\sigma} d_{2\sigma} + \text{H.c.}
\end{equation}
Furthermore, two nearby quantum dots are typically also coupled
capacitively
\cite{galpin2005,galpin2006b,mravlje2006,okazaki2011,nishikawa2013,amasha2013,PhysRevB.90.035119,
keller2014,filippone2014} leading to an interdot Coulomb interaction
term of the form 
\begin{equation*}
H_V = V n_1 n_2.
\end{equation*}

Most results presented in this work are calculated for 
\begin{equation}
U/D=0.27,\,
\Gamma/D=0.02,\,
\Delta/D=0.01,\,
V/D=0,
\end{equation}
while $\delta_i$ and $t$ will be variable. Exceptions to this
parameter setwill be clearly pointed out. For reference, the results
for the normal state ($\Delta=0$) are presented as Supplementary
information \cite{suppl}.

Since $U/\pi \Gamma \approx 4.3$, the dots are in the Kondo regime
near half-filling. This choice of parameters is experimentally
realistic. The Kondo temperature (according to Wilson's definition) of
each separate QD is given by \cite{krishna1980a,hewson}
\begin{equation}
\label{tk}
T_K=0.182 U \sqrt{\rho J_K} \exp\left[ - \frac{1}{\rho J_K} \right],
\end{equation}
where
\begin{equation}
\rho J_K = \frac{2\Gamma}{\pi} \left(
\frac{1}{U/2-\delta} + \frac{1}{U/2+\delta}
\right).
\end{equation}
At $\delta=0$ this gives $T_K \approx 10^{-4} \ll \Delta$. The ground
state of a single quantum dot would be a doublet, with the excited
singlet state at energy $\approx 0.6\Delta$ \cite{suppl}.

The method of choice for this class of problems is the numerical
renormalization group (NRG)
\cite{wilson1975,krishna1980a,krishna1980b,satori1992,sakai1993,
yoshioka1998,yoshioka2000,oguri2004josephson,bauer2007,hecht2008,bulla2008}
which is able to quantitatively reproduce the experimental results
\cite{Lim:2008bc,Deacon:2010jn,MartinRodero:2012fd,pillet2013,Kumar:2014cq},
but also provides additional detailed information about the system
properties which are difficult or impossible to measure. The
calculations were performed in the low-temperature limit with the
discretization parameter $\Lambda=4$, keeping $5000$ states at each
diagonalization step, with $N_z=2$ discretization meshes
\cite{oliveira1994,silva1996,paula1999,campo2005}. The only symmetry
in the problem is the SU(2) spin symmetry. The spectra are computed
using the density-matrix NRG algorithm \cite{hofstetter2000} with the
$N/N+1$ patching approach and with a mixed broadening scheme: for
$|\omega|>\Delta$ log-Gaussian broadening kernel with $\alpha=0.6$ is
used, while for $|\omega|<\Delta$ Gaussian broadening with
$\sigma=10^{-3}$ is used (this latter choice mimics non-zero lifetime
of resonances due to intrinsic relaxation processes at non-zero
temperatures \cite{franke2015} and due to the presence of the
tunneling probe \cite{zitko2015shiba}).

The spectral function of dot $i$ is defined as
\begin{equation}
A_i(\omega) = -\frac{1}{\pi} \Im G_i(\omega+i\delta),
\end{equation}
where $G_i(z) = \corr{d_{i\sigma} ; d_{i\sigma}^\dag }_z$ is the local
Green's function. Spectra are measurable by tunneling spectroscopy
using weakly coupled additional normal-state or superconducting probes
at finite bias voltage \cite{pillet2010,franke2011}. Finite bias
drives the system out of equilibrium. In this work, it is assumed that
the coupling is sufficiently weak ($\Gamma_N/\Gamma_S \ll 1$) that the
non-equilibrium effects may be neglected and that the transport
properties may be approximated using equilibrium spectral functions
(i.e., the system remains in the linear response regime). For stronger
coupling of the probe, different transport regimes become dominant
\cite{koerting2010,yamada2011,koga2013,franke2015}.

Another quantity of interest is the interdot spectral function
\begin{equation}
A_{12}(\omega) = -\frac{1}{\pi} \Im G_{12}(\omega+i\delta),
\end{equation}
where $G_{12}(z) = \corr{ d_{1\sigma}; d^\dag_{2\sigma} }_z$,
which quantifies interdot correlations of the excitations. If the
tunneling probe is coupled to both dots, the resulting spectrum
is proportional to
\begin{equation}
\begin{split}
&\Im\Bigl[|v_1|^2 G_1(z) + |v_2|^2 G_2(z) + \\
&\quad\quad\quad v_1^* v_2 G_{12}(z)
+ v_1 v_2^* G_{21}(z) \Bigr]_{z=\omega+i\delta},
\end{split}
\end{equation}
providing access to the information contained in $G_{12}$. Here
$v_1$ and $v_2$ are the tunnel couplings of the probe to either QD.
Such situation occurs in the experiments described in
Refs.~\onlinecite{pillet2010} and \onlinecite{pillet2013}, where the
segment of the carbon nanotube actually hosts two QDs and the probe is
attached near the center of the tube.

At qualitative level the structure of the sub-gap many-particle
spectra can be described in the superconducting atomic limit (also
known as the wide-gap limit)
\cite{PhysRevB.62.6687,vecino2003,bauer2007,meng2009}. It consists of
taking $\Delta$ to infinity at constant width of the conduction band.
This limit is different from the related zero-bandwidth limit
\cite{vecino2003,bergeret2007}, where $\Delta$ is kept constant and
the conduction-band width is reduced to zero (i.e., the band is
replaced by a single representative site). The wide-gap limit
eliminates the continuum of the quasiparticle levels above the gap and
the sole remaining effect of each lead is the proximity pairing term
of the form $\Gamma d^\dag_{i\uparrow} d^\dag_{i\downarrow} +
\text{H.c.}$. We thus obtain a two-site discrete model with the
following Hamiltonian:
\begin{equation}
\begin{split}
H &= \sum_i \delta_i (n_i-1) + \frac{U}{2} \sum_i (n_i-1)^2 \\
&+ \sum_i \left( \Gamma e^{i\phi_i} d^\dag_{i\uparrow} d^\dag_{i\downarrow} +
\text{H.c.} \right) - t \sum_\sigma \left(d^\dag_{1\sigma} d_{2\sigma}
+ \text{H.c.} \right),
\end{split}
\label{wg}
\end{equation}
which can be easily diagonalized. The Kondo correlations due to
quasiparticle states above the superconducting gap are fully excluded
in this description, thus the results are quantitatively quite
different compared to the accurate NRG calculations, especially when
the role of the Kondo effect is increased either by increasing
$\Gamma$ or by decreasing $\Delta$. Furthermore, this method tends to
overemphasize the BCS character of singlet states at the expense of
the Kondo character in the exact solution. Nevertheless, the
consecutive order of the lowest levels mostly match the observed
sub-gap Shiba states. A systematic method for improving perturbatively
upon the superconducting atomic limit has been described in
Ref.~\onlinecite{meng2009}. Some further details are in the
Supplementary information \cite{suppl}.

\section{Results}
\label{sec3}

\subsection{Left-right symmetric case, half-filling}

\begin{figure}
\centering
\includegraphics[width=0.45\textwidth]{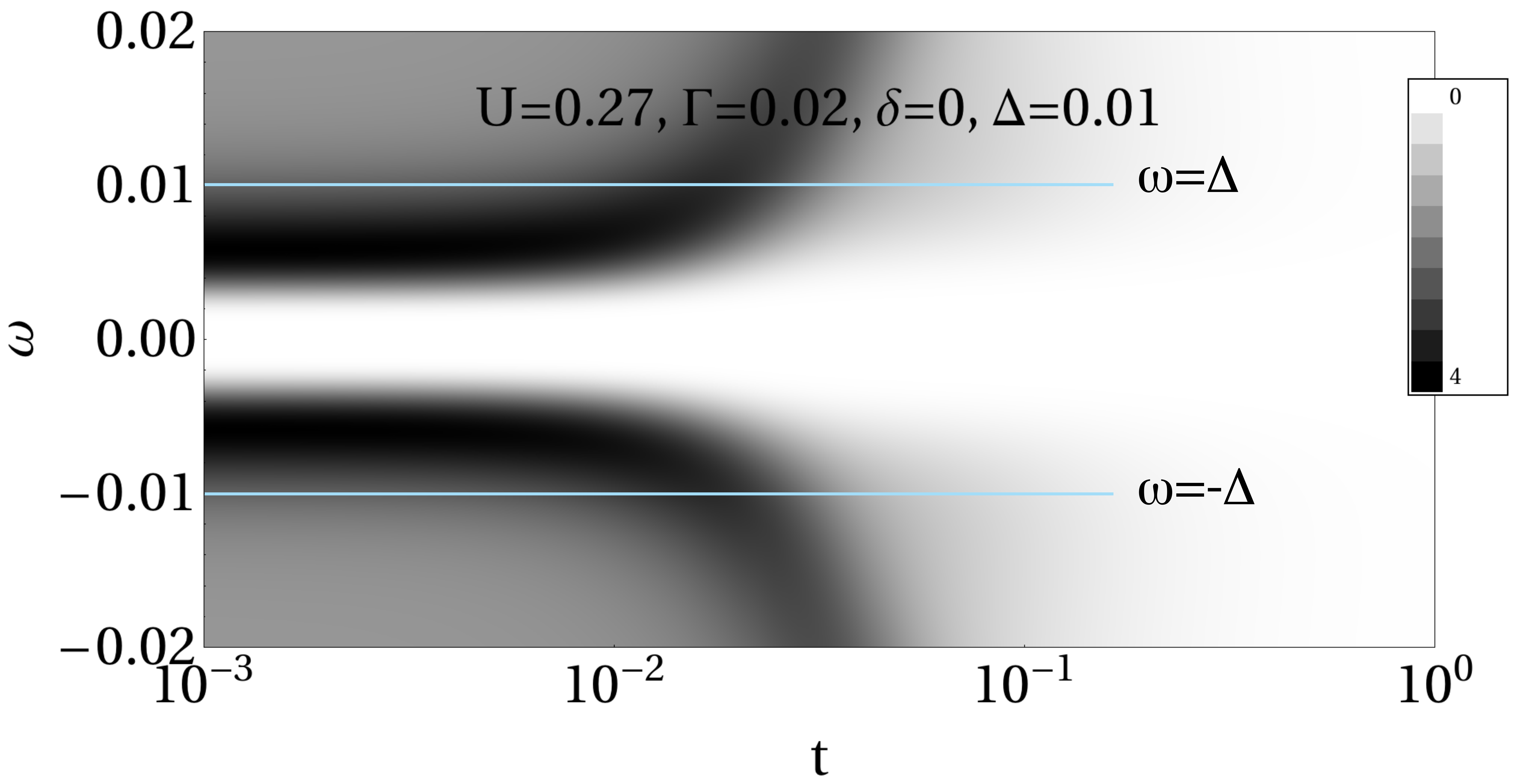}
\includegraphics[width=0.45\textwidth]{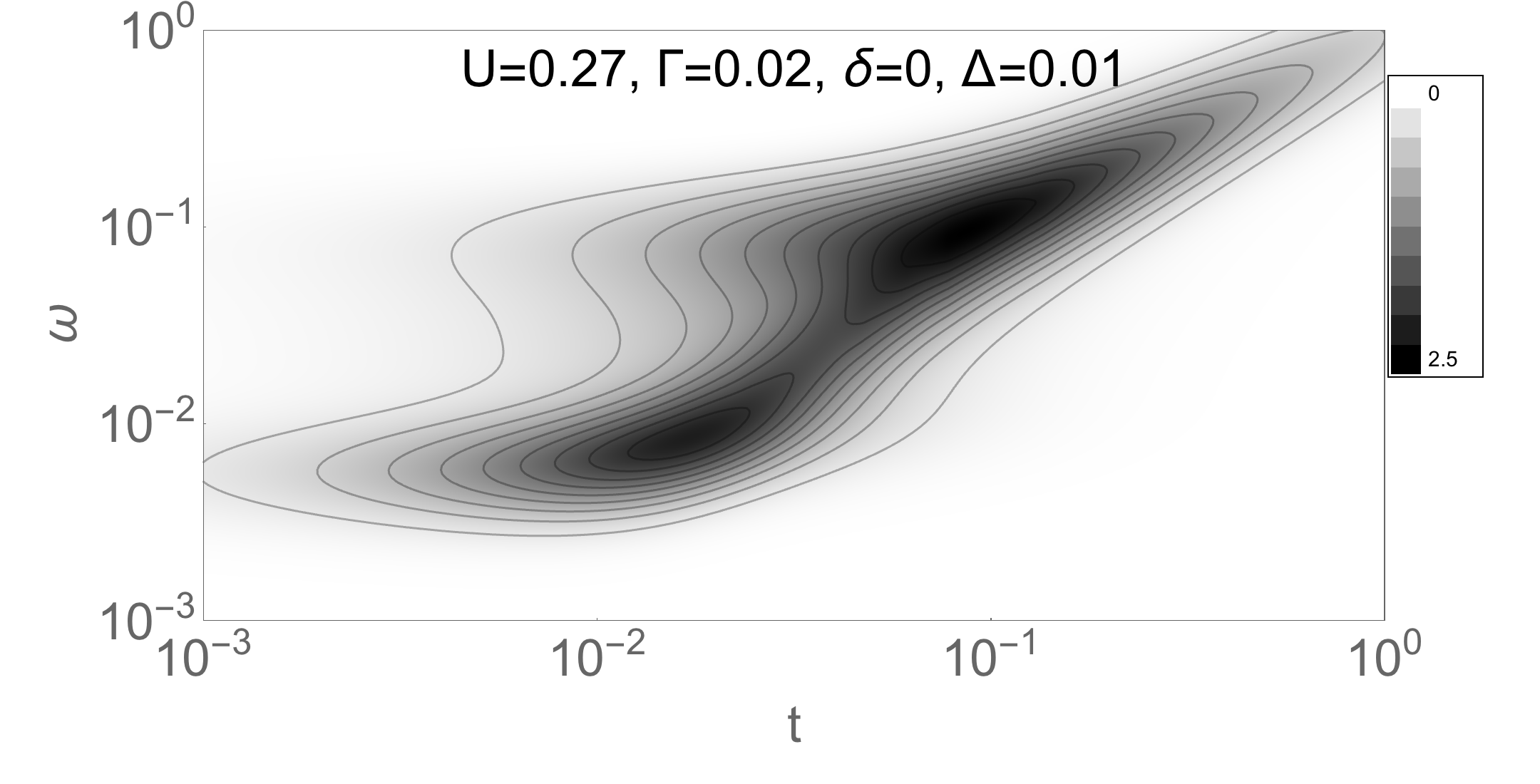}
\caption{(Color online) Left-right symmetric DQD system at half
filling. (a) Local spectral function $A_1(\omega)$. (b) Inter-dot
spectral function $A_{12}(\omega)$: we plot the positive frequency
part on a logarithmic scale.
}
\label{fig4}
\end{figure}

We first consider the case of two equal dots at half-filling,
$\delta_1=\delta_2=0$, see Fig.~\ref{fig4}. In panel (a) we plot the
local spectral function $A_1(\omega)=A_2(\omega)$ in the frequency
range inside and around the gap, $\omega \sim \Delta$. In panel (b) we
plot the inter-dot spectral function $A_{12}(\omega)$, which reveals
the transition from Kondo to antiferromagnetic regime at $t \sim
10^{-2}$, corresponding to $J\sim T_K$, then from the
antiferromagnetic to the molecular-orbital regime at $t \sim 10^{-1}$,
corresponding to $J \sim t$.

\begin{figure}
\centering
\includegraphics[width=0.45\textwidth]{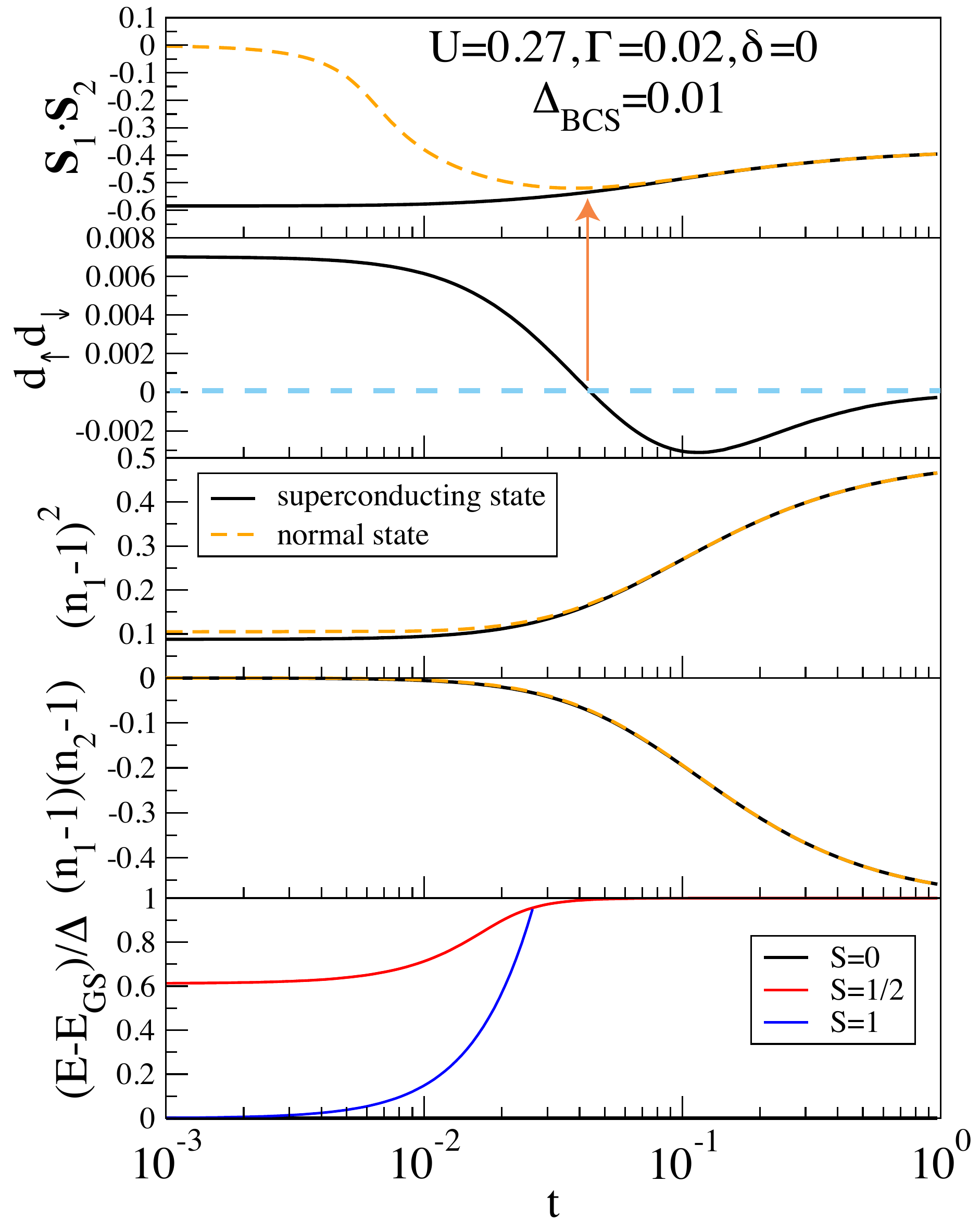}
\caption{(Color online) Expectation values and Shiba state energies in
the left-right symmetric DQD system at half filling. Results, where
applicable, are shown for both normal-state and superconducting leads.
The arrow indicates the point where the on-site pairing changes sign
and the spin-spin correlations in the superconducting and normal-state
cases start to deviate.
}
\label{fig5}
\end{figure}

For small $t$ the sub-gap peaks appear close to the energy $\Omega
\approx 0.6\Delta$ of the excitations in the single-dot case
\cite{suppl}. While for a single dot the ground-state is a doublet and
the excited state a singlet, for two QDs the lowest-lying
many-particle states are generated by combining two doublet states
into a singlet and a triplet that are split at low $t$ by $J \approx
4t^2/U$ due to the inter-dot superexchange coupling. The ground state
is always a singlet at half filling. The singlet-triplet transition is
not spectroscopically visible as it violates the $\Delta S_z = \pm
1/2$ sum rule. Its presence is, however, revealed by a direct
calculation of the many-particle spectrum using the numerical
renormalization group, see the bottom-most panel in Fig.~\ref{fig5}.

In the wide-gap limit, Eq.~\eqref{wg}, the triplet state
$\ket{T}=d^\dag_{1\uparrow}d^\dag_{2\uparrow}\ket{0}$ always has
zero energy. There are five singlet states, four of which have energy of
order $U$ for $t,\delta \ll U$. The remaining singlet state $\ket{S1}$
represents the inter-dot singlet and it has energy $\approx -4t^2/U$
for low $t$, while at high $t$ it is better described as the
molecular orbital state with two electrons in the bonding orbital. In
the wide-gap limit, the state $\ket{S1}$ is the ground state at
$\delta=0$ for all values of $t$.

The resonances in the spectral function $A_1(\omega)$ reveal the
excited doublet states. There are, in fact, two exactly degenerate
doublets. This degeneracy is broken in the presence of the flux and
away from half-filling as we show in the following subsections. In the
$t \to 0$ limit, these spectral peaks are exactly the same excitations
as the doublet-singlet Shiba resonances in the single-dot case. They
persist to finite $t$, although their energy is affected by the
inter-dot coupling. The shift of the resonance is quadratic for $t$
around $10^{-2}$. This is the region associated with the competition
between the Kondo screening and the superexchange ($T_K \sim J$)
\cite{suppl}. The NRG results show that that the doublet-singlet
energy difference is mostly driven by the downward shift of the
singlet ground state which undergoes significant changes as a function
of $t$, as also revealed by the evolution of the ground-state
expectation values, in particular the spin $\expv{\vc{S}_1 \cdot
\vc{S}_2}$ and pairing correlations $\expv{d_{1\uparrow}
d_{1\downarrow}}$, see Fig.~\ref{fig5}.

An important difference between the normal-state and superconducting
case is found for the spin correlation $\expv{\vc{S}_1\cdot\vc{S}_2}$
in the small $t$ limit. In the normal case, the Kondo screening wins
over the superexchange and each dot is screened individually, leading
to $\expv{\vc{S}_1 \cdot \vc{S}_2} \to 0$. In the superconducting
case, the Kondo screening cannot be completed due to the lack of
quasiparticles, thus the two unscreened moments are free to form a
tightly bound local singlet state and $\expv{\vc{S}_1 \cdot \vc{S}_2}$
attains values close to the saturation ($-0.6$). It is also revealing
to observe that the difference in $\expv{\vc{S}_1 \cdot \vc{S}_2}$
between the two cases starts to grow near the point where the pairing
correlations $\expv{d_{1\uparrow} d_{1\downarrow}}$ change sign, as
indicated by the arrow in Fig.~\ref{fig5}. This is yet another sign
that the Kondo screening becomes ineffective for $t$ lower than this
limiting value. Due to non-zero electron hopping in the model, this
passing through zero is a simple cross-over, not a quantum phase
transition.

In the opposite limit of very large $t$, the behavior can be described
in terms of molecular orbitals. In fact, from Fig.~\ref{fig5} it is
immediately obvious that in this limit there is little difference
between the normal-state and the superconducting case, since the
superconducting gap is positioned in the region of low spectral
density between the bonding and anti-bonding orbitals, thus it is
ineffective. This is also signalled by the pairing correlations
$\expv{d_{1\uparrow} d_{1\downarrow}}$ going to zero for large $t$.

\begin{figure}
\centering
\includegraphics[width=0.45\textwidth]{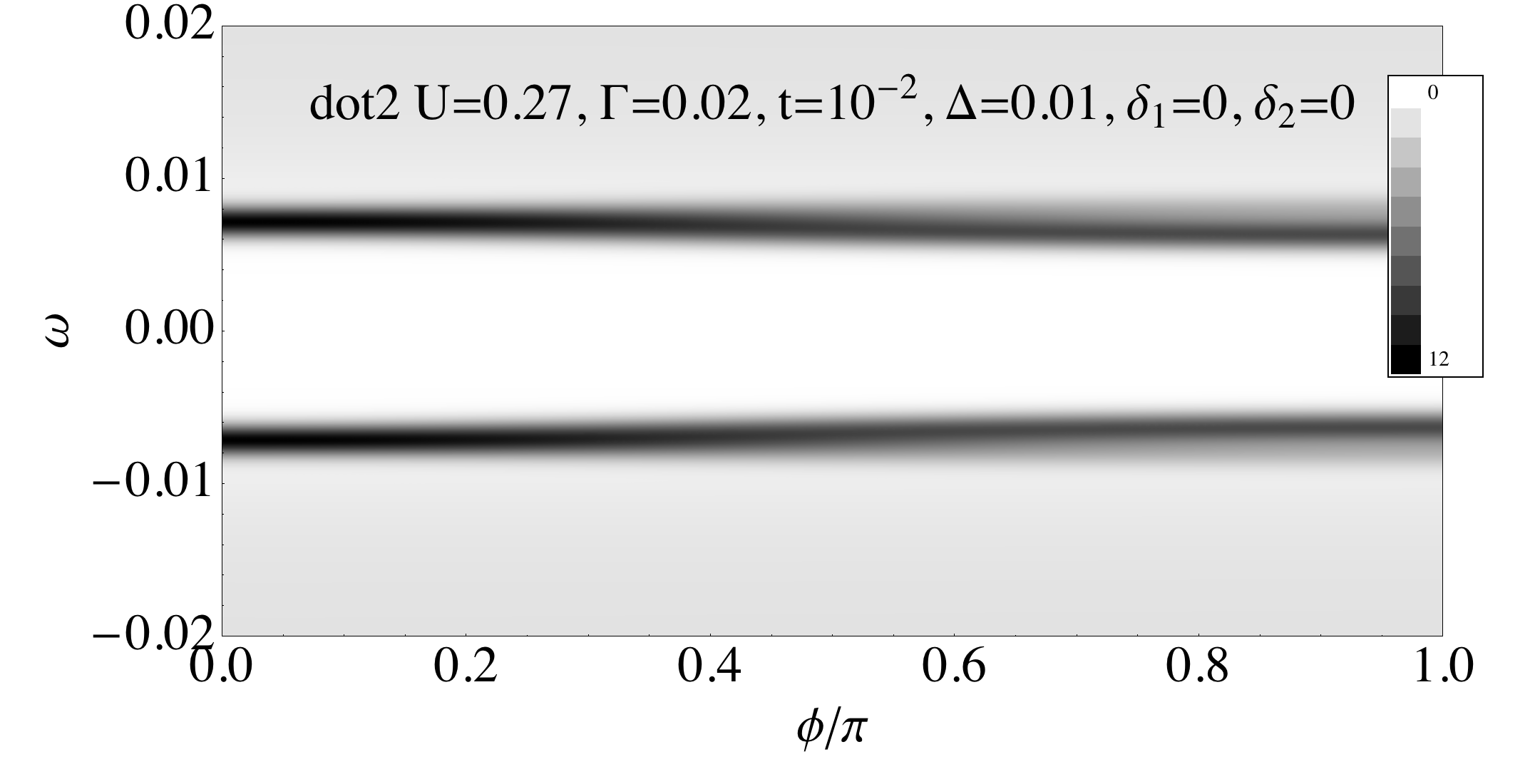}
\includegraphics[width=0.37\textwidth]{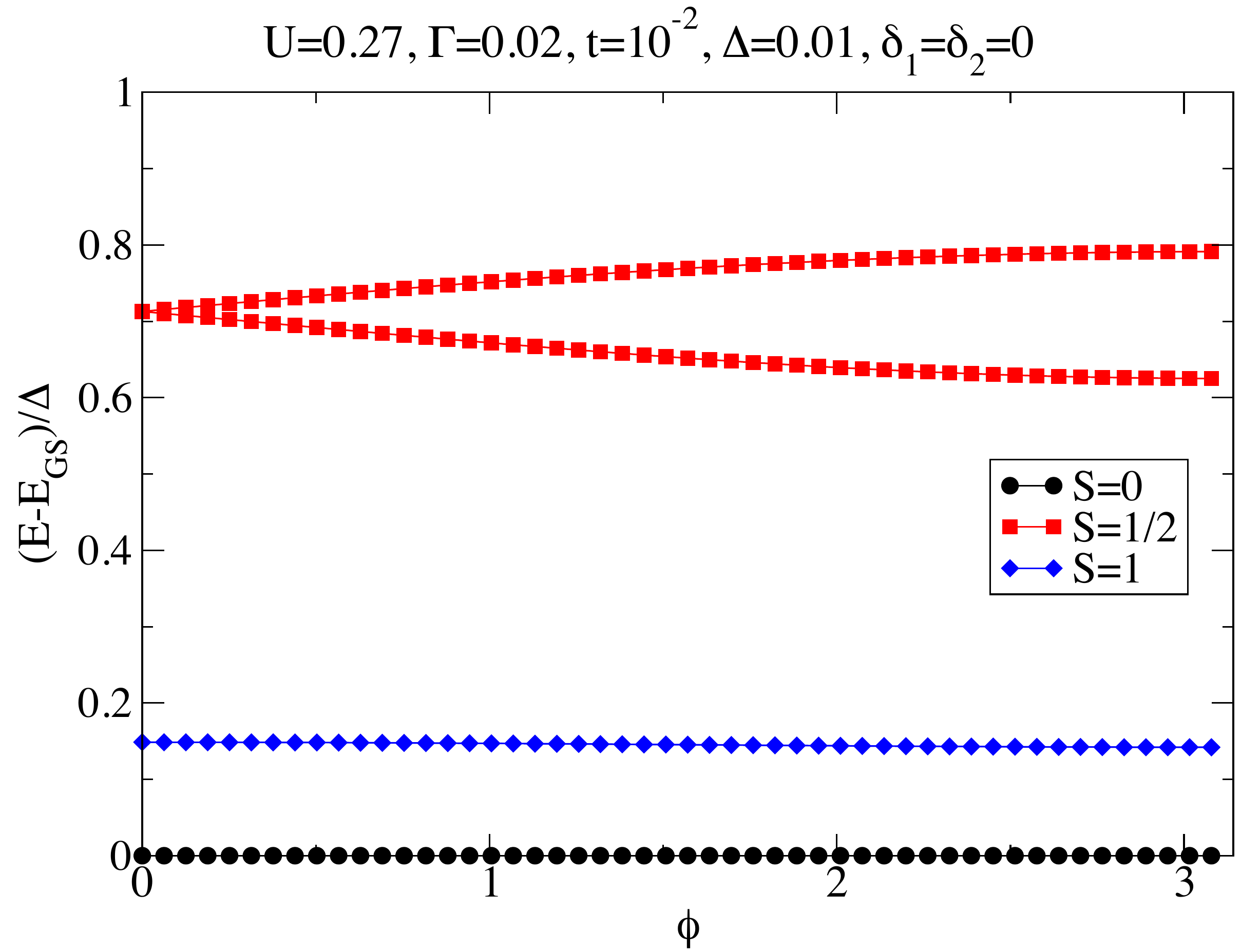}
\caption{(Color online) Spectral function $A_1(\omega)$ and
the diagram of the sub-gap Shiba states for $\delta=0$ as a function
of the superconducting phase difference $\phi$.}
\label{flux1}
\end{figure}

While the spin degrees of freedom are significantly affected by the
superconductivity, the charge fluctuations are approximately the same
in both normal and superconducting case, as evidenced by nearly
overlapping results for the inter-dot charge fluctuations
$\expv{(n_1-1)(n_2-1)}$ and only slightly reduced intra-dot charge
fluctuations $\expv{(n_1-1)^2}$ in the superconducting state in the
Kondo and antiferromagnetic regimes. The small reduction is due to the
opening of the gap in the density of states and the corresponding
reduction of electron hopping.

\subsubsection{Flux dependence}

We now study the effect of the difference in the superconducting phase
(i.e., flux through the ring), $\phi$. We show the results for an
intermediately strong $t=10^{-2}$ in Fig.~\ref{flux1}. The flux
induces a {\sl splitting of the doublet Shiba states} which are
otherwise degenerate at half filling. At small $\phi$ the splitting in
the superconducting atomic limit is
\begin{equation}
\frac{t \Gamma^2}{\Gamma^2+t^2}\, \phi.
\end{equation}
The NRG results also show proportionality to $t$ and $\phi$ when both
are small. The splitting is approximately sinusoidal and largest for
$\phi=\pi$. In the spectral function, the splitting is actually only
faintly visible because of the very unequal spectral weights: exactly
at half-filling, only the lower energy doublet state appears in the
spectra. The singlet-triplet splitting is
hardly affected by non-zero $\phi$, there is only a slight downward
trend in the energy difference, see the bottom panel in
Fig.~\ref{flux1}.

\subsubsection{$\Gamma$ dependence}
\label{secGamma}

\begin{figure}
\centering
\includegraphics[width=0.45\textwidth]{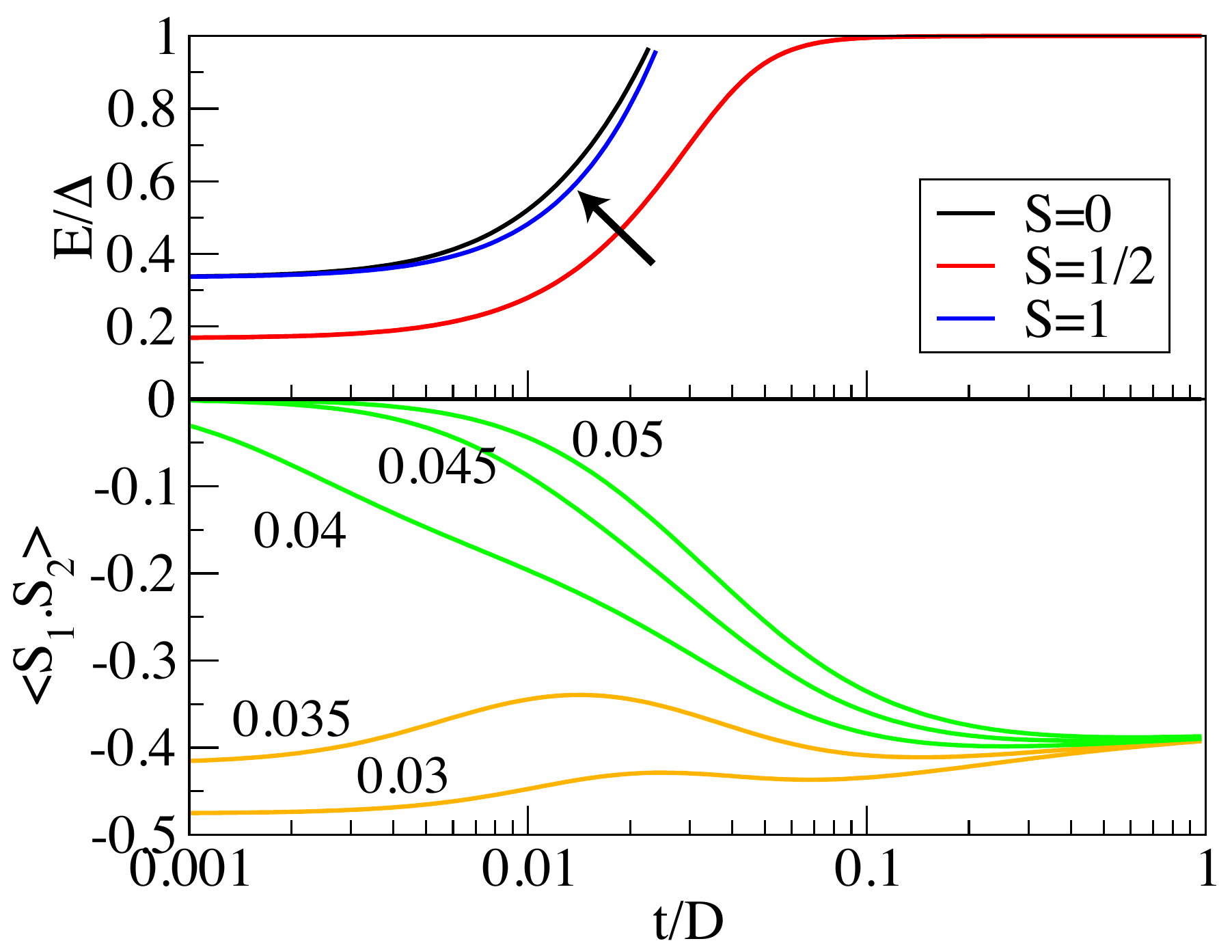}
\caption{(Color online) Dependence on the hybridization strength
$\Gamma$ in the left-right symmetric case at half filling. (a) Sub-gap
Shiba states for $\Gamma=0.045$. The arrow brings to attention that
the inter-dot triplet Shiba state is below the inter-dot singlet
state. (b) Spin correlation function $\expv{\vc{S}_1 \cdot
\vc{S}_2}(t)$ for a range of $\Gamma$ (indicated near the
corresponding curves) across the singlet-singlet cross-over at
$\Gamma_t \approx 0.04$. }
\label{figA}
\end{figure}

We now discuss the role of the hybridization $\Gamma$. In a single QD,
the doublet-singlet transition occurs for $\Gamma = \Gamma_t \approx
0.04$. In the DQD system, with increasing $\Gamma$ the doublet excited
states decrease in energy and at $\Gamma \approx 0.025$ an additional
excited singlet also enters the sub-gap region. This trend continues
until $\Gamma = \Gamma_t$, where in the limit $t\to0$ we find
degeneracy of five multiplets: two singlets, two (degenerate)
doublets, and the triplet. For $\Gamma > \Gamma_t$, the nature of the
singlet ground state changes, since it is associated with two separate
Kondo clouds, rather than with the inter-dot singlet induced by the
superexchange coupling. Consequently, in this range the triplet state
lies close to the {\sl excited} singlet state, rather than the ground
state, see Fig.~\ref{figA}(a) where the sub-gap states are plotted for
$\Gamma=0.0045 > \Gamma_t$. Curiously, the inter-dot triplet is {\sl
lower} in energy than the inter-dot singlet state, although the
superexchange coupling would lead to the opposite order. In the
wide-gap limit, the triplet is indeed above the singlet. A simple
interpretation is based on a three level model with two singlets and a
triplet:
\begin{equation}
\begin{pmatrix}
\epsilon_{S1} & 0 & \alpha \\
0 & \epsilon_{T} & 0 \\
\alpha & 0 & \epsilon_{S2}
\end{pmatrix}
\end{equation}
with $\epsilon_T-\epsilon_{S1}=J=4t^2/U$, such that the coupling
between the singlets $\alpha$ is (much) larger than $J$. In this case
a similar phenomenology is found, with the triplet state moving from a
value above the ground state singlet to a value below the excited
singlet as $\epsilon_{S2}-\epsilon_{S1}$ changes sign, as happens at
$\Gamma=\Gamma_t$. Numerical results indicate that $\alpha \propto t$,
thus the requirement $\alpha \gg J$ is fulfilled for the relevant
range of $t$.

A further difference between the $\Gamma < \Gamma_t$ and $\Gamma >
\Gamma_t$ regimes is manifest in the $t$-dependence of $\expv{\vc{S}_1
\cdot \vc{S}_2}$ on the approach to the decoupled-dot $t\to0$ limit,
see Fig.~\ref{figA}(b).

In the superconducting atomic limit, this transition corresponds to
the transition from the inter-dot singlet state $\ket{S1}$ with energy
$\approx -4t^2/U$ to the on-site-singlets state $\ket{S2}$ with energy
$\approx U-2\Gamma$. The wide-gap limit does not distinguish between
BCS and Kondo singlets, thus $\ket{S2}$ has characteristics of a state
with strong on-site BCS pairing, but at finite $\Delta$ the NRG
results show that the corresponding many-particle state should rather
be described as two separate Kondo clouds. It should also be noted
that this is a true transition only strictly at $t=0$. For finite $t$,
it is a smooth cross-over.

Irrespective of the value of $\Gamma$ and of the nature of the
ground-state singlet, we find that the excited singlet state always
monotonously increases in energy with $t$ and the spectral functions
are qualitatively always very similar. There are no quantum phase
transitions between singlets as a function of $t$. This issue is
discussed further in Sec.~\ref{sec4} where the model without particle
hopping but only exchange interaction is studied: that model has a
true quantum phase transition.

\subsubsection{$V$ dependence}

\begin{figure}
\centering
\includegraphics[width=0.45\textwidth]{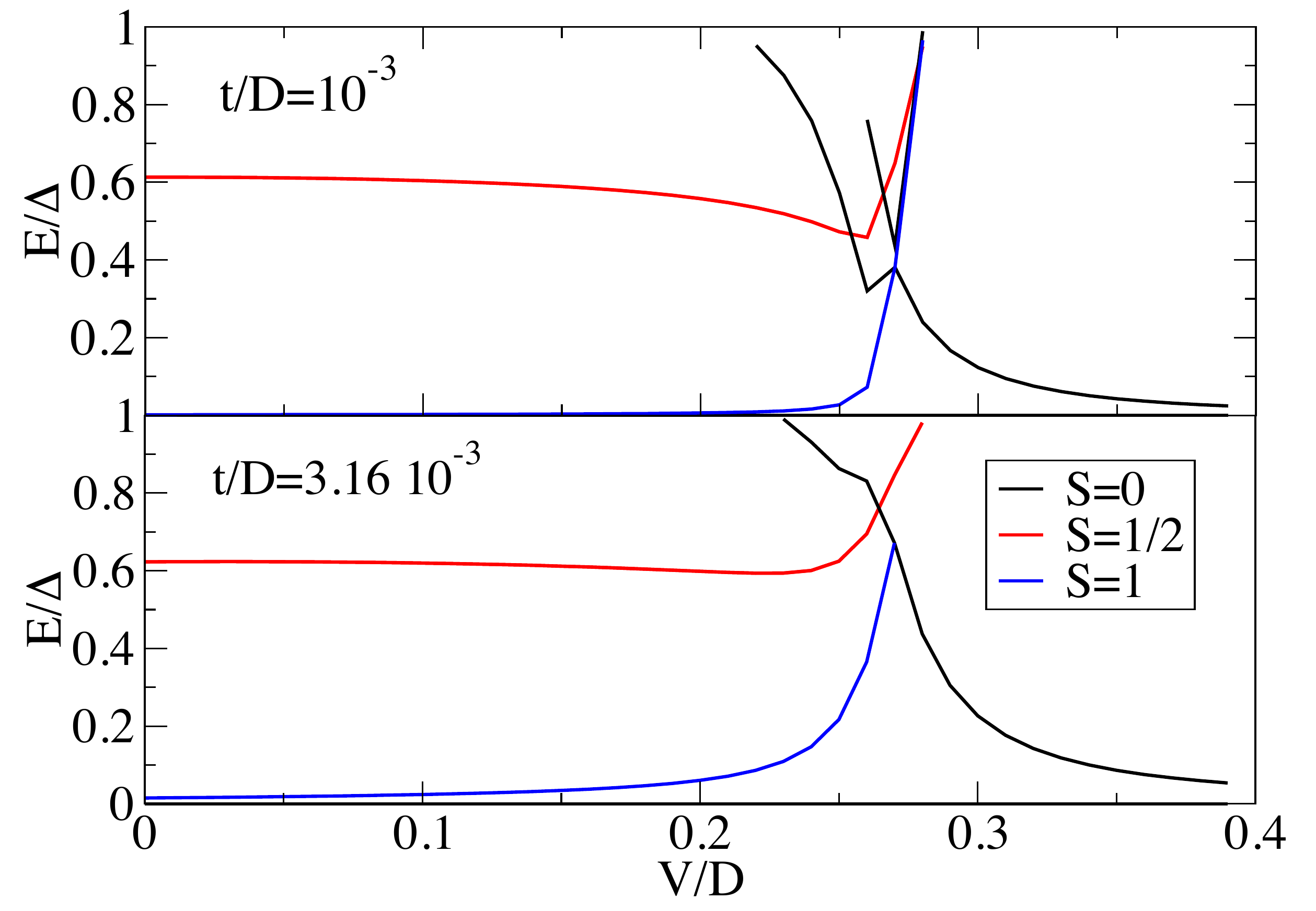}
\caption{(Color online) Dependence on the inter-dot capacitive coupling
at half filling.}
\label{fa}
\end{figure}

We now discuss the effects of the inter-dot capacitive coupling $V$ at
half filling. For moderate $V \lesssim U$, the effect of $V$ is only
quantitative and rather weak. As $V$ increases, there is an increase
in energy of the double-occupancy integer-spin Shiba states (seen
through a decreasing energy of the excitation to the odd-occupancy
half-spin, i.e., doublet, Shiba state), and an enhancement of the
exchange coupling $J$:
\begin{equation}
J = \frac{1}{2}\left[ \sqrt{(U-V)^2 +16 t^2} - (U-V) \right]
\approx \frac{4t^2}{U-V}.
\end{equation}
Both tendencies are well visible in the numerical results in
Fig.~\ref{fa}. At $V=U$, the system enters a phase with increased
symmetry and enhanced Kondo temperature in the normal state if the
inter-dot tunneling $t$ is small, while for $V>U$ the electrons tend
to localize on a single site and form ``local charge singlets'' in the
charge-ordering regime \cite{galpin2005,mravlje2005,vzporedne}. In the
superconducting case at finite $t$, these trends are visible as a
cross-over at $V \sim U$. Although there is no change of the ground
state, its nature changes significantly. For $V \gg U$, the lowest
energy Shiba state is characterized by near zero spin correlations,
$\expv{ \vc{S}_1 \cdot \vc{S}_2 } \sim 0$, large inter-dot charge
fluctuations, $\expv{(n_1-1)(n_2-1)} \to -1$, and negative on-site
pairing, $\expv{d_{i\uparrow}d_{i\downarrow}}<0$. These are indeed the
signatures of the charge-ordering regime. For large $V$, there is a
single Shiba excited state inside the gap, which is also of local
charge singlet kind. The excitation energy in fact corresponds to the
splitting between the $\frac{1}{\sqrt{2}} \left( \ket{2,0} \pm
\ket{0,2} \right)$ eigenstates, which is equal to the isospin exchange
coupling
\begin{equation}
J_\mathrm{iso} = \frac{4t^2}{V-U},
\end{equation}
and decreases with increasing $V$. Close to the cross-over point at $V
\sim U$, the behavior is particularly complex for small $t$; see the
case of $t/D=10^{-3}$, upper panel in Fig.~\ref{fa}. In this case we
even observe three sub-gap singlet states, which is a unique
occurence. Two of these are the local charge singlets and the third is
the inter-dot spin singlet. In the $t \to 0$ limit, these three
singlets would combine with the inter-dot spin triplet to form a
six-fold degenerate ground state with SU(4) symmetry.

\subsection{Left-right symmetric case, away from half-filling}

\begin{figure*}
\centering
\includegraphics[width=0.98\textwidth]{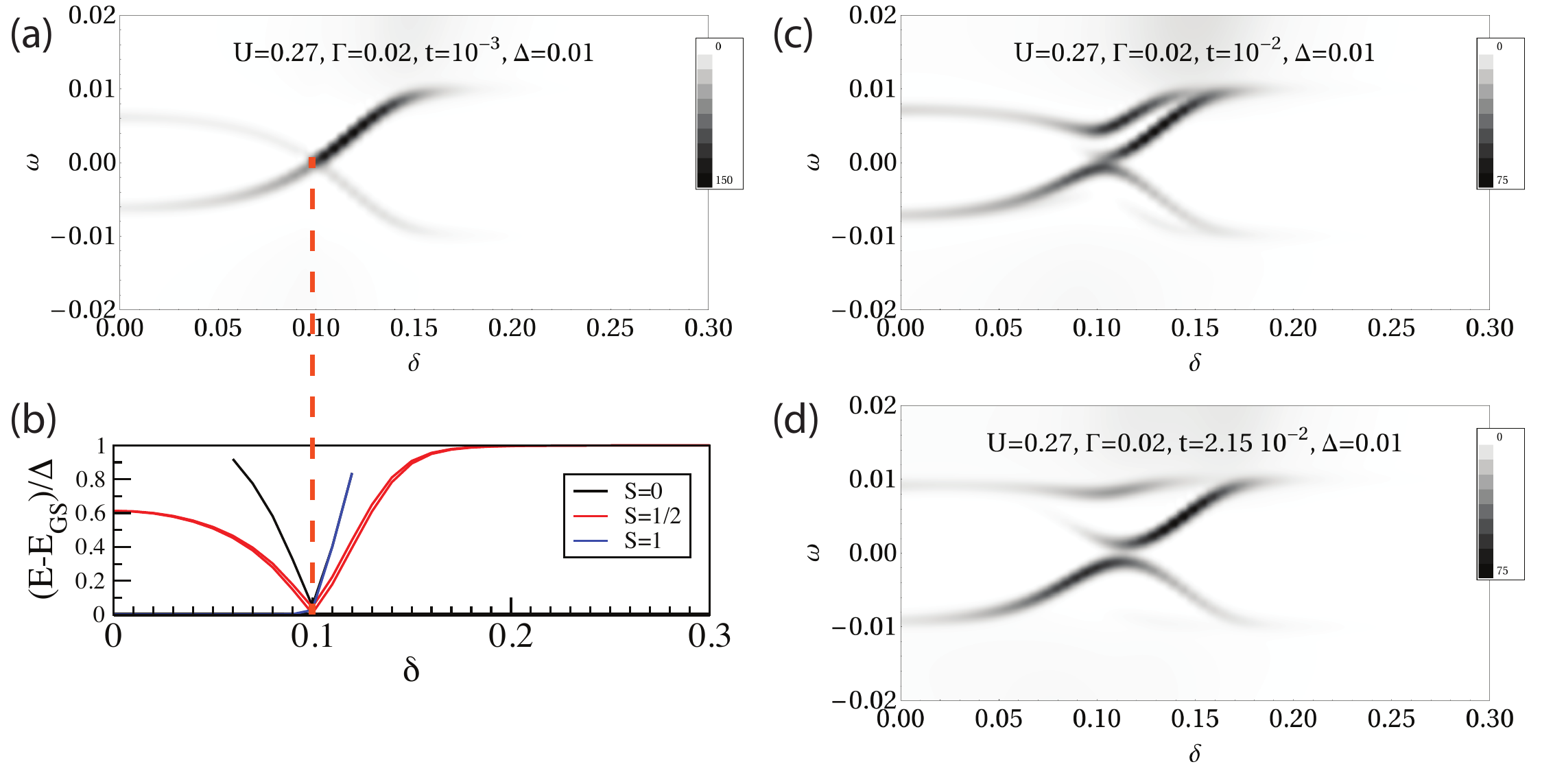}
\caption{(Color online) Spectral function $A_1(\omega)$ in the
left-right symmetric DQD system away from half-filling. (a) Small
inter-dot tunneling $t=10^{-3}$. (b) Sub-gap Shiba state diagram for 
$t=10^{-3}$. In the range $\delta \sim 0.1$, the $S=1/2$ and the
excited $S=0$ state have small but non-zero energy. (c,d)
Spectra at stronger inter-dot coupling. }
\label{fig6}
\end{figure*}

At small $t$ the spectra away from half-filling, Fig.~\ref{fig6}(a),
look similar to those in the single dot limit \cite{suppl}, however a
closer look at the sub-gap states reveals some interesting details,
Fig.~\ref{fig6}(b). At $\delta \sim 0.1$ there appears to be a
transition between two different singlet ground states, which is
actually a rapid cross-over. For $\delta < 0.1$, the ground state of a
single dot is a spin doublet Shiba state, thus the DQD systems has a
nearly degenerate singlet and triplet states for low $t$.  The lowest
spectroscopically observable excitations are the two doublet states
which decrease in energy $\Omega$ as $\delta$ increases; these
doublets correspond to one of the QDs having a quasiparticle bound to
it. As $\Omega$ drops below $0.5\Delta$, an additional singlet state
enters the sub-gap range. This state can be interpreted as the state
where both QDs have one quasiparticle attached each. The ground state
for $\delta>0.1$ is indeed of this type.

\begin{figure}
\centering
\includegraphics[width=0.45\textwidth]{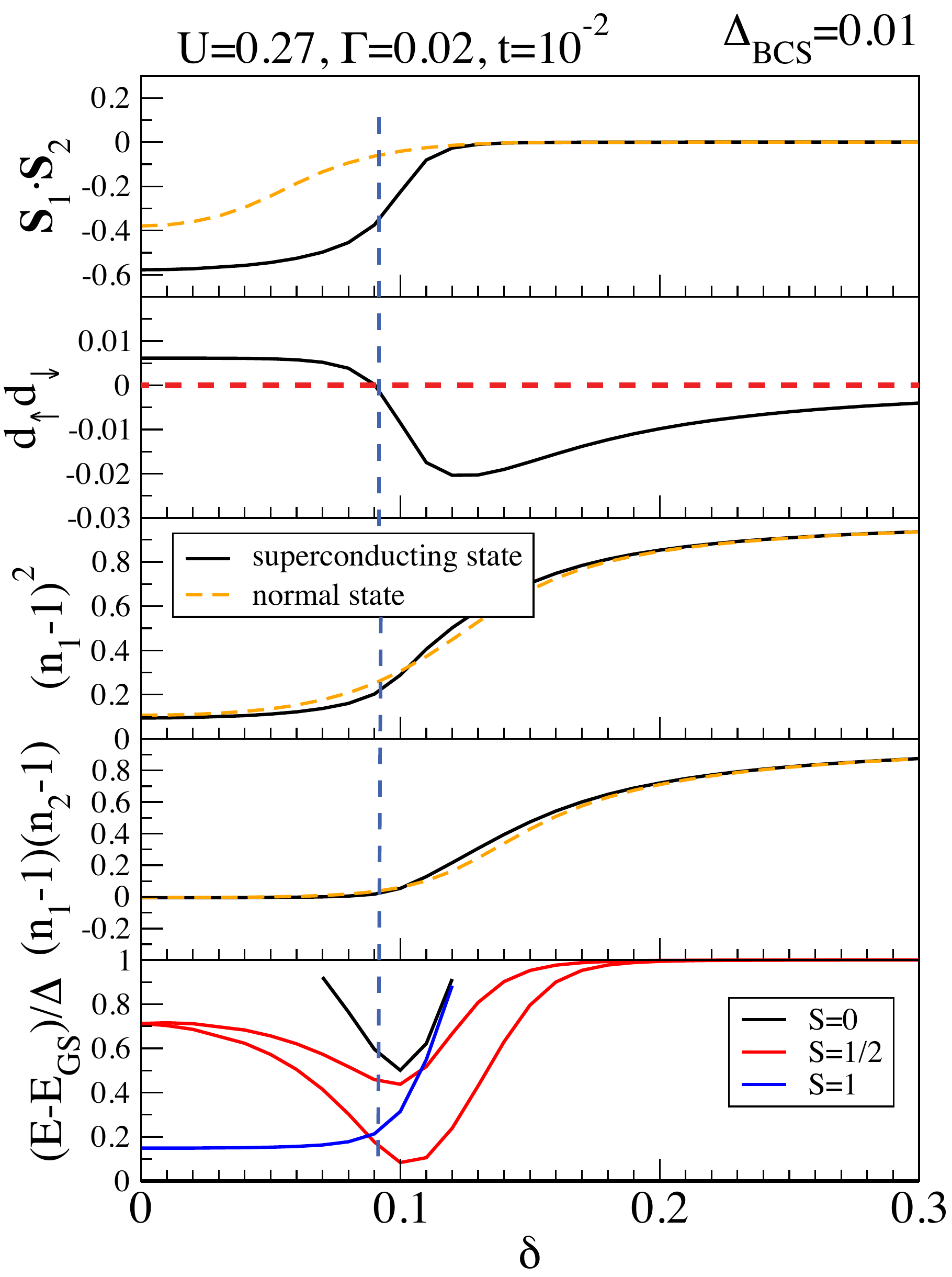}
\caption{(Color online) Expectation values in the left-right symmetric
DQD system away from half filling. The value of $t$ is the same as in
Fig.~\ref{fig6}(c). Note the position of the triplet state below the
excited singlet for $\delta > 0.1$.
}
\label{fig7b}
\end{figure}

In the cross-over region the excited singlet state occurs
approximately at the sum of energies of the two doublet Shiba states,
thus it is pushed together with these two states to higher energies as
$t$ increases. The spectra for larger $t$ are shown in
Fig.~\ref{fig6}(c,d). Finiteness of excitation energies to the doublet
states is confirmed by inspecting the spectrum of the sub-gap
excitations (bottom panel in Fig.~\ref{fig7b}). Despite the continuous
evolution of the ground state as a function of $\delta$, its nature is
not the same at $\delta=0$ as in the large $|\delta|$ limit, see other
panels in Fig.~\ref{fig7b}. Most notably, the pairing expectation
value $\expv{d_{1\uparrow}d_{1\downarrow}}$ has opposite signs. The
$|\delta| \lesssim 0.1$ ground state is the inter-dot singlet
$\ket{S1}$ formed by unscreened local moments. For $|\delta| \gtrsim
0.1$, the ground state is close to $\ket{S2}$, which can also be
thought of as two separate Kondo compensated states with
quasiparticles attached. The energy of $\ket{S2}$ away from half
filling evolves as $\approx U-2\Gamma-\delta^2/\Gamma$. In the
wide-gap limit it thus becomes important at $\delta$ of order $\sqrt{U
\Gamma}$ which is the charge fluctuation scale in the single-impurity
Anderson model. For very large $\delta$ the dots are unoccupied and
play no role, thus the system is a simple BCS superconductor. As
expected, the main component of $\ket{S2}$ for large $\delta$ is the
empty state $\ket{0}$.

It may be noted that the singlet ground-state for $\delta \geq 0.1$ is
actually similar in nature to the singlet ground state at $\delta=0$
at $\Gamma > \Gamma_t$ discussed in the previous subsection. One could
draw a phase diagram in the $(\Gamma/U,\delta/U)$ plane delineating
the parameter ranges where the single ground state is either of type
$\ket{S1}$ or $\ket{S2}$. In the $t\to 0$ limit, this line would be
well defined and would, in fact, coincide with the singlet-doublet QPT
line of the single QD problem. At $t\neq 0$, the QPT is however
replaced by a cross-over with a width roughly proportional to $t$.

\subsubsection{Doublet splitting}

While at $\delta=0$ the doublet excitations are degenerate, there is a
splitting induced by the inter-dot coupling of order proportional to
$\delta t$. This can be understood as the formation of molecular
orbitals by the quasiparticle attached to the dots and can be
understood within the superconducting atomic limit. The eigenvalues
of the doublet states are
\begin{equation}
\begin{split}
&\frac{1}{2}
\Bigl(
 U \pm 
 \bigl\{ 
  2 \bigl( 
   2t^2+2\Gamma^2+\delta_1^1+\delta_2^2  \\
   &\quad\quad
   \pm \sqrt{[(4t^2+(\delta_1-\delta_2)^2](\delta_1+\delta_2)^2)}
  \bigr) 
 \bigr\}^{1/2}
\Bigr),
\end{split}
\end{equation}
which for $\delta=\delta_1=\delta_2$ simplifies to
\begin{equation}
\frac{U}{2} \pm \sqrt{\Gamma^2+ (t \pm \delta)^2}.
\end{equation}
(The two $\pm$ signs in this expression are independent.) For
$t=\delta=0$, these energies are
\begin{equation}
\frac{U}{2} \pm \Gamma,
\end{equation}
each being doubly degenerate. These corresponding states are of the
form
\begin{equation}
\frac{1}{\sqrt{2}} d^\dag_{1\sigma} \left(1 \pm d^\dag_{2\uparrow}
d^\dag_{2\downarrow} \right) \ket{0},
\end{equation}
and the analogous states with $1\leftrightarrow 2$. The degeneracy is
lifted when {\sl both} $t$ and $\delta$ are non-zero, the splitting
being
\begin{equation}
\frac{2\delta t}{\Gamma}
\end{equation}
to lowest order in $t$ and $\delta$. This can be interpreted as the
bond formation ($t\neq0$), with bonding-antibonding splitting 
emerging only away from half-filling ($\delta \neq 0$) in the
superconducting case because one-particle and three-particle states
are mixed.

For intermediate inter-dot coupling, see the $t=10^{-2}$ case in
Fig.~\ref{fig6}(c), the splitting of the doublet Shiba states is
clearly visible. We notice that the asymmetry of the particle-like
($\omega>0$) and hole-like ($\omega<0$) transitions exists for both
branches. At larger $t=2.15\times10^{-2}$, the higher-energy doublet
state is already essentially merged with the continuum and is hardly
spectroscopically observable. For even larger $t\sim 0.1$ (not shown)
the system is in the molecular orbital regime and the two dots behave
as a single large quantum dot which undergoes Kondo screening when
$\delta$ is appropriately tuned and therefore manifests the
singlet-doublet transition.

We remark on the decreasing spectral weight of the sub-gap states as
they approach the gap edges at $\omega = \pm \Delta$, already observed
and discussed for the case of a single quantum dot in
Ref.~\onlinecite{bauer2007}. This is well visible in Fig.~\ref{fig6},
as well as in all other spectra in the following, and appears to be a
general property. The sub-gap states always merge with the continuum
in such a way that the energy evolves continuously.

\subsubsection{Flux dependence}

\begin{figure}
\centering
\includegraphics[width=0.45\textwidth]{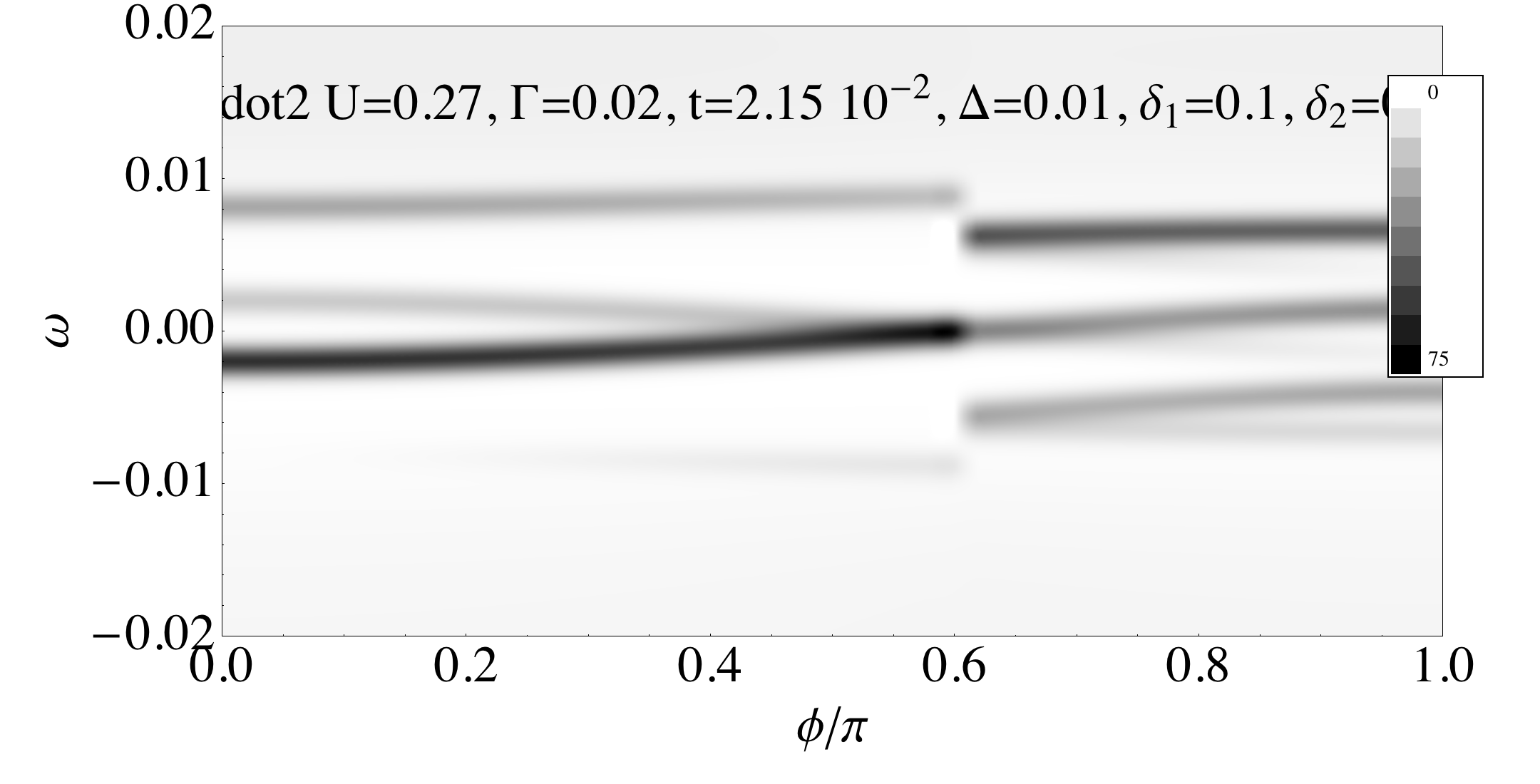}
\includegraphics[width=0.37\textwidth]{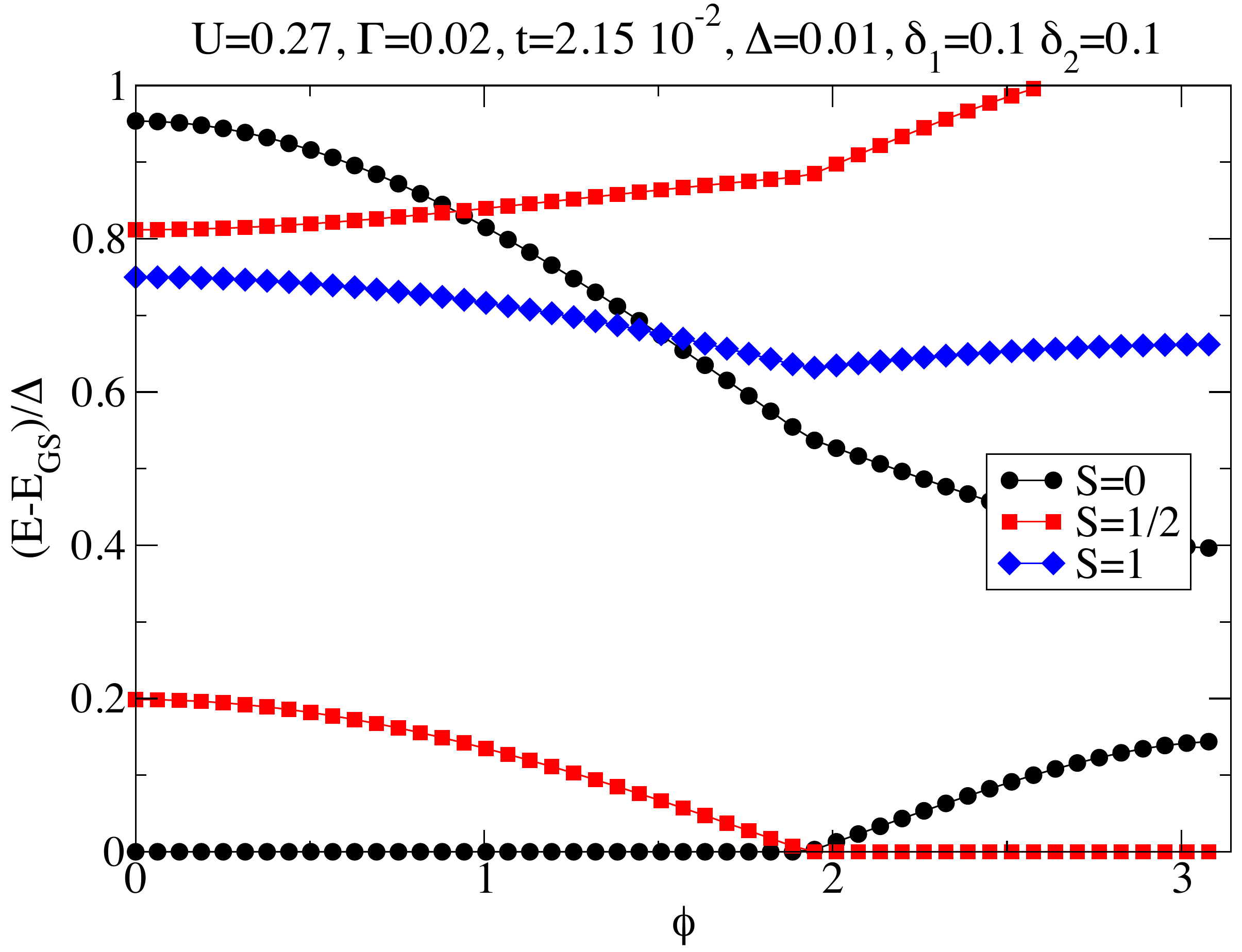}
\caption{(Color online) Spectral function $A(\omega)$ and the diagram
of the sub-gap Shiba states for $\delta_1=\delta_2=0.1$ as a function of
the superconducting phase difference $\phi$.}
\label{flux2}
\end{figure}

We now consider the flux dependence at $\delta \neq 0$. The main
effect for small $\delta$ and $t$ is some initial ($\phi=0$) splitting
of the doublet Shiba state, otherwise the results are rather similar
to those shown in Fig.~\ref{flux1}. For larger $\delta \approx 0.1$ in
the valence fluctuation region, the behavior is more intersting and
the flux can induce a quantum phase transition by strongly increasing
the splitting between the doublet states, see Fig.~\ref{flux2}. This
type of the phase transition will be studied in depth in the following
sections.
Here we comment on some other features: a) both doublet excitations
are now spectroscopically observable, unlike at half-filling, b) the
singlet-triplet splitting is now more significantly affected by
non-zero $\phi$, c) the second singlet state is also significantly
$\phi$ dependent (in fact, the two singlet states merge at $\phi=\pi$
in the $t \to 0$ limit). There is a significant spectral weight
redistribution across the phase transition, thus it should be easily
experimentally observable.

\subsubsection{$V$ dependence}

\begin{figure}
\centering
\includegraphics[width=0.45\textwidth]{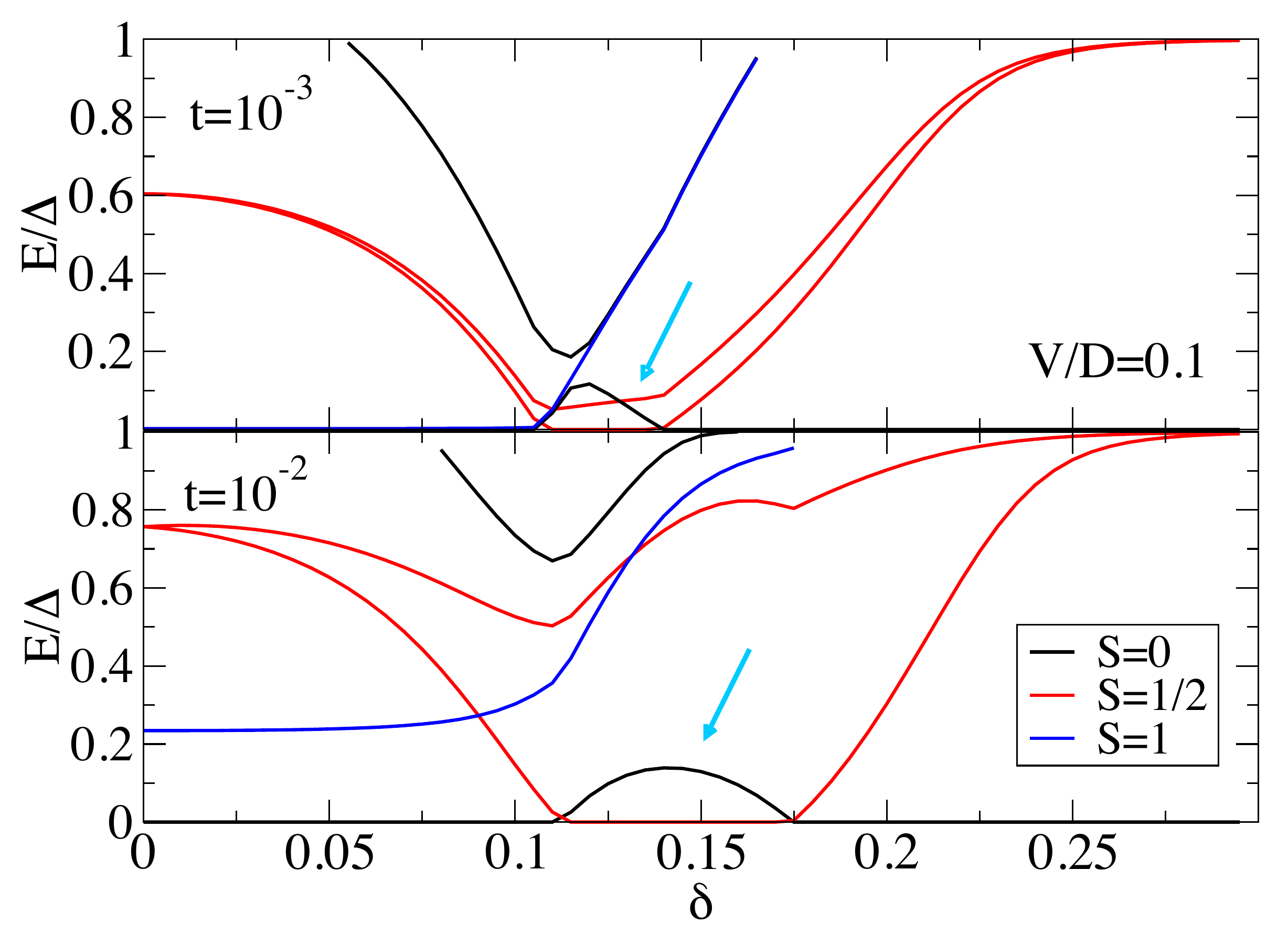}
\caption{(Color online) Sub-gap Shiba state diagram at finite
inter-dot capacitive coupling $V/D=0.1$. Top panel: $t=10^{-3}$, to be
compared with Figs.~\ref{fig6}(a) and (b). Bottom panel: $t=10^{-2}$,
to be compared with Figs.~\ref{fig6}(c) and \ref{fig7b}. The arrows
indicate the new feature: a region close to the VF point where the
inter-dot coupling induces a phase transition to the doublet state,
corresponding to the emergence of a local moment regime with a single
electron in the DQD (quarter filling).}
\label{fe}
\end{figure}

In the normal state, the inter-dot capacitive coupling leads to
non-trivial effects also at quarter filling 
\cite{amasha2013,filippone2014,PhysRevB.90.035119}. For sufficiently
large $V$ (which does not need to be comparable in value to $U$), a
local moment fixed point emerges in the valence fluctuation range,
with the occupancy pinned to one quarter. In this regime, the
degeneracy between the $\ket{0,\uparrow}$, $\ket{0,\downarrow}$,
$\ket{\uparrow,0}$, $\ket{\downarrow,0}$ states leads to a SU(4) Kondo
effect different from the one at half filling (which requires fine
tunning $V \approx U$). For exact degeneracy, there must be no
inter-dot tunneling, otherwise there is a splitting into bonding and
antibonding molecular orbitals and a regular SU(2) spin Kondo effect
is expected.

We find signatures of these effects also in the superconducting case.
The results for moderately large $V/D=0.1$ are shown in Fig.~\ref{fe}
for two values of inter-dot tunneling $t$. Compared to the $V=0$ case,
we find a range of $\delta$ where quarter filling is stabilized and
the ground state is a spin doublet. For $t=0$, both doublets would be
degenerate, but at finite $t$ we find a splitting of order $t$. 
In this range of $\delta$, both singlet and triplet Shiba states are
spectroscopically visible. 

\subsection{Generic case: unequal dots}

\begin{figure*}
\centering
\includegraphics[width=0.98\textwidth]{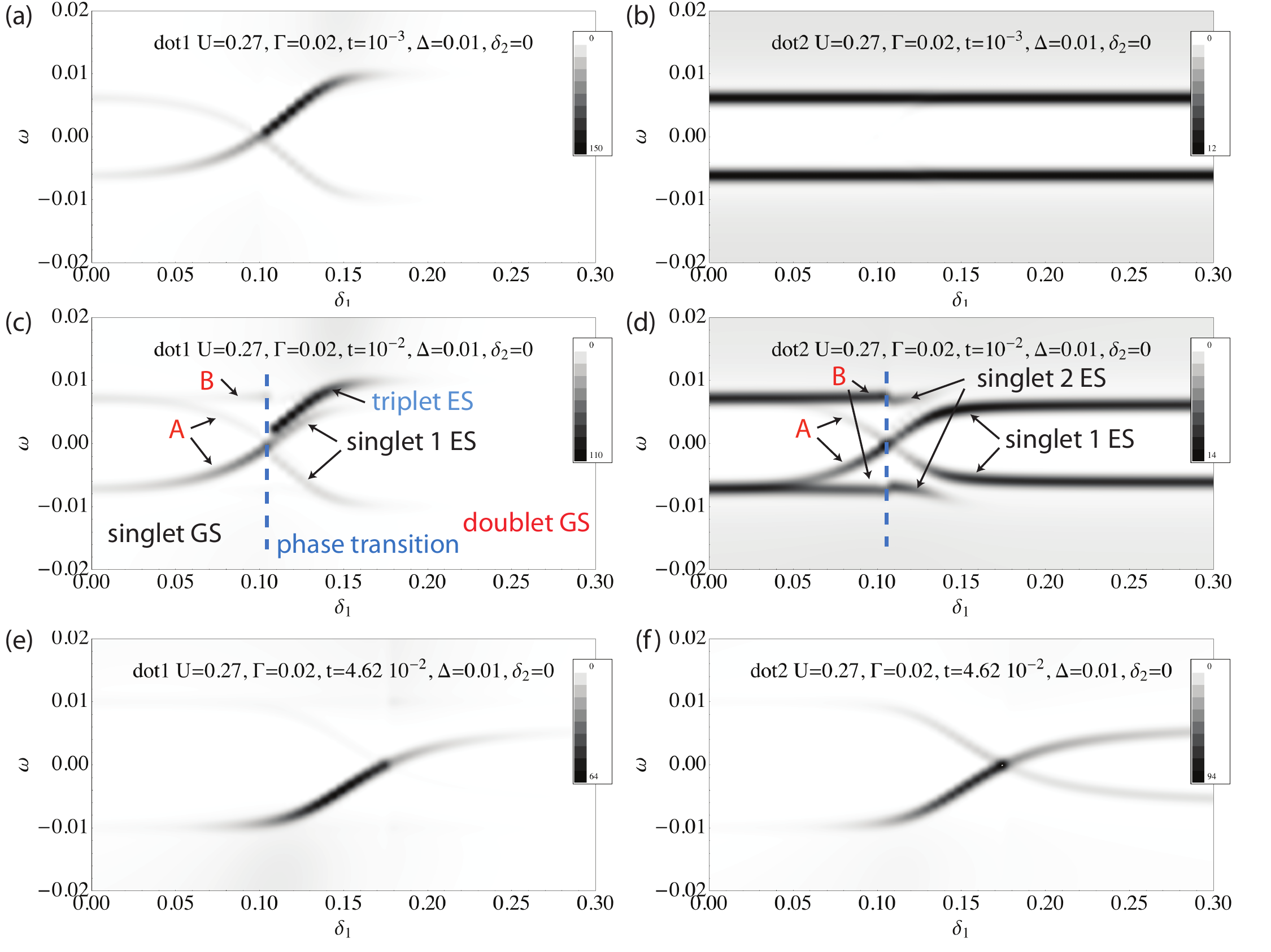}
\caption{(Color online) Spectral function on dot 1 (left panels) and
dot 2 (right panels) for a range of inter-dot couplings $t$. The
quantum dot 2 is kept at $\delta_2=0$, which is the {\sl Kondo
regime}. The arrows mark the different spectroscopically observable
excited states (ES).}
\label{fig9}
\end{figure*}

We fix the on-site energy of one of the dots ($\delta_2$) and vary the
other ($\delta_1$). We do so for different choices of $\delta_2$: at
$\delta_2=0$ (dot 2 in the Kondo regime) and $\delta_2=0.1$ (dot 2 in
the valence fluctuation regime).

\subsubsection{Dot 2 in the Kondo regime}

In Fig.~\ref{fig9} we plot the results for $\delta_2=0$. At small
$t=10^{-3}$, top row, the dots are nearly decoupled, thus the dot 1
shows the evolution vs. $\delta$ typical of a single QD, while the
spectrum on dot 2 shows peaks at constant energy of
$\Omega=0.6\Delta$, as expected for $\delta_2=0$. The inter-dot
effects become visible around $t=2\times10^{-3}$ in the form of weak
features of $A_1(\omega)$ mirrored in $A_2(\omega)$, which then
amplify and mix with the Shiba states of dot 2. By $t=10^{-2}$, center
row in Fig.~\ref{fig9}, the structure of the sub-gap states is already
quite complex and the superexchange $J$ needs to be invoked to explain
all features. The evolution of the spectral weights is non-trivial and
discontinuous at the point where the ground state changes from singlet
to doublet. 

\begin{figure}
\centering
\includegraphics[width=0.45\textwidth]{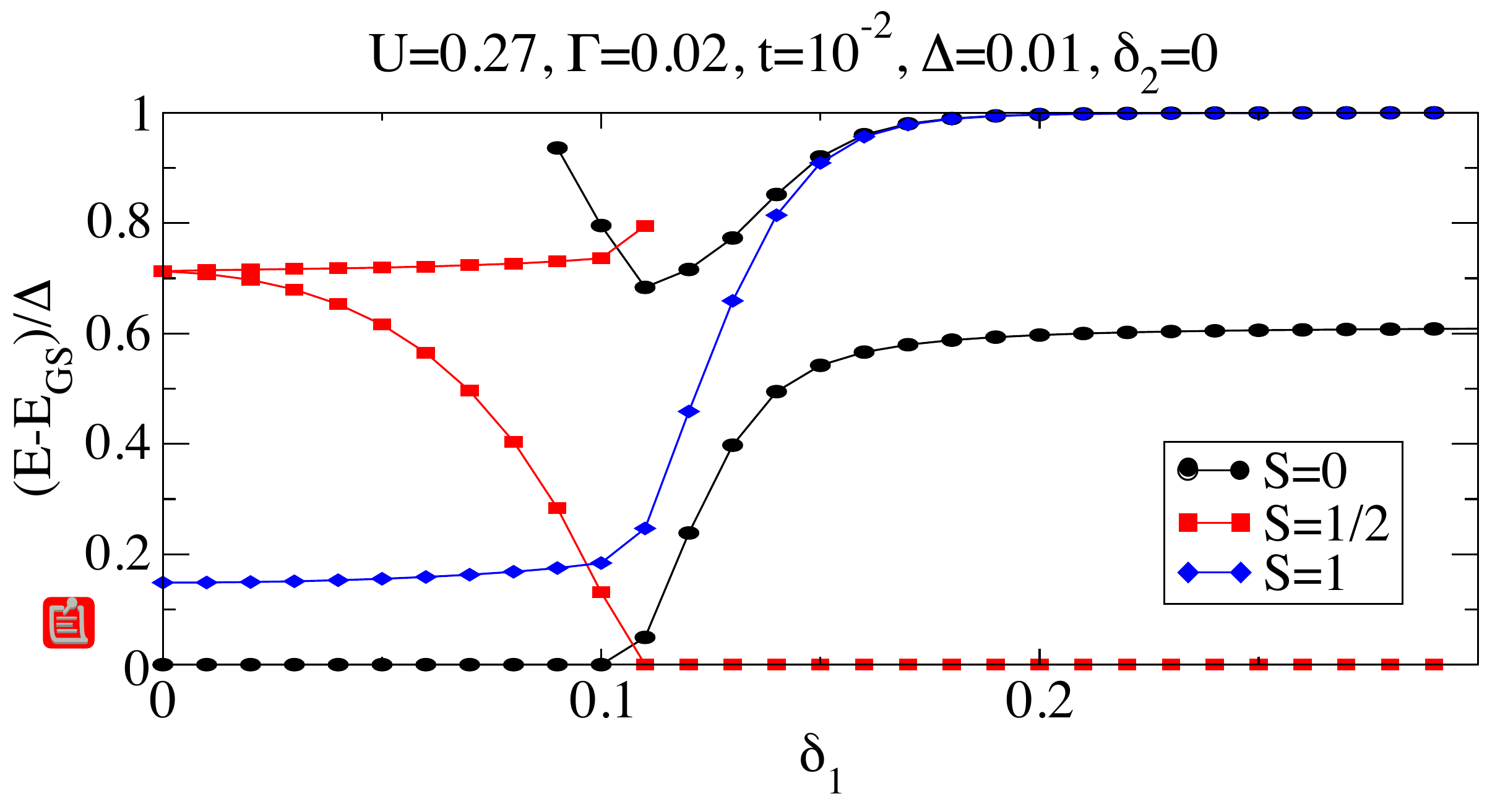}
\caption{(Color online) Sub-gap spectrum corresponding to
the center row in Fig.~\ref{fig9}.}
\label{fig10}
\end{figure}

As an aid in the interpretation, we show in Fig.~\ref{fig10} the
sub-gap many-particle spectrum. The transition occurs at $\delta_1
\approx 0.1$. For $\delta_1 < 0.1$, the spectrum is similar to that in
the $t\to 0$ limit: the ground state is a singlet, and with increasing
$\delta_1$ one of the doublet states comes down in energy and becomes
the new ground state. The effect of non-zero $t$ is visible in the
slightly increasing (vs. $\delta_1$) energy of the other doublet state
and in the singlet-triplet splitting.  The most interesting features
occur for $\delta_1 \gtrsim 0.1$.  The differences between
$A_1(\omega)$ and $A_2(\omega)$ are notable. In $A_2(\omega)$, the
most pronounced features correspond to the transitions from the
doublet GS to the two excited singlet states, while the transition to
the triplet state is not visible at all. In $A_1(\omega)$, on the
other hand, the dominant feature is the transition from the doublet GS
to the {\sl triplet excited state}, while the weights for the
transitions to the singlet excited states are much weaker. 
Since in the atomic limit the triplet state does not depend at all on
$\delta_i$, nor on $t$, and the lowest lying singlet state depends on
these parameters only very weakly, the differences between the two
spectra are predominantly due to the different site amplitudes of the
doublet wavefunction. The very strong transition on the particle side
($\omega>0$) of $A_1(\omega)$ is due to the charge depletion by high
on-site potential which leads to strong enhancement of the
matrix element for the particle addition. 

For $\delta_1 \gg 0.1$, the only resonance remaining inside the gap is
that corresponding to the Shiba state on dot 2: it is positioned at
$0.6 \Delta$ in the large $\delta_1$ limit where the dot 1 has little
effect on dot 2 in the sub-gap energy range. Other sub-gap states are
pushed into the continuum, since the dot 1 no longer has a magnetic
moment.  

For large $t=4.62\times 10^{-2}$, bottom row in Fig.~\ref{fig9}, we
enter the regime of a single effective QD, thus the variation of the
sub-gap state manifests in a similar way in both spectral functions,
except for the differences in spectral weights due to the different
on-site energies.

\begin{figure*}
\centering
\includegraphics[width=0.45\textwidth]{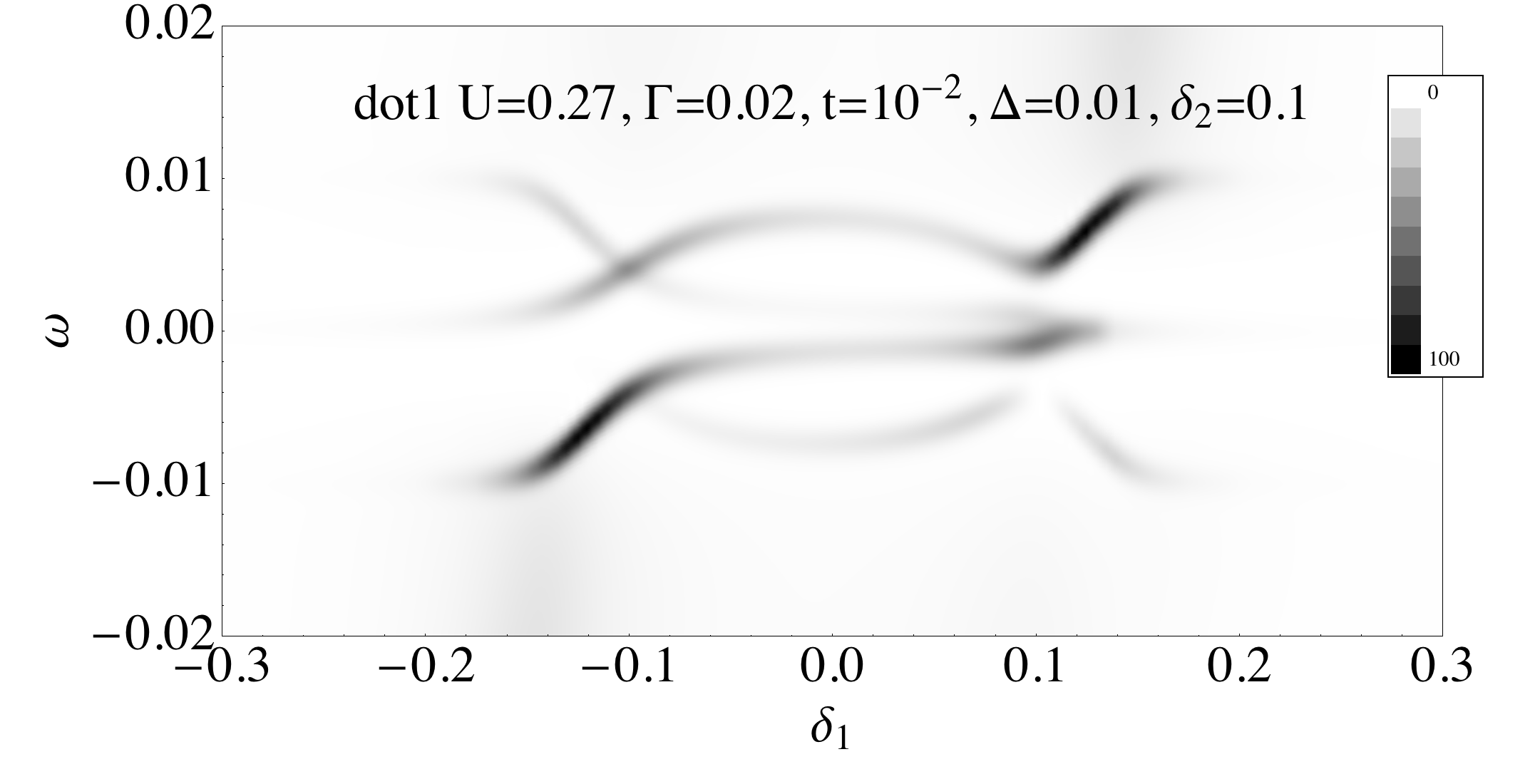}
\includegraphics[width=0.45\textwidth]{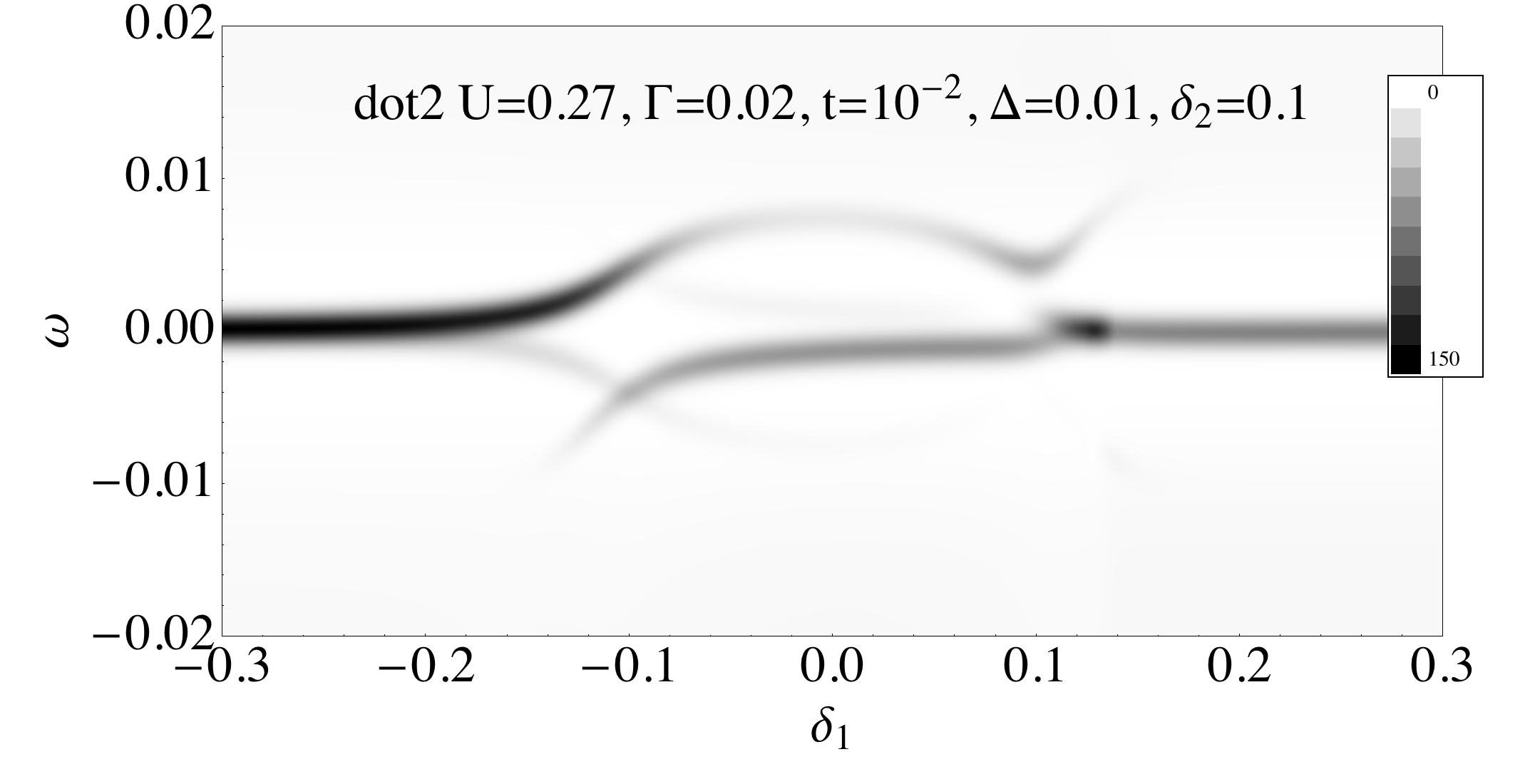}
\includegraphics[width=0.45\textwidth]{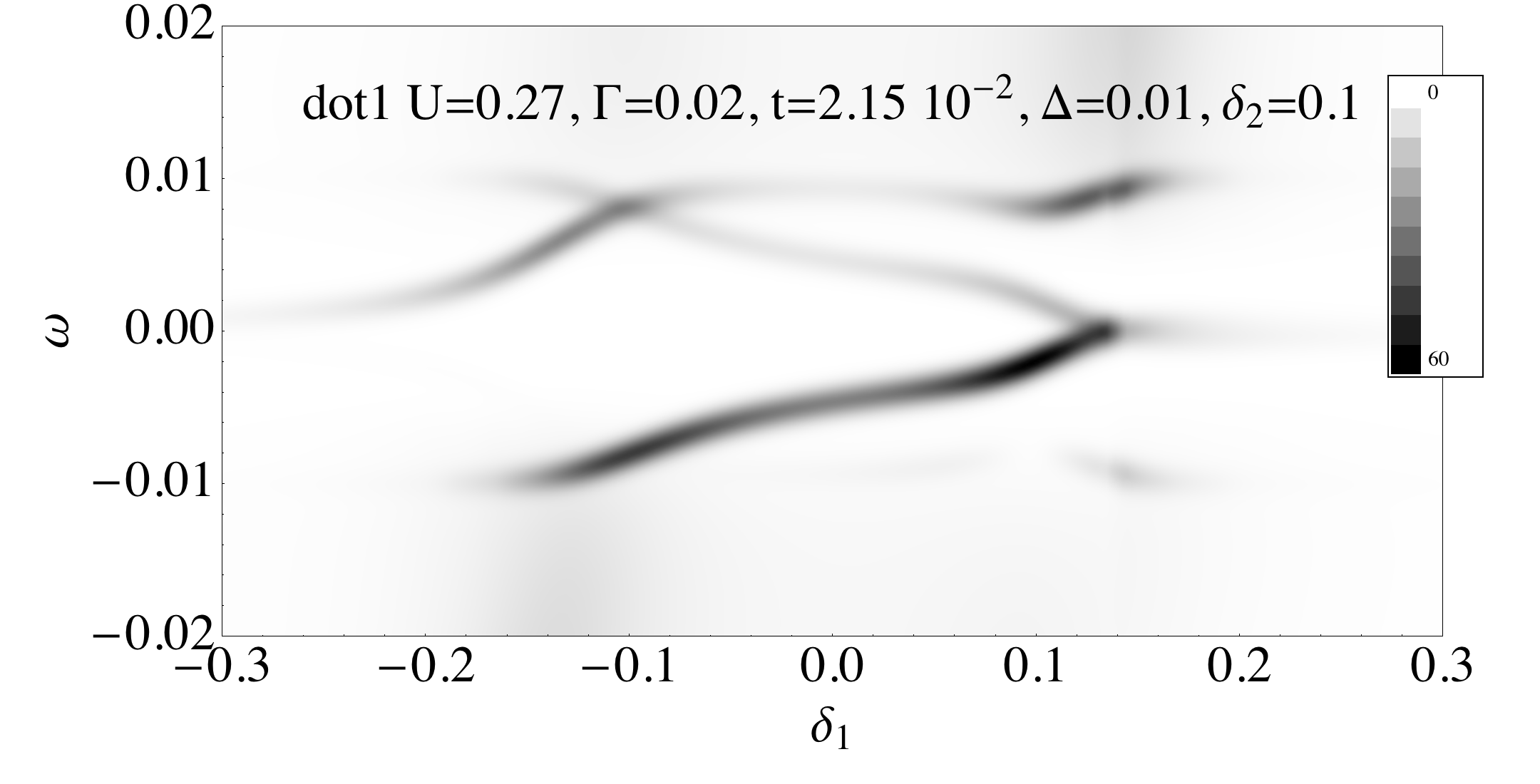}
\includegraphics[width=0.45\textwidth]{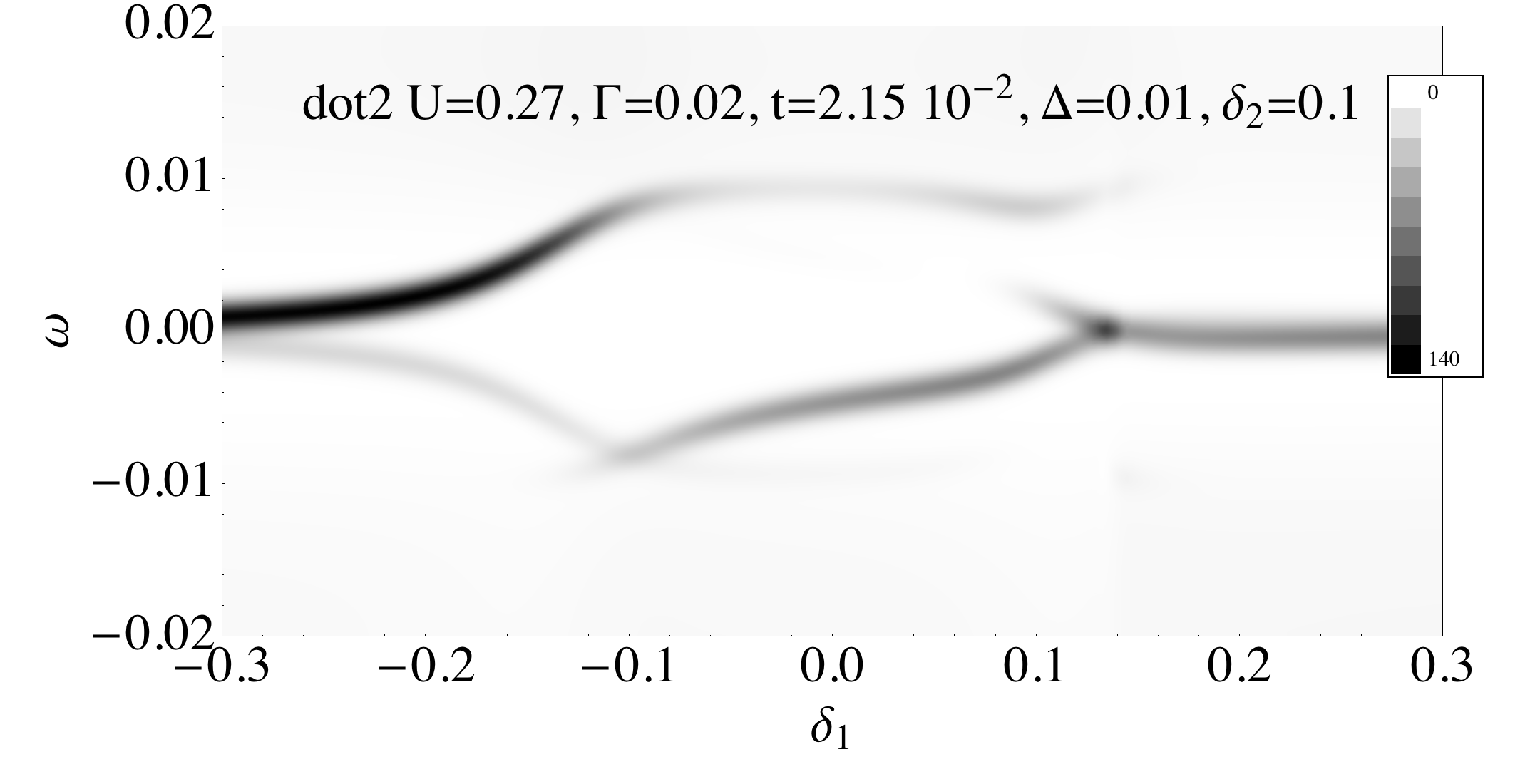}
\includegraphics[width=0.45\textwidth]{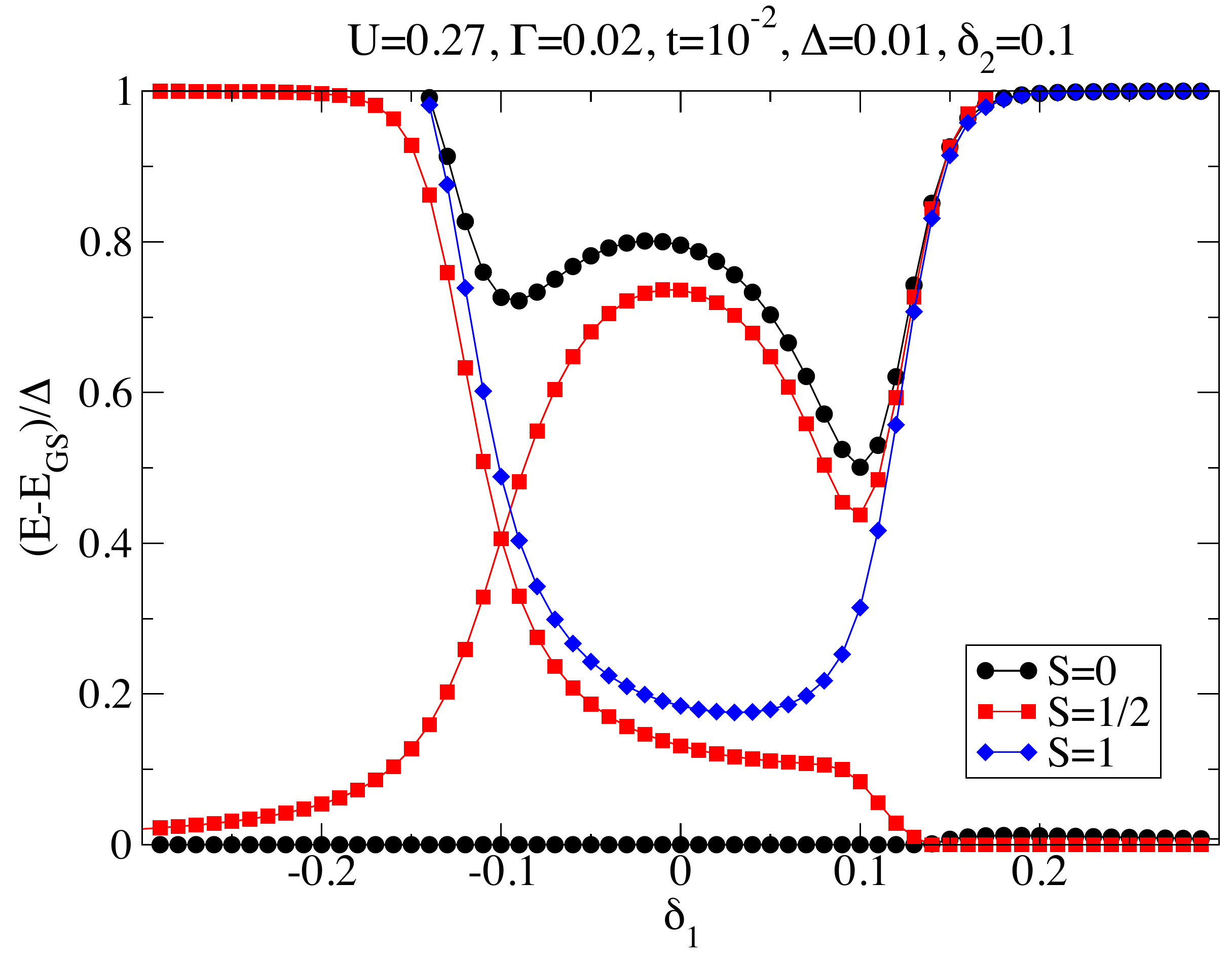}
\includegraphics[width=0.45\textwidth]{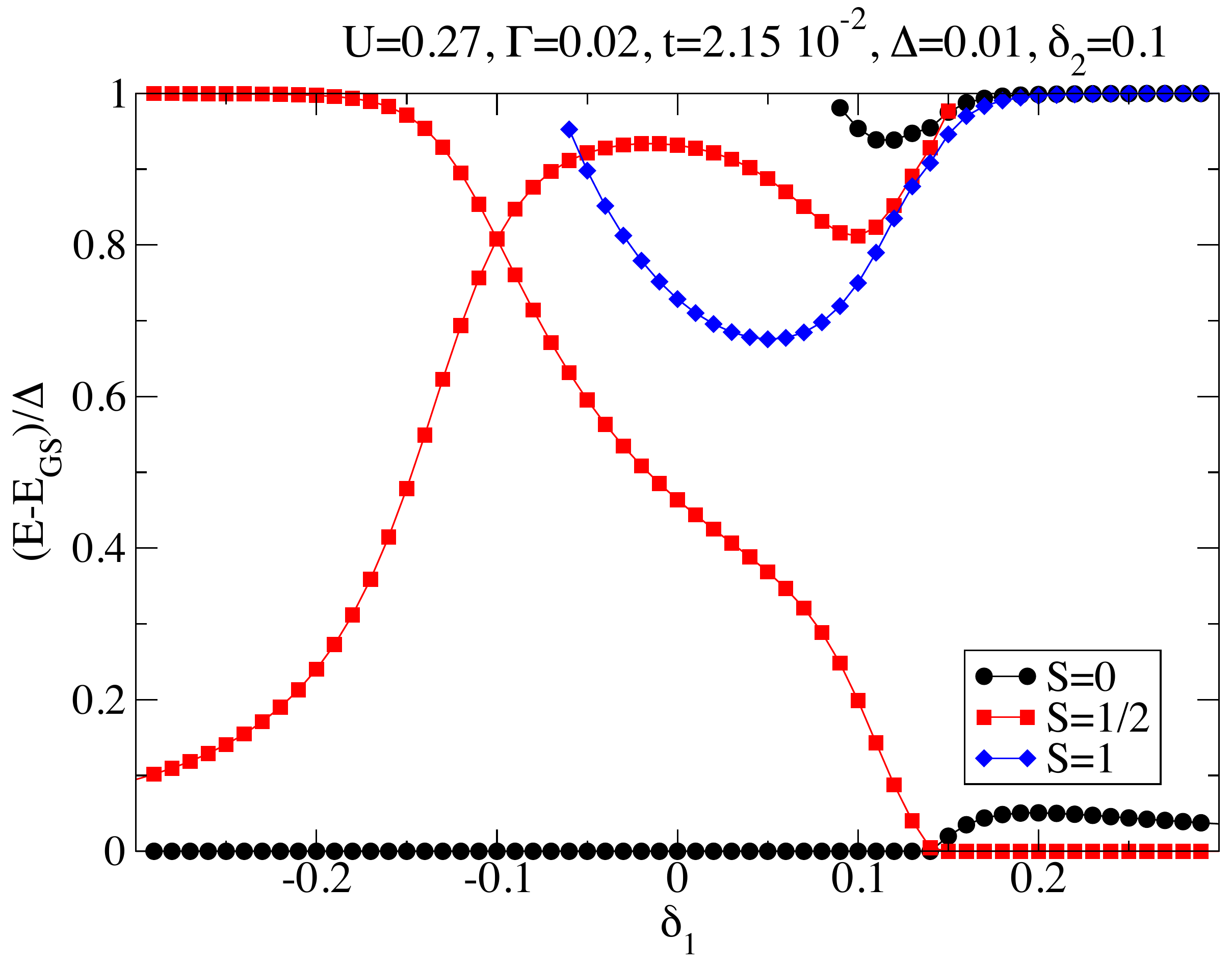}
\caption{(Color online) Spectral function on dot 1 (left panels) and
dot 2 (right panels) in the {\sl superconducting} state for a range of inter-dot
couplings $t$. The quantum dot 2 is kept at $\delta_2=0.1$, which is the
{\sl valence fluctuation regime}. The last row shows the corresponding
many-particle Shiba state diagrams.}
\label{fig12}
\end{figure*}

\subsubsection{Dot 2 in the valence fluctuation regime}

We now consider the case of $\delta_2=0.1$ where the second quantum
dot is in the valence fluctuation regime, see Fig.~\ref{fig12}. The
symmetry $\delta_1 \leftrightarrow -\delta_1$ is lost and we need to
consider the full range for $\delta_1$. The results for $\delta_1 \sim
0$ have been discussed in the previous subsection (with the roles of
$\delta_1$ and $\delta_2$ reversed), thus we focus on the cases where
$\delta_1$ is itself in the valence fluctuating range $|\delta_1|
\approx 0.1$.

For $\delta_1>0$, we find a transition to a doublet ground state at
$\delta_1 \approx 0.14$. For larger $\delta_1$, the excited singlet
state remains at low energies, while all other Shiba states move up in
energy and merge with the continuum. In this regime, the dot 1 is
fully depleted, thus the Shiba spectrum is determined by dot 2 which
is very close to the occupancy controlled singlet-doublet transition.
Consequently, this leads to a spectral resonance very close to
$\omega=0$ for all $\delta_1 \gtrsim 0.14$.

For $\delta_1<0$, the behavior is rather different. The dot 1 is
progressively filled, moving across the valence fluctuation point at
$\delta_1=-0.1$. The ground state is a singlet for all negative
$\delta_1$, but the doublet excitations undergo an interesting
evolution, namely at $\delta_1=-0.1$ we find a crossing of two
different types of the doublet excitations. This is the same type of
degeneracy of the doublet states as found at half-filling in the
left-right symmetric problems. The level crossing is thus a sign of a
symmetry which is established at $\delta_1=-\delta_2$, namely a
combination of reflection and particle-hole transformation, which
leaves the Hamiltonian invariant. Again, there is a spectral peak
close to $\omega=0$ for $\delta_1 < -0.1$. In this case, the residual
coupling to dot 2 leads to a different order of Shiba states, i.e.,
the ground state is a doublet, while the excited state is a singlet.
For large negative $\delta_1$ this order should eventually be reversed
again, because the excitation spectra for $\delta_1 \to \pm \infty$
are expected to be the same.

Finally, we briefly discuss the results for the case where one of the
dots (dot 2, for definitiveness) is driven to the empty orbital regime
by tuning $\delta_2$ to a large value well above $U/2+\Gamma \approx
0.15$. The system then essentially behaves as a single quantum dot;
this includes also the level diagram of the Shiba states. Even though
the dot 2 is unoccupied, at finite inter-dot coupling there are still 
some faint features from $A_1(\omega)$ being mirrored into the
spectral function $A_2(\omega)$ (results not shown). As $t$ grows one
can smoothly reach the regime of molecular orbitals. To conclude, for
large $\delta_2$ the DQD always effectively behaves as a single
quantum dot: for small $t$, the effective dot is the physical QD 1,
and for large $t$ it is the ``large'' QD made of molecular orbitals
spanning both dots. These two regimes are smoothly connected.

\section{Two-impurity Kondo ``criticality''}
\label{sec4}

In the DQD with normal-state leads the low-temperature
physics is governed by the two-impurity Kondo effect. The
non-Fermi-liquid (NFL) fixed point of this model is never actually
reached because the particle exchange between the leads is a relevant
perturbation (in the renormalization group sense) \cite{zarand2006},
but its presence still has important effects in a wide parameter
range. In Sec.~III~B we observed that at half filling the ground state
is the same singlet many-particle state for any value of $t$; this is
the case for any $\Gamma/U$ ratio. This continuity is the
superconducting counterpart of the cross-over behavior found in the
normal state. The presence of the two-impurity Kondo model fixed point
§is thus only felt through the non-trivial evolution of the
expectation values, in particular that of the spin-spin correlation
function $\expv{\vc{S}_1 \cdot \vc{S}_2}(t)$.

The NFL fixed point can be approached in the normal state arbitrarily
closely, however, if the particle exchange between the two channels is
suppressed while the magnetic exchange coupling is maintained
\cite{zarand2006}. We consider here how this scenario manifests in the
superconducting case through the properties of the sub-gap states. 
The inter-dot coupling term in the Hamiltonian is thus replaced by
\begin{equation}
H_{12} = J \vc{S}_1 \cdot \vc{S}_2,
\end{equation}
where $\vc{S}_i$ are the spin operators of the impurities.

\begin{figure}
\centering
\includegraphics[width=0.45\textwidth]{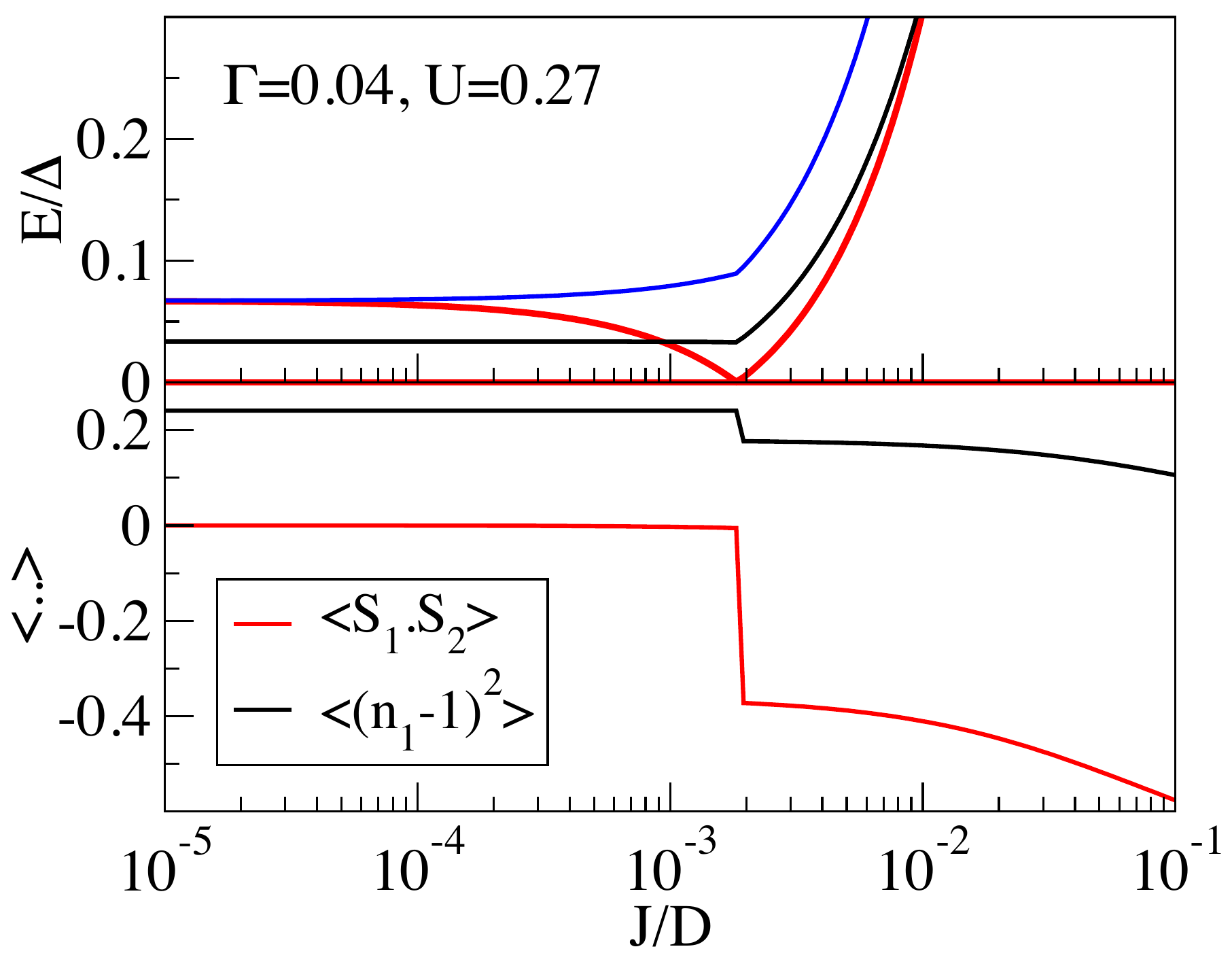}
\caption{(Color online) Two-impurity Kondo effect manifesting through
the quantum phase transition between two different many-particle
sub-gap singlet states in the model with exchange coupling $J$ between
the dots and no particle transfer ($t=0$).}
\label{critical}
\end{figure}

The results are shown in Fig.~\ref{critical} for a value of $\Gamma/U$
in the suitable range for a competition between the singlet ground
states (a) consisting of two separate Kondo clouds and (b) inter-site
singlet generated by the superexchange coupling $4t^2/U$. We indeed
observe a genuine quantum phase transition (level crossing) between
the two singlets at $J = J_c \approx 1.1\times 10^{-3}$. The state (a)
is characterized by near-zero spin correlations, $\expv{\vc{S}_1 \cdot
\vc{S}_2} \approx 0$, while the state (b) is an inter-site singlet
state with large spin-spin values. All other system properties also
change discontinuous across the transition (see the example of the
charge fluctuations, $\expv{(n_1-1)^2}$, also shown in
Fig.~\ref{critical}).

The excited singlet sub-gap state is not directly spectroscopically
observable in tunneling experiments. Nevertheless, the phase
transition does manifest as a visible spectral discontinuity: there is
a kink in the energy of the doublet excitation, and the spectral
weights have a jump (results not shown). Alternatively, the
radio-frequency spectroscopy could be used \cite{yao2014}. The
two-impurity Kondo effect quantum phase transition is thus in
principle spectroscopically observable, if only a system with
sufficiently suppressed particle exchange could be physically realized
\cite{zarand2006}. 

This section may be concluded with the following observation: in the
normal-state case, the two-impurity Kondo quantum phase transition is
a second-order transition with true quantum criticality, while in the
superconducting case it is a first order transition (i.e., level
crossing between two sub-gap singlet states that are separated from
the quasiparticle continuum). An interesting question for future work
is to explore the structure of the excitations in the continuum above
$\Delta$: are the Bogolioubov states formed out of Fermi liquid or
non-Fermi liquid quasiparticles? For $\Delta < T_K^{2IK}$, the latter
should be the case.

\section{Conclusion}

Double quantum dots are described by impurity models with properties
controlled by a number of fixed points. To each fixed point (regime)
it is possible to associate particular features observed in the
sub-gap spectra. In this work double dots were studied with
a quantitatively accurate method focusing on the regimes of strong
dependence on model parameters that could be targeted by future
experiments.  Qualitative trends can be reproduced in the
superconducting atomic limit which projects out the continua of the
quasiparticle states above $\Delta$ by taking the $\Delta \to \infty$
limit, while for quantitative correctness it is crucial to use a
non-perturbative technique such as the NRG, especially in the regimes
where the Kondo effect plays an important role. 

We now summarize the main experimentally verifiable predictions of
this work:
\begin{itemize}
\item the Kondo, antiferromagnetic and molecular-orbital regimes
at half-filling can be distinguished by the features in the spectral
function both inside and outside the gap;

\item the splitting of the doublet excited states induced by the
flux (difference of superconducting phases) is only visible away from
half-filling and is proportional to the inter-dot coupling;

\item in the large hybridization regime, the inter-dot triplet state
has lower energy than the inter-dot singlet state due to the level
repulsion between the singlet Shiba states;

\item at half-filling, the inter-dot capacitive coupling leads to
a cross-over at $V \sim U$ between spin-singlet and charge-singlet
regimes;

\item at quarter-filling, the inter-dot capacitive coupling generates
a quantum phase transition to a doublet ground state which corresponds
to the local-moment fixed point with four-fold degeneracy in the $t
\to 0$ limit (finite $t$ leads to a splitting of the doublet states); 

\item by tuning one dot to the Kondo regime and the other to the 
valence fluctuation regime, the ground state becomes a doublet and
both singlet and triplet excitations can be observed: the triplet
state have high spectral weight on the valence fluctuating dots, while
the singlet state is better observed on the half-filled dot in the
Kondo regime;

\item by tuning both dots to the valence fluctuation regime, one
fluctuating between zero and single occupancy, the other between
double and single occupancy, the system has combined reflection and
particle-hole transformation symmetry, leading to a crossing of the
excited doublet states.
\end{itemize}
The transitions which change the spin by $1/2$ are directly observable
in tunneling spectroscopy, while radiofrequency spectroscopy would
uncover further transitions. It should be stressed that some of the
most interesting features are present (or become observable) only away
from half-filling. The valence fluctuation and local moment regimes
near quarter filling are just as intriguing, if not more, as the Kondo
regime at half filling.

\begin{acknowledgments}
I acknowledge the support of the Slovenian Research Agency (ARRS)
under Program No. P1-0044. 
\end{acknowledgments}

\bibliography{paper}

\begin{thebibliography}{73}%
\makeatletter
\providecommand \@ifxundefined [1]{%
 \@ifx{#1\undefined}
}%
\providecommand \@ifnum [1]{%
 \ifnum #1\expandafter \@firstoftwo
 \else \expandafter \@secondoftwo
 \fi
}%
\providecommand \@ifx [1]{%
 \ifx #1\expandafter \@firstoftwo
 \else \expandafter \@secondoftwo
 \fi
}%
\providecommand \natexlab [1]{#1}%
\providecommand \enquote  [1]{``#1''}%
\providecommand \bibnamefont  [1]{#1}%
\providecommand \bibfnamefont [1]{#1}%
\providecommand \citenamefont [1]{#1}%
\providecommand \href@noop [0]{\@secondoftwo}%
\providecommand \href [0]{\begingroup \@sanitize@url \@href}%
\providecommand \@href[1]{\@@startlink{#1}\@@href}%
\providecommand \@@href[1]{\endgroup#1\@@endlink}%
\providecommand \@sanitize@url [0]{\catcode `\\12\catcode `\$12\catcode
  `\&12\catcode `\#12\catcode `\^12\catcode `\_12\catcode `\%12\relax}%
\providecommand \@@startlink[1]{}%
\providecommand \@@endlink[0]{}%
\providecommand \url  [0]{\begingroup\@sanitize@url \@url }%
\providecommand \@url [1]{\endgroup\@href {#1}{\urlprefix }}%
\providecommand \urlprefix  [0]{URL }%
\providecommand \Eprint [0]{\href }%
\providecommand \doibase [0]{http://dx.doi.org/}%
\providecommand \selectlanguage [0]{\@gobble}%
\providecommand \bibinfo  [0]{\@secondoftwo}%
\providecommand \bibfield  [0]{\@secondoftwo}%
\providecommand \translation [1]{[#1]}%
\providecommand \BibitemOpen [0]{}%
\providecommand \bibitemStop [0]{}%
\providecommand \bibitemNoStop [0]{.\EOS\space}%
\providecommand \EOS [0]{\spacefactor3000\relax}%
\providecommand \BibitemShut  [1]{\csname bibitem#1\endcsname}%
\let\auto@bib@innerbib\@empty
\bibitem [{\citenamefont {Giaever}(1960)}]{giaever1960}%
  \BibitemOpen
  \bibfield  {author} {\bibinfo {author} {\bibfnamefont {Ivar}\ \bibnamefont
  {Giaever}},\ }\bibfield  {title} {\enquote {\bibinfo {title} {Energy gap in
  superconductors measured by electron tunneling},}\ }\href@noop {} {\bibfield
  {journal} {\bibinfo  {journal} {Phys. Rev. Lett.}\ }\textbf {\bibinfo
  {volume} {5}},\ \bibinfo {pages} {147} (\bibinfo {year} {1960})}\BibitemShut
  {NoStop}%
\bibitem [{\citenamefont {Giaever}(1974)}]{giaever1974}%
  \BibitemOpen
  \bibfield  {author} {\bibinfo {author} {\bibfnamefont {I.}~\bibnamefont
  {Giaever}},\ }\bibfield  {title} {\enquote {\bibinfo {title} {Electron
  tunneling and superconductivity},}\ }\href@noop {} {\bibfield  {journal}
  {\bibinfo  {journal} {Rev. Mod. Phys.}\ }\textbf {\bibinfo {volume} {46}},\
  \bibinfo {pages} {245} (\bibinfo {year} {1974})}\BibitemShut {NoStop}%
\bibitem [{\citenamefont {Pan}\ \emph {et~al.}(2000)\citenamefont {Pan},
  \citenamefont {Hudson}, \citenamefont {Lang}, \citenamefont {Eisaki},
  \citenamefont {Uchida},\ and\ \citenamefont {Davis}}]{pan1999zn}%
  \BibitemOpen
  \bibfield  {author} {\bibinfo {author} {\bibfnamefont {S.~H.}\ \bibnamefont
  {Pan}}, \bibinfo {author} {\bibfnamefont {E.~W.}\ \bibnamefont {Hudson}},
  \bibinfo {author} {\bibfnamefont {K.~M.}\ \bibnamefont {Lang}}, \bibinfo
  {author} {\bibfnamefont {H.}~\bibnamefont {Eisaki}}, \bibinfo {author}
  {\bibfnamefont {S.}~\bibnamefont {Uchida}}, \ and\ \bibinfo {author}
  {\bibfnamefont {J.~C.}\ \bibnamefont {Davis}},\ }\bibfield  {title} {\enquote
  {\bibinfo {title} {Imaging the effects of individual zinc impurity atoms on
  superconductivity in {Bi$_2$Sr$_2$CaCu$_2$O$_{8+\delta}$}},}\ }\href@noop {}
  {\bibfield  {journal} {\bibinfo  {journal} {Nature}\ }\textbf {\bibinfo
  {volume} {403}},\ \bibinfo {pages} {746} (\bibinfo {year}
  {2000})}\BibitemShut {NoStop}%
\bibitem [{\citenamefont {Balatsky}\ \emph {et~al.}(2006)\citenamefont
  {Balatsky}, \citenamefont {Vekhter},\ and\ \citenamefont
  {Zhu}}]{balatsky2006}%
  \BibitemOpen
  \bibfield  {author} {\bibinfo {author} {\bibfnamefont {A.~V.}\ \bibnamefont
  {Balatsky}}, \bibinfo {author} {\bibfnamefont {I.}~\bibnamefont {Vekhter}}, \
  and\ \bibinfo {author} {\bibfnamefont {Jian-Xin}\ \bibnamefont {Zhu}},\
  }\bibfield  {title} {\enquote {\bibinfo {title} {Impurity-induced states in
  conventional and unconventional superconductors},}\ }\href@noop {} {\bibfield
   {journal} {\bibinfo  {journal} {Rev. Mod. Phys.}\ }\textbf {\bibinfo
  {volume} {78}},\ \bibinfo {pages} {373} (\bibinfo {year} {2006})}\BibitemShut
  {NoStop}%
\bibitem [{\citenamefont {Ji}\ \emph {et~al.}(2008)\citenamefont {Ji},
  \citenamefont {Zhang}, \citenamefont {Fu}, \citenamefont {Chen},
  \citenamefont {Ma}, \citenamefont {Li}, \citenamefont {Duan}, \citenamefont
  {Jia},\ and\ \citenamefont {Xue}}]{ji2008}%
  \BibitemOpen
  \bibfield  {author} {\bibinfo {author} {\bibfnamefont {S.~H.}\ \bibnamefont
  {Ji}}, \bibinfo {author} {\bibfnamefont {T.}~\bibnamefont {Zhang}}, \bibinfo
  {author} {\bibfnamefont {Y.~S.}\ \bibnamefont {Fu}}, \bibinfo {author}
  {\bibfnamefont {X.}~\bibnamefont {Chen}}, \bibinfo {author} {\bibfnamefont
  {Xu-Cun}\ \bibnamefont {Ma}}, \bibinfo {author} {\bibfnamefont
  {J.}~\bibnamefont {Li}}, \bibinfo {author} {\bibfnamefont {Wen-Hui}\
  \bibnamefont {Duan}}, \bibinfo {author} {\bibfnamefont {Jin-Feng}\
  \bibnamefont {Jia}}, \ and\ \bibinfo {author} {\bibfnamefont {Qi-Kun}\
  \bibnamefont {Xue}},\ }\bibfield  {title} {\enquote {\bibinfo {title}
  {High-resolution scanning tunneling spectroscopy of magnetic impurity induced
  bound states in the superconducting gap of {Pb} thin films},}\ }\href@noop {}
  {\bibfield  {journal} {\bibinfo  {journal} {Phys. Rev. Lett.}\ }\textbf
  {\bibinfo {volume} {100}},\ \bibinfo {pages} {226801} (\bibinfo {year}
  {2008})}\BibitemShut {NoStop}%
\bibitem [{\citenamefont {Iavarone}\ \emph {et~al.}(2010)\citenamefont
  {Iavarone}, \citenamefont {Karapetrov}, \citenamefont {Fedor}, \citenamefont
  {Rosenmann}, \citenamefont {Nishizaki},\ and\ \citenamefont
  {Kobayashi}}]{iavarone2010}%
  \BibitemOpen
  \bibfield  {author} {\bibinfo {author} {\bibfnamefont {M}~\bibnamefont
  {Iavarone}}, \bibinfo {author} {\bibfnamefont {G}~\bibnamefont {Karapetrov}},
  \bibinfo {author} {\bibfnamefont {J}~\bibnamefont {Fedor}}, \bibinfo {author}
  {\bibfnamefont {D}~\bibnamefont {Rosenmann}}, \bibinfo {author}
  {\bibfnamefont {T}~\bibnamefont {Nishizaki}}, \ and\ \bibinfo {author}
  {\bibfnamefont {N}~\bibnamefont {Kobayashi}},\ }\bibfield  {title} {\enquote
  {\bibinfo {title} {The local effect of magnetic impurities on
  superconductivity in {Co$_x$NbSe$_2$} and {Mn$_x$NbSe$_2$} single
  crystals},}\ }\href@noop {} {\bibfield  {journal} {\bibinfo  {journal} {J.
  Phys.: Condens. Matter}\ }\textbf {\bibinfo {volume} {22}},\ \bibinfo {pages}
  {015501} (\bibinfo {year} {2010})}\BibitemShut {NoStop}%
\bibitem [{\citenamefont {Ji}\ \emph {et~al.}(2010)\citenamefont {Ji},
  \citenamefont {Zhang}, \citenamefont {Fu}, \citenamefont {Chen},
  \citenamefont {Jia}, \citenamefont {Xue},\ and\ \citenamefont {Ma}}]{ji2010}%
  \BibitemOpen
  \bibfield  {author} {\bibinfo {author} {\bibfnamefont {Shuai-Hua}\
  \bibnamefont {Ji}}, \bibinfo {author} {\bibfnamefont {Tong}\ \bibnamefont
  {Zhang}}, \bibinfo {author} {\bibfnamefont {Ying-Shuang}\ \bibnamefont {Fu}},
  \bibinfo {author} {\bibfnamefont {Xi}~\bibnamefont {Chen}}, \bibinfo {author}
  {\bibfnamefont {Jin-Feng}\ \bibnamefont {Jia}}, \bibinfo {author}
  {\bibfnamefont {Qi-Kun}\ \bibnamefont {Xue}}, \ and\ \bibinfo {author}
  {\bibfnamefont {Xu-Cun}\ \bibnamefont {Ma}},\ }\bibfield  {title} {\enquote
  {\bibinfo {title} {Application of magnetic atom induced bound states in
  superconducting gap for chemical identification of single magnetic atoms},}\
  }\href@noop {} {\bibfield  {journal} {\bibinfo  {journal} {Appl. Phys.
  Lett.}\ }\textbf {\bibinfo {volume} {96}},\ \bibinfo {pages} {073113}
  (\bibinfo {year} {2010})}\BibitemShut {NoStop}%
\bibitem [{\citenamefont {Franke}\ \emph {et~al.}(2011)\citenamefont {Franke},
  \citenamefont {Schulze},\ and\ \citenamefont {Pascual}}]{franke2011}%
  \BibitemOpen
  \bibfield  {author} {\bibinfo {author} {\bibfnamefont {K.~J.}\ \bibnamefont
  {Franke}}, \bibinfo {author} {\bibfnamefont {G.}~\bibnamefont {Schulze}}, \
  and\ \bibinfo {author} {\bibfnamefont {J.~I.}\ \bibnamefont {Pascual}},\
  }\bibfield  {title} {\enquote {\bibinfo {title} {Competition of
  superconductivity phenomena and {Kondo} screening at the nanoscale},}\
  }\href@noop {} {\bibfield  {journal} {\bibinfo  {journal} {Science}\ }\textbf
  {\bibinfo {volume} {332}},\ \bibinfo {pages} {940} (\bibinfo {year}
  {2011})}\BibitemShut {NoStop}%
\bibitem [{\citenamefont {Ruby}\ \emph {et~al.}(2015)\citenamefont {Ruby},
  \citenamefont {Pientka}, \citenamefont {Peng}, \citenamefont {von Oppen},
  \citenamefont {Heinrich},\ and\ \citenamefont {Franke}}]{franke2015}%
  \BibitemOpen
  \bibfield  {author} {\bibinfo {author} {\bibfnamefont {M.}~\bibnamefont
  {Ruby}}, \bibinfo {author} {\bibfnamefont {F.}~\bibnamefont {Pientka}},
  \bibinfo {author} {\bibfnamefont {Y.}~\bibnamefont {Peng}}, \bibinfo {author}
  {\bibfnamefont {F.}~\bibnamefont {von Oppen}}, \bibinfo {author}
  {\bibfnamefont {B.~W.}\ \bibnamefont {Heinrich}}, \ and\ \bibinfo {author}
  {\bibfnamefont {K.~J.}\ \bibnamefont {Franke}},\ }\href@noop {} {\enquote
  {\bibinfo {title} {Tunneling processes into localized subgap states in
  superconductors},}\ }\bibinfo {howpublished} {arXiv:1502.05048} (\bibinfo
  {year} {2015})\BibitemShut {NoStop}%
\bibitem [{\citenamefont {Hewson}(1993)}]{hewson}%
  \BibitemOpen
  \bibfield  {author} {\bibinfo {author} {\bibfnamefont {A.~C.}\ \bibnamefont
  {Hewson}},\ }\href@noop {} {\emph {\bibinfo {title} {The Kondo Problem to
  Heavy-Fermions}}}\ (\bibinfo  {publisher} {Cambridge University Press,
  Cambridge},\ \bibinfo {year} {1993})\BibitemShut {NoStop}%
\bibitem [{\citenamefont {Shiba}(1968)}]{shiba1968}%
  \BibitemOpen
  \bibfield  {author} {\bibinfo {author} {\bibfnamefont {H.}~\bibnamefont
  {Shiba}},\ }\bibfield  {title} {\enquote {\bibinfo {title} {Classical spins
  in superconductors},}\ }\href@noop {} {\bibfield  {journal} {\bibinfo
  {journal} {Prog. Theor. Phys.}\ }\textbf {\bibinfo {volume} {40}},\ \bibinfo
  {pages} {435} (\bibinfo {year} {1968})}\BibitemShut {NoStop}%
\bibitem [{\citenamefont {Sakurai}(1970)}]{sakurai1970}%
  \BibitemOpen
  \bibfield  {author} {\bibinfo {author} {\bibfnamefont {Akio}\ \bibnamefont
  {Sakurai}},\ }\bibfield  {title} {\enquote {\bibinfo {title} {Comments on
  superconductors with magnetic impurities},}\ }\href@noop {} {\bibfield
  {journal} {\bibinfo  {journal} {Prog. Theor. Phys.}\ }\textbf {\bibinfo
  {volume} {44}},\ \bibinfo {pages} {1472} (\bibinfo {year}
  {1970})}\BibitemShut {NoStop}%
\bibitem [{\citenamefont {Zittartz}\ and\ \citenamefont
  {M\"uller-Hartmann}(1970)}]{zmha}%
  \BibitemOpen
  \bibfield  {author} {\bibinfo {author} {\bibfnamefont {J.}~\bibnamefont
  {Zittartz}}\ and\ \bibinfo {author} {\bibfnamefont {E.}~\bibnamefont
  {M\"uller-Hartmann}},\ }\bibfield  {title} {\enquote {\bibinfo {title}
  {Theory of magnetic impurities in superconductors. {I}, {Exact} solution of
  the {Nagaoka} equations},}\ }\href@noop {} {\bibfield  {journal} {\bibinfo
  {journal} {J. Physik}\ }\textbf {\bibinfo {volume} {232}},\ \bibinfo {pages}
  {11} (\bibinfo {year} {1970})}\BibitemShut {NoStop}%
\bibitem [{\citenamefont {Choi}\ \emph {et~al.}(2004)\citenamefont {Choi},
  \citenamefont {Lee}, \citenamefont {Kang},\ and\ \citenamefont
  {Belzig}}]{choi2004josephson}%
  \BibitemOpen
  \bibfield  {author} {\bibinfo {author} {\bibfnamefont {Mahn-Soo}\
  \bibnamefont {Choi}}, \bibinfo {author} {\bibfnamefont {Minchul}\
  \bibnamefont {Lee}}, \bibinfo {author} {\bibfnamefont {Kicheon}\ \bibnamefont
  {Kang}}, \ and\ \bibinfo {author} {\bibfnamefont {W.}~\bibnamefont
  {Belzig}},\ }\bibfield  {title} {\enquote {\bibinfo {title} {Kondo effect and
  josephson current through a quantum dot between two superconductors},}\
  }\href@noop {} {\bibfield  {journal} {\bibinfo  {journal} {Phys. Rev. B}\
  }\textbf {\bibinfo {volume} {70}},\ \bibinfo {pages} {020502} (\bibinfo
  {year} {2004})}\BibitemShut {NoStop}%
\bibitem [{\citenamefont {Bauer}\ \emph {et~al.}(2007)\citenamefont {Bauer},
  \citenamefont {Oguri},\ and\ \citenamefont {Hewson}}]{bauer2007}%
  \BibitemOpen
  \bibfield  {author} {\bibinfo {author} {\bibfnamefont {J.}~\bibnamefont
  {Bauer}}, \bibinfo {author} {\bibfnamefont {A.}~\bibnamefont {Oguri}}, \ and\
  \bibinfo {author} {\bibfnamefont {A.~C.}\ \bibnamefont {Hewson}},\ }\bibfield
   {title} {\enquote {\bibinfo {title} {Spectral properties of locally
  correlated electrons in a {Bardeen-Cooper-Schrieffer} superconductor},}\
  }\href@noop {} {\bibfield  {journal} {\bibinfo  {journal} {J. Phys.: Condens.
  Matter}\ }\textbf {\bibinfo {volume} {19}},\ \bibinfo {pages} {486211}
  (\bibinfo {year} {2007})}\BibitemShut {NoStop}%
\bibitem [{\citenamefont {Karrasch}\ \emph {et~al.}(2008)\citenamefont
  {Karrasch}, \citenamefont {Oguri},\ and\ \citenamefont
  {Meden}}]{karrasch2008}%
  \BibitemOpen
  \bibfield  {author} {\bibinfo {author} {\bibfnamefont {C.}~\bibnamefont
  {Karrasch}}, \bibinfo {author} {\bibfnamefont {A.}~\bibnamefont {Oguri}}, \
  and\ \bibinfo {author} {\bibfnamefont {V.}~\bibnamefont {Meden}},\ }\bibfield
   {title} {\enquote {\bibinfo {title} {Josephson current through a single
  {Anderson} impurity coupled to {BCS} leads},}\ }\href@noop {} {\bibfield
  {journal} {\bibinfo  {journal} {Phys. Rev. B}\ }\textbf {\bibinfo {volume}
  {77}},\ \bibinfo {pages} {024517} (\bibinfo {year} {2008})}\BibitemShut
  {NoStop}%
\bibitem [{\citenamefont {Moca}\ \emph {et~al.}(2008)\citenamefont {Moca},
  \citenamefont {Demler}, \citenamefont {Jank{\'o}},\ and\ \citenamefont
  {Zar{\'a}nd}}]{moca2008}%
  \BibitemOpen
  \bibfield  {author} {\bibinfo {author} {\bibfnamefont {C.~P.}\ \bibnamefont
  {Moca}}, \bibinfo {author} {\bibfnamefont {E.}~\bibnamefont {Demler}},
  \bibinfo {author} {\bibfnamefont {B.}~\bibnamefont {Jank{\'o}}}, \ and\
  \bibinfo {author} {\bibfnamefont {G.}~\bibnamefont {Zar{\'a}nd}},\ }\bibfield
   {title} {\enquote {\bibinfo {title} {Spin-resolved spectra of {Shiba}
  multiplets from {Mn} impurities in {MgB$_2$}},}\ }\href@noop {} {\bibfield
  {journal} {\bibinfo  {journal} {Phys. Rev. B}\ }\textbf {\bibinfo {volume}
  {77}},\ \bibinfo {pages} {174516} (\bibinfo {year} {2008})}\BibitemShut
  {NoStop}%
\bibitem [{\citenamefont {Meng}\ \emph {et~al.}(2009)\citenamefont {Meng},
  \citenamefont {Florens},\ and\ \citenamefont {Simon}}]{meng2009}%
  \BibitemOpen
  \bibfield  {author} {\bibinfo {author} {\bibfnamefont {Tobias}\ \bibnamefont
  {Meng}}, \bibinfo {author} {\bibfnamefont {Serge}\ \bibnamefont {Florens}}, \
  and\ \bibinfo {author} {\bibfnamefont {Pascal}\ \bibnamefont {Simon}},\
  }\bibfield  {title} {\enquote {\bibinfo {title} {Self-consistent description
  of {Andreev} bound states in {Josephson} quantum dot devices},}\ }\href@noop
  {} {\bibfield  {journal} {\bibinfo  {journal} {Phys. Rev. B}\ }\textbf
  {\bibinfo {volume} {79}},\ \bibinfo {pages} {224521} (\bibinfo {year}
  {2009})}\BibitemShut {NoStop}%
\bibitem [{\citenamefont {Franceschi}\ \emph {et~al.}(2010)\citenamefont
  {Franceschi}, \citenamefont {Kouwenhoven}, \citenamefont {Schönenberger},\
  and\ \citenamefont {Wernsdorfer}}]{hybrid2010}%
  \BibitemOpen
  \bibfield  {author} {\bibinfo {author} {\bibfnamefont {Silvano~De}\
  \bibnamefont {Franceschi}}, \bibinfo {author} {\bibfnamefont {Leo}\
  \bibnamefont {Kouwenhoven}}, \bibinfo {author} {\bibfnamefont {Christian}\
  \bibnamefont {Schönenberger}}, \ and\ \bibinfo {author} {\bibfnamefont
  {Wolfgang}\ \bibnamefont {Wernsdorfer}},\ }\bibfield  {title} {\enquote
  {\bibinfo {title} {Hybrid superconductor-quantum dot devices},}\ }\href@noop
  {} {\bibfield  {journal} {\bibinfo  {journal} {Nat. Nanotechnology}\ }\textbf
  {\bibinfo {volume} {5}},\ \bibinfo {pages} {703} (\bibinfo {year}
  {2010})}\BibitemShut {NoStop}%
\bibitem [{\citenamefont {Maurand}\ \emph {et~al.}(2012)\citenamefont
  {Maurand}, \citenamefont {Meng}, \citenamefont {Bonet}, \citenamefont
  {Florens}, \citenamefont {Marty},\ and\ \citenamefont
  {Wernsdorfer}}]{maurand2012}%
  \BibitemOpen
  \bibfield  {author} {\bibinfo {author} {\bibfnamefont {Romain}\ \bibnamefont
  {Maurand}}, \bibinfo {author} {\bibfnamefont {Tobias}\ \bibnamefont {Meng}},
  \bibinfo {author} {\bibfnamefont {Edgar}\ \bibnamefont {Bonet}}, \bibinfo
  {author} {\bibfnamefont {Serge}\ \bibnamefont {Florens}}, \bibinfo {author}
  {\bibfnamefont {La\"etitia}\ \bibnamefont {Marty}}, \ and\ \bibinfo {author}
  {\bibfnamefont {Wolfgang}\ \bibnamefont {Wernsdorfer}},\ }\bibfield  {title}
  {\enquote {\bibinfo {title} {First-order 0-$\pi$ quantum phase transition in
  the {Kondo} regime of a superconducting carbon-nanotube quantum dot},}\
  }\href@noop {} {\bibfield  {journal} {\bibinfo  {journal} {Phys. Rev. X}\
  }\textbf {\bibinfo {volume} {2}},\ \bibinfo {pages} {011009} (\bibinfo {year}
  {2012})}\BibitemShut {NoStop}%
\bibitem [{\citenamefont {Pillet}\ \emph {et~al.}(2010)\citenamefont {Pillet},
  \citenamefont {Quay}, \citenamefont {Morin}, \citenamefont {Bena},
  \citenamefont {Yeyati},\ and\ \citenamefont {Joyez}}]{pillet2010}%
  \BibitemOpen
  \bibfield  {author} {\bibinfo {author} {\bibfnamefont {J.-D.}\ \bibnamefont
  {Pillet}}, \bibinfo {author} {\bibfnamefont {C.~H.~L.}\ \bibnamefont {Quay}},
  \bibinfo {author} {\bibfnamefont {P.}~\bibnamefont {Morin}}, \bibinfo
  {author} {\bibfnamefont {C.}~\bibnamefont {Bena}}, \bibinfo {author}
  {\bibfnamefont {A.~Levy}\ \bibnamefont {Yeyati}}, \ and\ \bibinfo {author}
  {\bibfnamefont {P.}~\bibnamefont {Joyez}},\ }\bibfield  {title} {\enquote
  {\bibinfo {title} {Andreev bound states in supercurrent-carrying carbon
  nanotubes revealed},}\ }\href@noop {} {\bibfield  {journal} {\bibinfo
  {journal} {Nat. Physics}\ }\textbf {\bibinfo {volume} {6}},\ \bibinfo {pages}
  {965} (\bibinfo {year} {2010})}\BibitemShut {NoStop}%
\bibitem [{\citenamefont {Lee}\ \emph {et~al.}(2014)\citenamefont {Lee},
  \citenamefont {Jiang}, \citenamefont {Houzet}, \citenamefont {Aguado},
  \citenamefont {Lieber},\ and\ \citenamefont {De~Franceschi}}]{lee2014}%
  \BibitemOpen
  \bibfield  {author} {\bibinfo {author} {\bibfnamefont {E.~J.~H}\ \bibnamefont
  {Lee}}, \bibinfo {author} {\bibfnamefont {X.}~\bibnamefont {Jiang}}, \bibinfo
  {author} {\bibfnamefont {M.}~\bibnamefont {Houzet}}, \bibinfo {author}
  {\bibfnamefont {R.}~\bibnamefont {Aguado}}, \bibinfo {author} {\bibfnamefont
  {C.~M.}\ \bibnamefont {Lieber}}, \ and\ \bibinfo {author} {\bibfnamefont
  {S.}~\bibnamefont {De~Franceschi}},\ }\bibfield  {title} {\enquote {\bibinfo
  {title} {Spin-resolved {Andreev} levels and parity crossings in hybrid
  superconductor-semiconductor nanostructures},}\ }\href@noop {} {\bibfield
  {journal} {\bibinfo  {journal} {Nature Nanotech.}\ }\textbf {\bibinfo
  {volume} {9}},\ \bibinfo {pages} {79} (\bibinfo {year} {2014})}\BibitemShut
  {NoStop}%
\bibitem [{\citenamefont {Pillet}\ \emph {et~al.}(2013)\citenamefont {Pillet},
  \citenamefont {Joyez}, \citenamefont {\v{Z}itko},\ and\ \citenamefont
  {Goffman}}]{pillet2013}%
  \BibitemOpen
  \bibfield  {author} {\bibinfo {author} {\bibfnamefont {J.-D.}\ \bibnamefont
  {Pillet}}, \bibinfo {author} {\bibfnamefont {P.}~\bibnamefont {Joyez}},
  \bibinfo {author} {\bibfnamefont {R.}~\bibnamefont {\v{Z}itko}}, \ and\
  \bibinfo {author} {\bibfnamefont {M.~F.}\ \bibnamefont {Goffman}},\
  }\bibfield  {title} {\enquote {\bibinfo {title} {Tunneling spectroscopy of a
  single quantum dot coupled to a superconductor: From {Kondo} ridge to
  {Andreev} bound states},}\ }\href@noop {} {\bibfield  {journal} {\bibinfo
  {journal} {Phys. Rev. B}\ }\textbf {\bibinfo {volume} {88}},\ \bibinfo
  {pages} {045101} (\bibinfo {year} {2013})}\BibitemShut {NoStop}%
\bibitem [{\citenamefont {Mart{\'\i}n-Rodero}\ and\ \citenamefont
  {Levy~Yeyati}(2011)}]{rodero2011}%
  \BibitemOpen
  \bibfield  {author} {\bibinfo {author} {\bibfnamefont {A.}~\bibnamefont
  {Mart{\'\i}n-Rodero}}\ and\ \bibinfo {author} {\bibfnamefont
  {A.}~\bibnamefont {Levy~Yeyati}},\ }\bibfield  {title} {\enquote {\bibinfo
  {title} {Josephson and {Andreev} transport through quantum dots},}\
  }\href@noop {} {\bibfield  {journal} {\bibinfo  {journal} {Advances in
  Physics}\ }\textbf {\bibinfo {volume} {60}},\ \bibinfo {pages} {899--958}
  (\bibinfo {year} {2011})}\BibitemShut {NoStop}%
\bibitem [{\citenamefont {Georges}\ and\ \citenamefont
  {Meir}(1999)}]{georges1999}%
  \BibitemOpen
  \bibfield  {author} {\bibinfo {author} {\bibfnamefont {A.}~\bibnamefont
  {Georges}}\ and\ \bibinfo {author} {\bibfnamefont {Y.}~\bibnamefont {Meir}},\
  }\bibfield  {title} {\enquote {\bibinfo {title} {Electronic correlations in
  transport through coupled quantum dots},}\ }\href@noop {} {\bibfield
  {journal} {\bibinfo  {journal} {Phys. Rev. Lett.}\ }\textbf {\bibinfo
  {volume} {82}},\ \bibinfo {pages} {3508} (\bibinfo {year}
  {1999})}\BibitemShut {NoStop}%
\bibitem [{\citenamefont {Aguado}\ and\ \citenamefont
  {Langreth}(2000)}]{aguado2000}%
  \BibitemOpen
  \bibfield  {author} {\bibinfo {author} {\bibfnamefont {Ramon}\ \bibnamefont
  {Aguado}}\ and\ \bibinfo {author} {\bibfnamefont {David~C.}\ \bibnamefont
  {Langreth}},\ }\bibfield  {title} {\enquote {\bibinfo {title}
  {Out-of-equilibrium {Kondo} effect in double quantum dots},}\ }\href@noop {}
  {\bibfield  {journal} {\bibinfo  {journal} {Phys. Rev. Lett.}\ }\textbf
  {\bibinfo {volume} {85}},\ \bibinfo {pages} {1946} (\bibinfo {year}
  {2000})}\BibitemShut {NoStop}%
\bibitem [{\citenamefont {Aono}\ and\ \citenamefont {Eto}(2001)}]{aono2001}%
  \BibitemOpen
  \bibfield  {author} {\bibinfo {author} {\bibfnamefont {T.}~\bibnamefont
  {Aono}}\ and\ \bibinfo {author} {\bibfnamefont {M.}~\bibnamefont {Eto}},\
  }\bibfield  {title} {\enquote {\bibinfo {title} {Kondo effect in coupled
  quantum dots under magnetic fields},}\ }\href@noop {} {\bibfield  {journal}
  {\bibinfo  {journal} {Phys. Rev. B}\ }\textbf {\bibinfo {volume} {64}},\
  \bibinfo {pages} {073307} (\bibinfo {year} {2001})}\BibitemShut {NoStop}%
\bibitem [{\citenamefont {L\'opez}\ \emph {et~al.}(2002)\citenamefont
  {L\'opez}, \citenamefont {Aguado},\ and\ \citenamefont
  {Platero}}]{lopez2002}%
  \BibitemOpen
  \bibfield  {author} {\bibinfo {author} {\bibfnamefont {R.}~\bibnamefont
  {L\'opez}}, \bibinfo {author} {\bibfnamefont {R.}~\bibnamefont {Aguado}}, \
  and\ \bibinfo {author} {\bibfnamefont {G.}~\bibnamefont {Platero}},\
  }\bibfield  {title} {\enquote {\bibinfo {title} {Nonequilibrium transport
  through double quantum dots: Kondo effect versus antiferromagnetic
  coupling},}\ }\href@noop {} {\bibfield  {journal} {\bibinfo  {journal} {Phys.
  Rev. Lett.}\ }\textbf {\bibinfo {volume} {89}},\ \bibinfo {pages} {136802}
  (\bibinfo {year} {2002})}\BibitemShut {NoStop}%
\bibitem [{\citenamefont {Simon}\ \emph {et~al.}(2005)\citenamefont {Simon},
  \citenamefont {L\'opez},\ and\ \citenamefont {Oreg}}]{simon2005}%
  \BibitemOpen
  \bibfield  {author} {\bibinfo {author} {\bibfnamefont {P.}~\bibnamefont
  {Simon}}, \bibinfo {author} {\bibfnamefont {R.}~\bibnamefont {L\'opez}}, \
  and\ \bibinfo {author} {\bibfnamefont {Y.}~\bibnamefont {Oreg}},\ }\bibfield
  {title} {\enquote {\bibinfo {title} {{Ruderman-Kittel-Kasuya-Yosida} and
  magnetic-field interactions in coupled {Kondo} quantum dots},}\ }\href@noop
  {} {\bibfield  {journal} {\bibinfo  {journal} {Phys. Rev. Lett.}\ }\textbf
  {\bibinfo {volume} {94}},\ \bibinfo {pages} {086602} (\bibinfo {year}
  {2005})}\BibitemShut {NoStop}%
\bibitem [{\citenamefont {Lee}\ \emph {et~al.}(2010)\citenamefont {Lee},
  \citenamefont {Choi}, \citenamefont {Lopez}, \citenamefont {Aguado},
  \citenamefont {Martinek},\ and\ \citenamefont {\v{Z}itko}}]{revisited}%
  \BibitemOpen
  \bibfield  {author} {\bibinfo {author} {\bibfnamefont {M.}~\bibnamefont
  {Lee}}, \bibinfo {author} {\bibfnamefont {M.-S.}\ \bibnamefont {Choi}},
  \bibinfo {author} {\bibfnamefont {R.}~\bibnamefont {Lopez}}, \bibinfo
  {author} {\bibfnamefont {R.}~\bibnamefont {Aguado}}, \bibinfo {author}
  {\bibfnamefont {J.}~\bibnamefont {Martinek}}, \ and\ \bibinfo {author}
  {\bibfnamefont {R.}~\bibnamefont {\v{Z}itko}},\ }\bibfield  {title} {\enquote
  {\bibinfo {title} {The two-impurity {Anderson} model revisited: Competition
  between {Kondo} effect and reservoir-mediated superexchange in double quantum
  dots},}\ }\href@noop {} {\bibfield  {journal} {\bibinfo  {journal} {Phys.
  Rev. B}\ }\textbf {\bibinfo {volume} {81}},\ \bibinfo {pages} {121311(R)}
  (\bibinfo {year} {2010})}\BibitemShut {NoStop}%
\bibitem [{\citenamefont {Bergeret}\ \emph {et~al.}(2006)\citenamefont
  {Bergeret}, \citenamefont {Yeyati},\ and\ \citenamefont
  {Mart{\'\i}n-Rodero}}]{bergeret2006}%
  \BibitemOpen
  \bibfield  {author} {\bibinfo {author} {\bibfnamefont {F.}~\bibnamefont
  {Bergeret}}, \bibinfo {author} {\bibfnamefont {A.}~\bibnamefont {Yeyati}}, \
  and\ \bibinfo {author} {\bibfnamefont {A.}~\bibnamefont
  {Mart{\'\i}n-Rodero}},\ }\bibfield  {title} {\enquote {\bibinfo {title}
  {Interplay between {Josephson} effect and magnetic interactions in double
  quantum dots},}\ }\href@noop {} {\bibfield  {journal} {\bibinfo  {journal}
  {Phys. Rev. B}\ }\textbf {\bibinfo {volume} {74}},\ \bibinfo {pages} {132505}
  (\bibinfo {year} {2006})}\BibitemShut {NoStop}%
\bibitem [{\citenamefont {\v{Z}itko}\ \emph {et~al.}(2010)\citenamefont
  {\v{Z}itko}, \citenamefont {Lee}, \citenamefont {Lopez}, \citenamefont
  {Aguado},\ and\ \citenamefont {Choi}}]{zitko2010}%
  \BibitemOpen
  \bibfield  {author} {\bibinfo {author} {\bibfnamefont {R.}~\bibnamefont
  {\v{Z}itko}}, \bibinfo {author} {\bibfnamefont {M.}~\bibnamefont {Lee}},
  \bibinfo {author} {\bibfnamefont {R.}~\bibnamefont {Lopez}}, \bibinfo
  {author} {\bibfnamefont {R.}~\bibnamefont {Aguado}}, \ and\ \bibinfo {author}
  {\bibfnamefont {M.-S.}\ \bibnamefont {Choi}},\ }\bibfield  {title} {\enquote
  {\bibinfo {title} {Josephson current in strongly correlated double quantum
  dots},}\ }\href@noop {} {\bibfield  {journal} {\bibinfo  {journal} {Phys.
  Rev. Lett.}\ }\textbf {\bibinfo {volume} {105}},\ \bibinfo {pages} {116803}
  (\bibinfo {year} {2010})}\BibitemShut {NoStop}%
\bibitem [{\citenamefont {\v{Z}itko}\ \emph {et~al.}(2011)\citenamefont
  {\v{Z}itko}, \citenamefont {Bodensiek},\ and\ \citenamefont
  {Pruschke}}]{dqdscaniso}%
  \BibitemOpen
  \bibfield  {author} {\bibinfo {author} {\bibfnamefont {R.}~\bibnamefont
  {\v{Z}itko}}, \bibinfo {author} {\bibfnamefont {O.}~\bibnamefont
  {Bodensiek}}, \ and\ \bibinfo {author} {\bibfnamefont {Th.}\ \bibnamefont
  {Pruschke}},\ }\bibfield  {title} {\enquote {\bibinfo {title} {Magnetic
  anisotropy effects on quantum impurities in superconducting host},}\
  }\href@noop {} {\bibfield  {journal} {\bibinfo  {journal} {Phys. Rev. B}\
  }\textbf {\bibinfo {volume} {83}},\ \bibinfo {pages} {054512} (\bibinfo
  {year} {2011})}\BibitemShut {NoStop}%
\bibitem [{\citenamefont {Yao}\ \emph {et~al.}(2014)\citenamefont {Yao},
  \citenamefont {Moca}, \citenamefont {Weymann}, \citenamefont {Sau},
  \citenamefont {Lukin}, \citenamefont {Demler},\ and\ \citenamefont
  {Zarand}}]{yao2014}%
  \BibitemOpen
  \bibfield  {author} {\bibinfo {author} {\bibfnamefont {N.~Y.}\ \bibnamefont
  {Yao}}, \bibinfo {author} {\bibfnamefont {C.~P.}\ \bibnamefont {Moca}},
  \bibinfo {author} {\bibfnamefont {I.}~\bibnamefont {Weymann}}, \bibinfo
  {author} {\bibfnamefont {J.~D.}\ \bibnamefont {Sau}}, \bibinfo {author}
  {\bibfnamefont {M.~D.}\ \bibnamefont {Lukin}}, \bibinfo {author}
  {\bibfnamefont {E.~A.}\ \bibnamefont {Demler}}, \ and\ \bibinfo {author}
  {\bibfnamefont {G.}~\bibnamefont {Zarand}},\ }\bibfield  {title} {\enquote
  {\bibinfo {title} {Phase diagram and excitations of a {Shiba} molecule},}\
  }\href@noop {} {\bibfield  {journal} {\bibinfo  {journal} {Phys. Rev. B}\
  }\textbf {\bibinfo {volume} {90}},\ \bibinfo {pages} {241108(R)} (\bibinfo
  {year} {2014})}\BibitemShut {NoStop}%
\bibitem [{\citenamefont {Galpin}\ \emph {et~al.}(2005)\citenamefont {Galpin},
  \citenamefont {Logan},\ and\ \citenamefont {Krishnamurthy}}]{galpin2005}%
  \BibitemOpen
  \bibfield  {author} {\bibinfo {author} {\bibfnamefont {M.~R.}\ \bibnamefont
  {Galpin}}, \bibinfo {author} {\bibfnamefont {D.~E.}\ \bibnamefont {Logan}}, \
  and\ \bibinfo {author} {\bibfnamefont {H.~R.}\ \bibnamefont
  {Krishnamurthy}},\ }\bibfield  {title} {\enquote {\bibinfo {title} {Quantum
  phase transition in capacitively coupled double quantum dot},}\ }\href@noop
  {} {\bibfield  {journal} {\bibinfo  {journal} {Phys. Rev. Lett.}\ }\textbf
  {\bibinfo {volume} {94}},\ \bibinfo {pages} {186406} (\bibinfo {year}
  {2005})}\BibitemShut {NoStop}%
\bibitem [{\citenamefont {Galpin}\ \emph {et~al.}(2006)\citenamefont {Galpin},
  \citenamefont {Logan},\ and\ \citenamefont {Krishnamurthy}}]{galpin2006b}%
  \BibitemOpen
  \bibfield  {author} {\bibinfo {author} {\bibfnamefont {Martin~R.}\
  \bibnamefont {Galpin}}, \bibinfo {author} {\bibfnamefont {David~E.}\
  \bibnamefont {Logan}}, \ and\ \bibinfo {author} {\bibfnamefont {H~R}\
  \bibnamefont {Krishnamurthy}},\ }\bibfield  {title} {\enquote {\bibinfo
  {title} {Dynamics of capacitively coupled double quantum dots},}\ }\href@noop
  {} {\bibfield  {journal} {\bibinfo  {journal} {J. Phys.: Condens. Matter}\
  }\textbf {\bibinfo {volume} {18}},\ \bibinfo {pages} {6571} (\bibinfo {year}
  {2006})}\BibitemShut {NoStop}%
\bibitem [{\citenamefont {Mravlje}\ \emph {et~al.}(2006)\citenamefont
  {Mravlje}, \citenamefont {Ram{\v s}ak},\ and\ \citenamefont
  {Rejec}}]{mravlje2006}%
  \BibitemOpen
  \bibfield  {author} {\bibinfo {author} {\bibfnamefont {J.}~\bibnamefont
  {Mravlje}}, \bibinfo {author} {\bibfnamefont {A.}~\bibnamefont {Ram{\v
  s}ak}}, \ and\ \bibinfo {author} {\bibfnamefont {T.}~\bibnamefont {Rejec}},\
  }\bibfield  {title} {\enquote {\bibinfo {title} {Kondo effect in double
  quantum dots with interdot repulsion},}\ }\href@noop {} {\bibfield  {journal}
  {\bibinfo  {journal} {Phys. Rev. B}\ }\textbf {\bibinfo {volume} {73}},\
  \bibinfo {pages} {241305(R)} (\bibinfo {year} {2006})}\BibitemShut {NoStop}%
\bibitem [{\citenamefont {Okazaki}\ \emph {et~al.}(2011)\citenamefont
  {Okazaki}, \citenamefont {Sasaki},\ and\ \citenamefont
  {Muraki}}]{okazaki2011}%
  \BibitemOpen
  \bibfield  {author} {\bibinfo {author} {\bibfnamefont {Yuma}\ \bibnamefont
  {Okazaki}}, \bibinfo {author} {\bibfnamefont {Satoshi}\ \bibnamefont
  {Sasaki}}, \ and\ \bibinfo {author} {\bibfnamefont {Koji}\ \bibnamefont
  {Muraki}},\ }\bibfield  {title} {\enquote {\bibinfo {title} {Spin-orbital
  kondo effect in a parallel double quantum dot},}\ }\href@noop {} {\bibfield
  {journal} {\bibinfo  {journal} {Phys. Rev. B}\ }\textbf {\bibinfo {volume}
  {84}},\ \bibinfo {pages} {161305} (\bibinfo {year} {2011})}\BibitemShut
  {NoStop}%
\bibitem [{\citenamefont {Nishikawa}\ \emph {et~al.}(2013)\citenamefont
  {Nishikawa}, \citenamefont {Hewson}, \citenamefont {Crow},\ and\
  \citenamefont {Bauer}}]{nishikawa2013}%
  \BibitemOpen
  \bibfield  {author} {\bibinfo {author} {\bibfnamefont {Yunori}\ \bibnamefont
  {Nishikawa}}, \bibinfo {author} {\bibfnamefont {Alex~C.}\ \bibnamefont
  {Hewson}}, \bibinfo {author} {\bibfnamefont {Daniel J.~G.}\ \bibnamefont
  {Crow}}, \ and\ \bibinfo {author} {\bibfnamefont {Johannes}\ \bibnamefont
  {Bauer}},\ }\bibfield  {title} {\enquote {\bibinfo {title} {Analysis of
  low-energy response and possible emergent su(4) kondo state in a double
  quantum dot},}\ }\href@noop {} {\bibfield  {journal} {\bibinfo  {journal}
  {Phys. Rev. B}\ }\textbf {\bibinfo {volume} {88}},\ \bibinfo {pages} {245130}
  (\bibinfo {year} {2013})}\BibitemShut {NoStop}%
\bibitem [{\citenamefont {Amasha}\ \emph {et~al.}(2013)\citenamefont {Amasha},
  \citenamefont {Keller}, \citenamefont {Rau}, \citenamefont {Carmi},
  \citenamefont {Katine}, \citenamefont {Shtrikman}, \citenamefont {Oreg},\
  and\ \citenamefont {Goldhaber-Gordon}}]{amasha2013}%
  \BibitemOpen
  \bibfield  {author} {\bibinfo {author} {\bibfnamefont {S.}~\bibnamefont
  {Amasha}}, \bibinfo {author} {\bibfnamefont {A.~J.}\ \bibnamefont {Keller}},
  \bibinfo {author} {\bibfnamefont {I.~G.}\ \bibnamefont {Rau}}, \bibinfo
  {author} {\bibfnamefont {A.}~\bibnamefont {Carmi}}, \bibinfo {author}
  {\bibfnamefont {J.~A.}\ \bibnamefont {Katine}}, \bibinfo {author}
  {\bibfnamefont {Hadas}\ \bibnamefont {Shtrikman}}, \bibinfo {author}
  {\bibfnamefont {Y.}~\bibnamefont {Oreg}}, \ and\ \bibinfo {author}
  {\bibfnamefont {D.}~\bibnamefont {Goldhaber-Gordon}},\ }\bibfield  {title}
  {\enquote {\bibinfo {title} {Pseudospin-resolved transport spectroscopy of
  the kondo effect in a double quantum dot},}\ }\href@noop {} {\bibfield
  {journal} {\bibinfo  {journal} {Phys. Rev. Lett.}\ }\textbf {\bibinfo
  {volume} {110}},\ \bibinfo {pages} {046604} (\bibinfo {year}
  {2013})}\BibitemShut {NoStop}%
\bibitem [{\citenamefont {Ruiz-Tijerina}\ \emph {et~al.}(2014)\citenamefont
  {Ruiz-Tijerina}, \citenamefont {Vernek},\ and\ \citenamefont
  {Ulloa}}]{PhysRevB.90.035119}%
  \BibitemOpen
  \bibfield  {author} {\bibinfo {author} {\bibfnamefont {David~A.}\
  \bibnamefont {Ruiz-Tijerina}}, \bibinfo {author} {\bibfnamefont
  {E.}~\bibnamefont {Vernek}}, \ and\ \bibinfo {author} {\bibfnamefont
  {Sergio~E.}\ \bibnamefont {Ulloa}},\ }\bibfield  {title} {\enquote {\bibinfo
  {title} {Capacitive interactions and kondo effect tuning in double quantum
  impurity systems},}\ }\href@noop {} {\bibfield  {journal} {\bibinfo
  {journal} {Phys. Rev. B}\ }\textbf {\bibinfo {volume} {90}},\ \bibinfo
  {pages} {035119} (\bibinfo {year} {2014})}\BibitemShut {NoStop}%
\bibitem [{\citenamefont {Keller}\ \emph {et~al.}(2014)\citenamefont {Keller},
  \citenamefont {Amasha}, \citenamefont {Weymann}, \citenamefont {Moca},
  \citenamefont {Rau}, \citenamefont {Katine}, \citenamefont {Shtrikman},
  \citenamefont {Zar\'and},\ and\ \citenamefont
  {Goldhaber-Gordon}}]{keller2014}%
  \BibitemOpen
  \bibfield  {author} {\bibinfo {author} {\bibfnamefont {A.~J.}\ \bibnamefont
  {Keller}}, \bibinfo {author} {\bibfnamefont {S.}~\bibnamefont {Amasha}},
  \bibinfo {author} {\bibfnamefont {I.}~\bibnamefont {Weymann}}, \bibinfo
  {author} {\bibfnamefont {C.~P.}\ \bibnamefont {Moca}}, \bibinfo {author}
  {\bibfnamefont {I.~G.}\ \bibnamefont {Rau}}, \bibinfo {author} {\bibfnamefont
  {J.~A.}\ \bibnamefont {Katine}}, \bibinfo {author} {\bibfnamefont
  {H.}~\bibnamefont {Shtrikman}}, \bibinfo {author} {\bibfnamefont
  {G.}~\bibnamefont {Zar\'and}}, \ and\ \bibinfo {author} {\bibfnamefont
  {D.}~\bibnamefont {Goldhaber-Gordon}},\ }\bibfield  {title} {\enquote
  {\bibinfo {title} {Emergent su(4) kondo physics in a spin–charge-entangled
  double quantum dot},}\ }\href@noop {} {\bibfield  {journal} {\bibinfo
  {journal} {Nat. Physics}\ }\textbf {\bibinfo {volume} {10}},\ \bibinfo
  {pages} {145} (\bibinfo {year} {2014})}\BibitemShut {NoStop}%
\bibitem [{\citenamefont {Filippone}\ \emph {et~al.}(2014)\citenamefont
  {Filippone}, \citenamefont {Moca}, \citenamefont {Zar\'and},\ and\
  \citenamefont {Mora}}]{filippone2014}%
  \BibitemOpen
  \bibfield  {author} {\bibinfo {author} {\bibfnamefont {Michele}\ \bibnamefont
  {Filippone}}, \bibinfo {author} {\bibfnamefont {C.~P.}\ \bibnamefont {Moca}},
  \bibinfo {author} {\bibfnamefont {Gergely}\ \bibnamefont {Zar\'and}}, \ and\
  \bibinfo {author} {\bibfnamefont {Christophe}\ \bibnamefont {Mora}},\
  }\bibfield  {title} {\enquote {\bibinfo {title} {Kondo temperature of su(4)
  symmetric quantum dots},}\ }\href@noop {} {\bibfield  {journal} {\bibinfo
  {journal} {Phys. Rev. B}\ }\textbf {\bibinfo {volume} {90}},\ \bibinfo
  {pages} {121406} (\bibinfo {year} {2014})}\BibitemShut {NoStop}%
\bibitem [{sup()}]{suppl}%
  \BibitemOpen
  \href@noop {} {\enquote {\bibinfo {title} {Supplementary information},}\
  }\BibitemShut {NoStop}%
\bibitem [{\citenamefont {Krishna-murthy}\ \emph
  {et~al.}(1980{\natexlab{a}})\citenamefont {Krishna-murthy}, \citenamefont
  {Wilkins},\ and\ \citenamefont {Wilson}}]{krishna1980a}%
  \BibitemOpen
  \bibfield  {author} {\bibinfo {author} {\bibfnamefont {H.~R.}\ \bibnamefont
  {Krishna-murthy}}, \bibinfo {author} {\bibfnamefont {J.~W.}\ \bibnamefont
  {Wilkins}}, \ and\ \bibinfo {author} {\bibfnamefont {K.~G.}\ \bibnamefont
  {Wilson}},\ }\bibfield  {title} {\enquote {\bibinfo {title}
  {Renormalization-group approach to the {Anderson} model of dilute magnetic
  alloys. {I.} {S}tatic properties for the symmetric case},}\ }\href@noop {}
  {\bibfield  {journal} {\bibinfo  {journal} {Phys. Rev. B}\ }\textbf {\bibinfo
  {volume} {21}},\ \bibinfo {pages} {1003} (\bibinfo {year}
  {1980}{\natexlab{a}})}\BibitemShut {NoStop}%
\bibitem [{\citenamefont {Wilson}(1975)}]{wilson1975}%
  \BibitemOpen
  \bibfield  {author} {\bibinfo {author} {\bibfnamefont {K.~G.}\ \bibnamefont
  {Wilson}},\ }\bibfield  {title} {\enquote {\bibinfo {title} {The
  renormalization group: {Critical} phenomena and the {Kondo} problem},}\
  }\href@noop {} {\bibfield  {journal} {\bibinfo  {journal} {Rev. Mod. Phys.}\
  }\textbf {\bibinfo {volume} {47}},\ \bibinfo {pages} {773} (\bibinfo {year}
  {1975})}\BibitemShut {NoStop}%
\bibitem [{\citenamefont {Krishna-murthy}\ \emph
  {et~al.}(1980{\natexlab{b}})\citenamefont {Krishna-murthy}, \citenamefont
  {Wilkins},\ and\ \citenamefont {Wilson}}]{krishna1980b}%
  \BibitemOpen
  \bibfield  {author} {\bibinfo {author} {\bibfnamefont {H.~R.}\ \bibnamefont
  {Krishna-murthy}}, \bibinfo {author} {\bibfnamefont {J.~W.}\ \bibnamefont
  {Wilkins}}, \ and\ \bibinfo {author} {\bibfnamefont {K.~G.}\ \bibnamefont
  {Wilson}},\ }\bibfield  {title} {\enquote {\bibinfo {title}
  {Renormalization-group approach to the {Anderson} model of dilute magnetic
  alloys. {II.} {S}tatic properties for the asymmetric case},}\ }\href@noop {}
  {\bibfield  {journal} {\bibinfo  {journal} {Phys. Rev. B}\ }\textbf {\bibinfo
  {volume} {21}},\ \bibinfo {pages} {1044} (\bibinfo {year}
  {1980}{\natexlab{b}})}\BibitemShut {NoStop}%
\bibitem [{\citenamefont {Satori}\ \emph {et~al.}(1992)\citenamefont {Satori},
  \citenamefont {Shiba}, \citenamefont {Sakai},\ and\ \citenamefont
  {Shimizu}}]{satori1992}%
  \BibitemOpen
  \bibfield  {author} {\bibinfo {author} {\bibfnamefont {Koji}\ \bibnamefont
  {Satori}}, \bibinfo {author} {\bibfnamefont {Hiroyuki}\ \bibnamefont
  {Shiba}}, \bibinfo {author} {\bibfnamefont {Osamu}\ \bibnamefont {Sakai}}, \
  and\ \bibinfo {author} {\bibfnamefont {Yukihiro}\ \bibnamefont {Shimizu}},\
  }\bibfield  {title} {\enquote {\bibinfo {title} {Numerical renormalization
  group study of magnetic impurities in superconductors},}\ }\href@noop {}
  {\bibfield  {journal} {\bibinfo  {journal} {J. Phys. Soc. Japan}\ }\textbf
  {\bibinfo {volume} {61}},\ \bibinfo {pages} {3239} (\bibinfo {year}
  {1992})}\BibitemShut {NoStop}%
\bibitem [{\citenamefont {Sakai}\ \emph {et~al.}(1993)\citenamefont {Sakai},
  \citenamefont {Shimizu}, \citenamefont {Shiba},\ and\ \citenamefont
  {Satori}}]{sakai1993}%
  \BibitemOpen
  \bibfield  {author} {\bibinfo {author} {\bibfnamefont {Osamu}\ \bibnamefont
  {Sakai}}, \bibinfo {author} {\bibfnamefont {Yukihiro}\ \bibnamefont
  {Shimizu}}, \bibinfo {author} {\bibfnamefont {Hiroyuki}\ \bibnamefont
  {Shiba}}, \ and\ \bibinfo {author} {\bibfnamefont {Koji}\ \bibnamefont
  {Satori}},\ }\bibfield  {title} {\enquote {\bibinfo {title} {Numerical
  renormalization group study of magnetic impurities in supercoductors. {II.}
  {D}ynamical excitations spectra and spatial variation of the order
  parameter},}\ }\href@noop {} {\bibfield  {journal} {\bibinfo  {journal} {J.
  Phys. Soc. Japan}\ }\textbf {\bibinfo {volume} {62}},\ \bibinfo {pages}
  {3181} (\bibinfo {year} {1993})}\BibitemShut {NoStop}%
\bibitem [{\citenamefont {Yoshioka}\ and\ \citenamefont
  {Ohashi}(1998)}]{yoshioka1998}%
  \BibitemOpen
  \bibfield  {author} {\bibinfo {author} {\bibfnamefont {Tomoki}\ \bibnamefont
  {Yoshioka}}\ and\ \bibinfo {author} {\bibfnamefont {Yoji}\ \bibnamefont
  {Ohashi}},\ }\bibfield  {title} {\enquote {\bibinfo {title} {Ground state
  properties and localized excited states around a magnetic impurity described
  by the anisotropic $s$-$d$ interaction in superconductivity},}\ }\href@noop
  {} {\bibfield  {journal} {\bibinfo  {journal} {J. Phys. Soc. Japan}\ }\textbf
  {\bibinfo {volume} {67}},\ \bibinfo {pages} {1332} (\bibinfo {year}
  {1998})}\BibitemShut {NoStop}%
\bibitem [{\citenamefont {Yoshioka}\ and\ \citenamefont
  {Ohashi}(2000)}]{yoshioka2000}%
  \BibitemOpen
  \bibfield  {author} {\bibinfo {author} {\bibfnamefont {Tomoki}\ \bibnamefont
  {Yoshioka}}\ and\ \bibinfo {author} {\bibfnamefont {Yoji}\ \bibnamefont
  {Ohashi}},\ }\bibfield  {title} {\enquote {\bibinfo {title} {Numerical
  renormalization group studies on single impurity anderson model in
  superconductivity: a unified treatment of magnetic, nonmagnetic impurities,
  and resonance scattering},}\ }\href@noop {} {\bibfield  {journal} {\bibinfo
  {journal} {J. Phys. Soc. Japan}\ }\textbf {\bibinfo {volume} {69}},\ \bibinfo
  {pages} {1812} (\bibinfo {year} {2000})}\BibitemShut {NoStop}%
\bibitem [{\citenamefont {Oguri}\ \emph {et~al.}(2004)\citenamefont {Oguri},
  \citenamefont {Tanaka},\ and\ \citenamefont {Hewson}}]{oguri2004josephson}%
  \BibitemOpen
  \bibfield  {author} {\bibinfo {author} {\bibfnamefont {Akira}\ \bibnamefont
  {Oguri}}, \bibinfo {author} {\bibfnamefont {Yoshihide}\ \bibnamefont
  {Tanaka}}, \ and\ \bibinfo {author} {\bibfnamefont {A.~C.}\ \bibnamefont
  {Hewson}},\ }\bibfield  {title} {\enquote {\bibinfo {title} {Quantum phase
  transition in a minimal model for the {Kondo} effect in a {Josephson}
  junction},}\ }\href@noop {} {\bibfield  {journal} {\bibinfo  {journal} {J.
  Phys. Soc. Japan}\ }\textbf {\bibinfo {volume} {73}},\ \bibinfo {pages}
  {2494} (\bibinfo {year} {2004})}\BibitemShut {NoStop}%
\bibitem [{\citenamefont {Hecht}\ \emph {et~al.}(2008)\citenamefont {Hecht},
  \citenamefont {Weichselbaum}, \citenamefont {von Delft},\ and\ \citenamefont
  {Bulla}}]{hecht2008}%
  \BibitemOpen
  \bibfield  {author} {\bibinfo {author} {\bibfnamefont {T.}~\bibnamefont
  {Hecht}}, \bibinfo {author} {\bibfnamefont {A.}~\bibnamefont {Weichselbaum}},
  \bibinfo {author} {\bibfnamefont {J.}~\bibnamefont {von Delft}}, \ and\
  \bibinfo {author} {\bibfnamefont {R.}~\bibnamefont {Bulla}},\ }\bibfield
  {title} {\enquote {\bibinfo {title} {Numerical renormalization group
  calculation of near-gap peaks in spectral functions of the {Anderson} model
  with superconducting leads},}\ }\href@noop {} {\bibfield  {journal} {\bibinfo
   {journal} {J. Phys. Condens. Mat.}\ }\textbf {\bibinfo {volume} {20}},\
  \bibinfo {pages} {275213} (\bibinfo {year} {2008})}\BibitemShut {NoStop}%
\bibitem [{\citenamefont {Bulla}\ \emph {et~al.}(2008)\citenamefont {Bulla},
  \citenamefont {Costi},\ and\ \citenamefont {Pruschke}}]{bulla2008}%
  \BibitemOpen
  \bibfield  {author} {\bibinfo {author} {\bibfnamefont {Ralf}\ \bibnamefont
  {Bulla}}, \bibinfo {author} {\bibfnamefont {Theo}\ \bibnamefont {Costi}}, \
  and\ \bibinfo {author} {\bibfnamefont {Thomas}\ \bibnamefont {Pruschke}},\
  }\bibfield  {title} {\enquote {\bibinfo {title} {The numerical
  renormalization group method for quantum impurity systems},}\ }\href@noop {}
  {\bibfield  {journal} {\bibinfo  {journal} {Rev. Mod. Phys.}\ }\textbf
  {\bibinfo {volume} {80}},\ \bibinfo {pages} {395} (\bibinfo {year}
  {2008})}\BibitemShut {NoStop}%
\bibitem [{\citenamefont {Lim}\ and\ \citenamefont {Choi}(2008)}]{Lim:2008bc}%
  \BibitemOpen
  \bibfield  {author} {\bibinfo {author} {\bibfnamefont {Jong~Soo}\
  \bibnamefont {Lim}}\ and\ \bibinfo {author} {\bibfnamefont {Mahn-Soo}\
  \bibnamefont {Choi}},\ }\bibfield  {title} {\enquote {\bibinfo {title}
  {{Andreev bound states in the Kondo quantum dots coupled to superconducting
  leads}},}\ }\href@noop {} {\bibfield  {journal} {\bibinfo  {journal} {Journal
  of Physics: Condensed Matter}\ }\textbf {\bibinfo {volume} {20}},\ \bibinfo
  {pages} {415225} (\bibinfo {year} {2008})}\BibitemShut {NoStop}%
\bibitem [{\citenamefont {Deacon}\ \emph {et~al.}(2010)\citenamefont {Deacon},
  \citenamefont {Tanaka}, \citenamefont {Oiwa}, \citenamefont {Sakano},
  \citenamefont {Yoshida}, \citenamefont {Shibata}, \citenamefont {Hirakawa},\
  and\ \citenamefont {Tarucha}}]{Deacon:2010jn}%
  \BibitemOpen
  \bibfield  {author} {\bibinfo {author} {\bibfnamefont {R~S}\ \bibnamefont
  {Deacon}}, \bibinfo {author} {\bibfnamefont {Y}~\bibnamefont {Tanaka}},
  \bibinfo {author} {\bibfnamefont {A}~\bibnamefont {Oiwa}}, \bibinfo {author}
  {\bibfnamefont {R}~\bibnamefont {Sakano}}, \bibinfo {author} {\bibfnamefont
  {K}~\bibnamefont {Yoshida}}, \bibinfo {author} {\bibfnamefont
  {K}~\bibnamefont {Shibata}}, \bibinfo {author} {\bibfnamefont
  {K}~\bibnamefont {Hirakawa}}, \ and\ \bibinfo {author} {\bibfnamefont
  {S}~\bibnamefont {Tarucha}},\ }\bibfield  {title} {\enquote {\bibinfo {title}
  {{Interplay of Kondo and superconducting correlations in the nonequilibrium
  Andreev transport through a quantum dot}},}\ }\href@noop {} {\bibfield
  {journal} {\bibinfo  {journal} {Physical Review Letters}\ }\textbf {\bibinfo
  {volume} {104}},\ \bibinfo {pages} {076805} (\bibinfo {year}
  {2010})}\BibitemShut {NoStop}%
\bibitem [{\citenamefont {Mart{\'\i}n-Rodero}\ and\ \citenamefont
  {Levy~Yeyati}(2012)}]{MartinRodero:2012fd}%
  \BibitemOpen
  \bibfield  {author} {\bibinfo {author} {\bibfnamefont {A}~\bibnamefont
  {Mart{\'\i}n-Rodero}}\ and\ \bibinfo {author} {\bibfnamefont {A}~\bibnamefont
  {Levy~Yeyati}},\ }\bibfield  {title} {\enquote {\bibinfo {title} {{The
  Andreev states of a superconducting quantum dot: mean field versus exact
  numerical results}},}\ }\href@noop {} {\bibfield  {journal} {\bibinfo
  {journal} {Journal of Physics: Condensed Matter}\ }\textbf {\bibinfo {volume}
  {24}},\ \bibinfo {pages} {385303} (\bibinfo {year} {2012})}\BibitemShut
  {NoStop}%
\bibitem [{\citenamefont {Kumar}\ \emph {et~al.}(2014)\citenamefont {Kumar},
  \citenamefont {Gaim}, \citenamefont {Steininger}, \citenamefont {Yeyati},
  \citenamefont {Mart{\'\i}n-Rodero}, \citenamefont {H{\"u}ttel},\ and\
  \citenamefont {Strunk}}]{Kumar:2014cq}%
  \BibitemOpen
  \bibfield  {author} {\bibinfo {author} {\bibfnamefont {A}~\bibnamefont
  {Kumar}}, \bibinfo {author} {\bibfnamefont {M}~\bibnamefont {Gaim}}, \bibinfo
  {author} {\bibfnamefont {D}~\bibnamefont {Steininger}}, \bibinfo {author}
  {\bibfnamefont {A~Levy}\ \bibnamefont {Yeyati}}, \bibinfo {author}
  {\bibfnamefont {A}~\bibnamefont {Mart{\'\i}n-Rodero}}, \bibinfo {author}
  {\bibfnamefont {A~K}\ \bibnamefont {H{\"u}ttel}}, \ and\ \bibinfo {author}
  {\bibfnamefont {C}~\bibnamefont {Strunk}},\ }\bibfield  {title} {\enquote
  {\bibinfo {title} {{Temperature dependence of Andreev spectra in a
  superconducting carbon nanotube quantum dot}},}\ }\href@noop {} {\bibfield
  {journal} {\bibinfo  {journal} {Physical Review B}\ }\textbf {\bibinfo
  {volume} {89}},\ \bibinfo {pages} {075428} (\bibinfo {year}
  {2014})}\BibitemShut {NoStop}%
\bibitem [{\citenamefont {Oliveira}\ and\ \citenamefont
  {Oliveira}(1994)}]{oliveira1994}%
  \BibitemOpen
  \bibfield  {author} {\bibinfo {author} {\bibfnamefont {W.~C.}\ \bibnamefont
  {Oliveira}}\ and\ \bibinfo {author} {\bibfnamefont {L.~N.}\ \bibnamefont
  {Oliveira}},\ }\bibfield  {title} {\enquote {\bibinfo {title} {Generalized
  numerical renormalization-group method to calculate the thermodynamical
  properties of impurities in metals},}\ }\href@noop {} {\bibfield  {journal}
  {\bibinfo  {journal} {Phys. Rev. B}\ }\textbf {\bibinfo {volume} {49}},\
  \bibinfo {pages} {11986} (\bibinfo {year} {1994})}\BibitemShut {NoStop}%
\bibitem [{\citenamefont {Silva}\ \emph {et~al.}(1996)\citenamefont {Silva},
  \citenamefont {Lima}, \citenamefont {Oliveira}, \citenamefont {Mello},
  \citenamefont {Oliveira},\ and\ \citenamefont {Wilkins}}]{silva1996}%
  \BibitemOpen
  \bibfield  {author} {\bibinfo {author} {\bibfnamefont {J.~B.}\ \bibnamefont
  {Silva}}, \bibinfo {author} {\bibfnamefont {W.~L.~C.}\ \bibnamefont {Lima}},
  \bibinfo {author} {\bibfnamefont {W.~C.}\ \bibnamefont {Oliveira}}, \bibinfo
  {author} {\bibfnamefont {J.~L.~N.}\ \bibnamefont {Mello}}, \bibinfo {author}
  {\bibfnamefont {L.~N.}\ \bibnamefont {Oliveira}}, \ and\ \bibinfo {author}
  {\bibfnamefont {J.~W.}\ \bibnamefont {Wilkins}},\ }\bibfield  {title}
  {\enquote {\bibinfo {title} {Particle-hole assymetry in the two-impurity
  {Kondo} model},}\ }\href@noop {} {\bibfield  {journal} {\bibinfo  {journal}
  {Phys. Rev. Lett.}\ }\textbf {\bibinfo {volume} {76}},\ \bibinfo {pages}
  {275} (\bibinfo {year} {1996})}\BibitemShut {NoStop}%
\bibitem [{\citenamefont {Paula}\ \emph {et~al.}(1999)\citenamefont {Paula},
  \citenamefont {Silva},\ and\ \citenamefont {Oliveira}}]{paula1999}%
  \BibitemOpen
  \bibfield  {author} {\bibinfo {author} {\bibfnamefont {C.~A.}\ \bibnamefont
  {Paula}}, \bibinfo {author} {\bibfnamefont {M.~F.}\ \bibnamefont {Silva}}, \
  and\ \bibinfo {author} {\bibfnamefont {L.~N.}\ \bibnamefont {Oliveira}},\
  }\bibfield  {title} {\enquote {\bibinfo {title} {Low-energy spectral density
  for the {Alexander}-{Anderson} model},}\ }\href@noop {} {\bibfield  {journal}
  {\bibinfo  {journal} {Phys. Rev. B}\ }\textbf {\bibinfo {volume} {59}},\
  \bibinfo {pages} {85} (\bibinfo {year} {1999})}\BibitemShut {NoStop}%
\bibitem [{\citenamefont {Campo}\ and\ \citenamefont
  {Oliveira}(2005)}]{campo2005}%
  \BibitemOpen
  \bibfield  {author} {\bibinfo {author} {\bibfnamefont {V.~L.}\ \bibnamefont
  {Campo}}\ and\ \bibinfo {author} {\bibfnamefont {L.~N.}\ \bibnamefont
  {Oliveira}},\ }\bibfield  {title} {\enquote {\bibinfo {title} {Alternative
  discretization in the numerical renormalization group},}\ }\href@noop {}
  {\bibfield  {journal} {\bibinfo  {journal} {Phys. Rev. B}\ }\textbf {\bibinfo
  {volume} {72}},\ \bibinfo {pages} {104432} (\bibinfo {year}
  {2005})}\BibitemShut {NoStop}%
\bibitem [{\citenamefont {Hofstetter}(2000)}]{hofstetter2000}%
  \BibitemOpen
  \bibfield  {author} {\bibinfo {author} {\bibfnamefont {Walter}\ \bibnamefont
  {Hofstetter}},\ }\bibfield  {title} {\enquote {\bibinfo {title} {Generalized
  numerical renormalization group for dynamical quantities},}\ }\href@noop {}
  {\bibfield  {journal} {\bibinfo  {journal} {Phys. Rev. Lett.}\ }\textbf
  {\bibinfo {volume} {85}},\ \bibinfo {pages} {1508} (\bibinfo {year}
  {2000})}\BibitemShut {NoStop}%
\bibitem [{\citenamefont {\v{Z}itko}\ \emph {et~al.}(2015)\citenamefont
  {\v{Z}itko}, \citenamefont {Lim}, \citenamefont {Lopez},\ and\ \citenamefont
  {Aguado}}]{zitko2015shiba}%
  \BibitemOpen
  \bibfield  {author} {\bibinfo {author} {\bibfnamefont {R.}~\bibnamefont
  {\v{Z}itko}}, \bibinfo {author} {\bibfnamefont {Jong~Soo}\ \bibnamefont
  {Lim}}, \bibinfo {author} {\bibfnamefont {Rosa}\ \bibnamefont {Lopez}}, \
  and\ \bibinfo {author} {\bibfnamefont {Ramon}\ \bibnamefont {Aguado}},\
  }\bibfield  {title} {\enquote {\bibinfo {title} {Shiba states and zero-bias
  anomalies in the hybrid normal-superconductor {Anderson} model},}\
  }\href@noop {} {\bibfield  {journal} {\bibinfo  {journal} {Phys. Rev. B}\
  }\textbf {\bibinfo {volume} {91}},\ \bibinfo {pages} {045441} (\bibinfo
  {year} {2015})}\BibitemShut {NoStop}%
\bibitem [{\citenamefont {Koerting}\ \emph {et~al.}(2010)\citenamefont
  {Koerting}, \citenamefont {Andersen}, \citenamefont {Flensberg},\ and\
  \citenamefont {Paaske}}]{koerting2010}%
  \BibitemOpen
  \bibfield  {author} {\bibinfo {author} {\bibfnamefont {V.}~\bibnamefont
  {Koerting}}, \bibinfo {author} {\bibfnamefont {B.~M.}\ \bibnamefont
  {Andersen}}, \bibinfo {author} {\bibfnamefont {K.}~\bibnamefont {Flensberg}},
  \ and\ \bibinfo {author} {\bibfnamefont {J.}~\bibnamefont {Paaske}},\
  }\bibfield  {title} {\enquote {\bibinfo {title} {Nonequilibrium transport via
  spin-induced subgap states in superconductor/quantum dot/normal metal
  cotunnel junctions},}\ }\href@noop {} {\bibfield  {journal} {\bibinfo
  {journal} {Phys. Rev. B}\ }\textbf {\bibinfo {volume} {82}},\ \bibinfo
  {pages} {245108} (\bibinfo {year} {2010})}\BibitemShut {NoStop}%
\bibitem [{\citenamefont {Yamada}\ \emph {et~al.}(2011)\citenamefont {Yamada},
  \citenamefont {Tanaka},\ and\ \citenamefont {Kawakami}}]{yamada2011}%
  \BibitemOpen
  \bibfield  {author} {\bibinfo {author} {\bibfnamefont {Y.}~\bibnamefont
  {Yamada}}, \bibinfo {author} {\bibfnamefont {Y.}~\bibnamefont {Tanaka}}, \
  and\ \bibinfo {author} {\bibfnamefont {N.}~\bibnamefont {Kawakami}},\
  }\bibfield  {title} {\enquote {\bibinfo {title} {Interplay of {Kondo} and
  superconducting correlations in the nonequilibrium {Andreev} transport
  through a quantum dot},}\ }\href@noop {} {\bibfield  {journal} {\bibinfo
  {journal} {Phys. Rev. B}\ }\textbf {\bibinfo {volume} {84}},\ \bibinfo
  {pages} {075484} (\bibinfo {year} {2011})}\BibitemShut {NoStop}%
\bibitem [{\citenamefont {Koga}(2013)}]{koga2013}%
  \BibitemOpen
  \bibfield  {author} {\bibinfo {author} {\bibfnamefont {Akihisa}\ \bibnamefont
  {Koga}},\ }\bibfield  {title} {\enquote {\bibinfo {title} {Quantum monte
  carlo study of nonequilibrium transport through a quantum dot coupled to
  normal and superconducting leads},}\ }\href@noop {} {\bibfield  {journal}
  {\bibinfo  {journal} {Phys. Rev. B}\ }\textbf {\bibinfo {volume} {87}},\
  \bibinfo {pages} {115409} (\bibinfo {year} {2013})}\BibitemShut {NoStop}%
\bibitem [{\citenamefont {Rozhkov}\ and\ \citenamefont
  {Arovas}(2000)}]{PhysRevB.62.6687}%
  \BibitemOpen
  \bibfield  {author} {\bibinfo {author} {\bibfnamefont {A.}~\bibnamefont
  {Rozhkov}}\ and\ \bibinfo {author} {\bibfnamefont {Daniel}\ \bibnamefont
  {Arovas}},\ }\bibfield  {title} {\enquote {\bibinfo {title}
  {Interacting-impurity {Josephson} junction: Variational wave functions and
  slave-boson mean-field theory},}\ }\href@noop {} {\bibfield  {journal}
  {\bibinfo  {journal} {Phys. Rev. B}\ }\textbf {\bibinfo {volume} {62}},\
  \bibinfo {pages} {6687--6691} (\bibinfo {year} {2000})}\BibitemShut {NoStop}%
\bibitem [{\citenamefont {Vecino}\ \emph {et~al.}(2003)\citenamefont {Vecino},
  \citenamefont {Mart{\'\i}n-Rodero},\ and\ \citenamefont
  {Yeyati}}]{vecino2003}%
  \BibitemOpen
  \bibfield  {author} {\bibinfo {author} {\bibfnamefont {E}~\bibnamefont
  {Vecino}}, \bibinfo {author} {\bibfnamefont {A}~\bibnamefont
  {Mart{\'\i}n-Rodero}}, \ and\ \bibinfo {author} {\bibfnamefont
  {A}~\bibnamefont {Yeyati}},\ }\bibfield  {title} {\enquote {\bibinfo {title}
  {{Josephson current through a correlated quantum level: Andreev states and
  $\pi$ junction behavior}},}\ }\href@noop {} {\bibfield  {journal} {\bibinfo
  {journal} {Phys. Rev. B}\ }\textbf {\bibinfo {volume} {68}},\ \bibinfo
  {pages} {035105} (\bibinfo {year} {2003})}\BibitemShut {NoStop}%
\bibitem [{\citenamefont {Bergeret}\ \emph {et~al.}(2007)\citenamefont
  {Bergeret}, \citenamefont {Yeyati},\ and\ \citenamefont
  {Mart{\'\i}n-Rodero}}]{bergeret2007}%
  \BibitemOpen
  \bibfield  {author} {\bibinfo {author} {\bibfnamefont {F}~\bibnamefont
  {Bergeret}}, \bibinfo {author} {\bibfnamefont {A}~\bibnamefont {Yeyati}}, \
  and\ \bibinfo {author} {\bibfnamefont {A}~\bibnamefont
  {Mart{\'\i}n-Rodero}},\ }\bibfield  {title} {\enquote {\bibinfo {title}
  {Josephson effect through a quantum dot array},}\ }\href@noop {} {\bibfield
  {journal} {\bibinfo  {journal} {Phys. Rev. B}\ }\textbf {\bibinfo {volume}
  {76}},\ \bibinfo {pages} {174510} (\bibinfo {year} {2007})}\BibitemShut
  {NoStop}%
\bibitem [{\citenamefont {Mravlje}\ \emph {et~al.}(2005)\citenamefont
  {Mravlje}, \citenamefont {Ram{\v s}ak},\ and\ \citenamefont
  {Rejec}}]{mravlje2005}%
  \BibitemOpen
  \bibfield  {author} {\bibinfo {author} {\bibfnamefont {J.}~\bibnamefont
  {Mravlje}}, \bibinfo {author} {\bibfnamefont {A.}~\bibnamefont {Ram{\v
  s}ak}}, \ and\ \bibinfo {author} {\bibfnamefont {T.}~\bibnamefont {Rejec}},\
  }\bibfield  {title} {\enquote {\bibinfo {title} {Conductance of deformable
  molecules with interaction},}\ }\href@noop {} {\bibfield  {journal} {\bibinfo
   {journal} {Phys. Rev. B}\ }\textbf {\bibinfo {volume} {72}},\ \bibinfo
  {pages} {121403(R)} (\bibinfo {year} {2005})}\BibitemShut {NoStop}%
\bibitem [{\citenamefont {{\v Z}itko}\ and\ \citenamefont {Bon{\v
  c}a}(2006)}]{vzporedne}%
  \BibitemOpen
  \bibfield  {author} {\bibinfo {author} {\bibfnamefont {Rok}\ \bibnamefont
  {{\v Z}itko}}\ and\ \bibinfo {author} {\bibfnamefont {Janez}\ \bibnamefont
  {Bon{\v c}a}},\ }\bibfield  {title} {\enquote {\bibinfo {title}
  {Multi-impurity anderson model for quantum dots coupled in parallel},}\
  }\href@noop {} {\bibfield  {journal} {\bibinfo  {journal} {Phys. Rev. B}\
  }\textbf {\bibinfo {volume} {74}},\ \bibinfo {pages} {045312} (\bibinfo
  {year} {2006})}\BibitemShut {NoStop}%
\bibitem [{\citenamefont {Zar{\'a}nd}\ \emph {et~al.}(2006)\citenamefont
  {Zar{\'a}nd}, \citenamefont {Chung}, \citenamefont {Simon},\ and\
  \citenamefont {Vojta}}]{zarand2006}%
  \BibitemOpen
  \bibfield  {author} {\bibinfo {author} {\bibfnamefont {Gergely}\ \bibnamefont
  {Zar{\'a}nd}}, \bibinfo {author} {\bibfnamefont {Chung-Hou}\ \bibnamefont
  {Chung}}, \bibinfo {author} {\bibfnamefont {Pascal}\ \bibnamefont {Simon}}, \
  and\ \bibinfo {author} {\bibfnamefont {Matthias}\ \bibnamefont {Vojta}},\
  }\bibfield  {title} {\enquote {\bibinfo {title} {Quantum criticality in a
  double quantum-dot system},}\ }\href@noop {} {\bibfield  {journal} {\bibinfo
  {journal} {Phys. Rev. Lett.}\ }\textbf {\bibinfo {volume} {97}},\ \bibinfo
  {pages} {166802} (\bibinfo {year} {2006})}\BibitemShut {NoStop}%
\end{thebibliography}%

\end{document}